\documentclass[reprint,
preprintnumbers,
nofootinbib,
amsmath,amssymb,
aps,
prd,
superscriptaddress,
]{revtex4-2}

\usepackage{graphicx}
\usepackage[range-phrase=\text{--}]{siunitx}
\usepackage{braket}
\usepackage{bm}
\usepackage[hidelinks]{hyperref}
\usepackage[all]{hypcap}
\usepackage{xcolor}
\usepackage{enumitem}
\usepackage{multirow}
\usepackage{orcidlink}

\hypersetup{
	colorlinks,
	linkcolor={red!50!black},
	citecolor={blue!50!black},
	urlcolor={blue!80!black}
}

\newcommand{\order}[1]{\mathcal{O}(#1)}

\DeclareMathOperator{\tr}{tr}

\begin{document}

\preprint{ADP26-09/T1306}

\title{Understanding the structure of nucleon excitations from their wavefunctions}

\author{Jackson A. Mickley \orcidlink{0000-0001-5294-2823}}
\affiliation{Centre for the Subatomic Structure of Matter, Department of Physics, Adelaide University, South Australia 5005, Australia}

\author{Waseem Kamleh \orcidlink{0000-0002-6177-5366}}
\affiliation{Centre for the Subatomic Structure of Matter, Department of Physics, Adelaide University, South Australia 5005, Australia}

\author{Derek B. Leinweber \orcidlink{0000-0002-4745-6027}}
\affiliation{Centre for the Subatomic Structure of Matter, Department of Physics, Adelaide University, South Australia 5005, Australia}

\author{Finn M. Stokes \orcidlink{0000-0003-1763-8847}}
\affiliation{Centre for the Subatomic Structure of Matter, Department of Physics, Adelaide University, South Australia 5005, Australia}
\affiliation{J{\"u}lich Supercomputing Centre, Institute for Advanced Simulation, Forschungszentrum J{\"u}lich, J{\"u}lich D-52425, Germany}

\begin{abstract}
Relativistic wavefunctions of nucleon excitations are scrutinised to understand their node structure and the underlying role of local interpolating fields in generating the nucleon spectrum. In addressing quark model perspectives, approximately 4000 propagators are employed on the heaviest PACS-CS ensemble at \(m_\pi \simeq \qty{702}{\MeV}\). We examine the ground and four lowest-lying excited states at zero momentum for both positive- and negative-parity spectra, where the proton's \(d\)-quark wavefunction is calculated about the two \(u\) quarks at the origin. This is achieved using two local interpolating fields that each carry the quantum numbers of the nucleon but with differing spin-flavour structures, one of which vanishes in the nonrelativistic limit. We find that two distinct types of wavefunction nodes are manifest:\ ``superposition nodes'' formed through a linear combination of interpolating fields, and novel ``built-in nodes'' that are fundamentally built in to the \(s\)-wave Dirac components of an individual interpolating field. These are investigated qualitatively through visualisations in the form of both volume and surface renderings, and quantitatively by the calculation of radial wavefunctions. Combined, these findings build a comprehensive picture of the single-particle nucleon spectrum and how its properties derive from fundamental lattice operators.
\end{abstract}

\maketitle

\section{Introduction} \label{sec:intro}
The emergent nonperturbative phenomena of quantum chromodynamics (QCD), including confinement and dynamical chiral symmetry breaking, continue to remain the subject of extensive theoretical studies. Of particular interest is the hadron spectrum, in which both confinement and chiral symmetry breaking play a role, with the latter generating the split between the positive-parity and negative-parity spectra. Although the masses and quantum numbers associated to each state are readily accessible through experiment, it is left to theory to provide insight into many of the fundamental properties and structures that underlie the observed spectrum.

One avenue that can be explored in this regard is to analyse the wavefunctions of individual quarks bound within a hadron. A probability distribution can be calculated for bound states through the Bethe-Salpeter amplitude~\cite{Salpeter:1951sz}, although the resulting wavefunction is not unique. One definition uses a gauge-invariant amplitude in which the quarks are connected via appropriate parallel transport operators~\cite{Roberts:2013ipa}. In this case, an average would ideally be taken over all possible paths connecting the quarks to remove the explicit path dependence. Another possibility, which we adopt, is to simply fix the gauge. The choice of Landau gauge is advantageous owing to its smoothness properties, and the resulting wavefunctions are described well by constituent quark models~\cite{Roberts:2013ipa}.

The upper Dirac components of low-lying positive-parity nucleon wavefunctions have previously been examined in Ref.~\cite{Roberts:2013oea}. This was carried out using exclusively the standard scalar-diquark nucleon interpolating field, which has a nonrelativistic reduction. By repeating this interpolator at four levels of gauge-invariant Gaussian smearing, their wavefunctions were found to have a typical node structure, starting at zero nodes in the ground state and gaining one node per excitation.

In this work, we expand upon these results with a larger variational basis that includes both scalar-diquark and pseudoscalar-diquark spin-flavour combinations for nucleon interpolating fields, each at five different smearing levels. Inclusion of the pseudoscalar-diquark operator allows for unique relativistic aspects of hadron structure to manifest. This operator brings the small lower components of the Dirac spinors into play and vanishes in the nonrelativistic limit. The varying properties of these local interpolating fields are dissected, with a particular emphasis on their roles in building both the positive- and negative-parity single-particle spectra.

Novel techniques are introduced to access the relativistic components of the nucleon wavefunction. These are demonstrated by qualitatively examining the structure of the wavefunctions through visualisations, which are shown for both the upper and, for the first time, lower spinor components of the wavefunctions. This is followed by an explicit quantitative calculation of the radial wavefunction dependence. These methods are further applied to a pioneering study of the wavefunctions for negative-parity excitations.

This paper is structured as follows. In Sec.~\ref{sec:methods} we detail our lattice correlation-matrix variational analysis, along with the resulting mass spectrum and subsequent calculation of the wavefunctions. We present our investigation into the positive- and negative-parity spectra throughout Secs.~\ref{sec:positiveparity} and \ref{sec:negativeparity}, respectively. These include visualisations of the quark probability distributions, followed by a detailed quantitative analysis of their node structure and radial wavefunction dependence. To explain the findings, Sec.~\ref{sec:interpolatingfields} contains a thorough examination of the properties of individual interpolating fields and how they combine to create the energy eigenstates. Finally, we summarise our main findings in Sec.~\ref{sec:conclusion}.

\section{Lattice calculations} \label{sec:methods}
We start by detailing the lattice methods that form the basis of our correlation-matrix analysis and wavefunction calculations. Our results utilise the \((2+1)\)-flavour dynamical ensembles of the PACS-CS collaboration~\cite{PACS-CS:2008bkb}, available through the ILDG~\cite{Beckett:2009cb}. These ensembles have a lattice volume of \(32^3 \times 64\) and are generated with the Iwasaki renormalisation-group-improved action~\cite{Iwasaki:1983iya, Iwasaki:1996sn} at \(\beta = 1.90\) for the gauge sector, and nonperturbatively \(\order{a}\)-improved Wilson quarks~\cite{Sheikholeslami:1985ij} with improvement coefficient \(c_\mathrm{SW}=1.715\)~\cite{CP-PACS:2005igb}.

For this work, we select the heaviest PACS-CS ensemble with pion mass \(m_\pi \simeq \qty{0.702}{\GeV}\) and lattice spacing \(a \simeq \qty{0.0907}{fm}\)~\cite{PACS-CS:2008bkb}. This ensemble is the most relevant for our primary focus of resolving and understanding quark-model-like states, while also being the least susceptible to finite volume effects. Ten random sources are employed for each of the 399 configurations in this ensemble, culminating in \(\simeq 4000\) propagators for our correlation-matrix analysis.

\subsection{Variational analysis} \label{subsec:variationalanalysis}
The calculation of hadronic wavefunctions starts with the momentum-projected two-point correlation function,
\begin{equation} \label{eq:Gji}
	G_{ji}(\mathbf{p},\tau) = \sum_{\mathbf{x}} e^{-i\mathbf{p}\cdot\mathbf{x}} \Braket{\Omega | \mathcal{T}\{\chi_j(x) \, \bar{\chi}_i(0)\} | \Omega} \,,
\end{equation}
where \(\{\chi_i\}\) are the hadronic interpolating fields. Herein, we will solely consider zero momentum, \(\mathbf{p} = \mathbf{0}\). Nonzero momenta are the subject of ongoing work. At zero momentum, states of definite parity can be isolated through a spinor trace with the appropriate parity-projection operator,
\begin{gather} \label{eq:Gji+-}
	G_{ji}^\pm(\tau) = \tr \left(\Gamma^\pm \, G_{ji}(\mathbf{0},\tau) \right) \,, \\
	\Gamma^\pm = \frac{1}{2}\left(\gamma_4 \pm \mathbb{I}\right) \,.
\end{gather}

There are three local interpolating fields that describe the quantum numbers of the proton and which can be used to construct a correlation matrix~\cite{Chung:1981cc, Brommel:2003jm},
\begin{align}
	\chi_1(x) &= \epsilon_{abc} \left(u^T_a(x) \, C \, \gamma_5 \, d_b(x)\right) u_c(x) \,, \label{eq:chi1} \\
	\chi_2(x) &= \epsilon_{abc} \left(u^T_a(x) \, C \, d_b(x)\right) \gamma_5 \, u_c(x) \,, \label{eq:chi2} \\
	\chi_4(x) &= \epsilon_{abc} \left(u^T_a(x) \, C \, \gamma_5 \, \gamma_4 \, d_b(x)\right) u_c(x) \,, \label{eq:chi4}
\end{align}
where \(a,b,c\) denote colour indices. The first of these, \(\chi_1\), is the standard scalar-diquark interpolator which has a nonrelativistic reduction that strongly overlaps with the \(\mathrm{SU}(6)\) quark-model wavefunction. The second interpolating field, \(\chi_2\), positions the \(\gamma_5\) in front of the uncontracted quark field \(u_c(x)\) to reflect a pseudoscalar-diquark structure. This vanishes in the nonrelativistic limit, and couples strongly to higher-energy states~\cite{Leinweber:1994nm, Mahbub:2010jz, Mahbub:2013ala}. Finally, the third interpolator, \(\chi_4\), is the time component of the local spin-\(\frac{3}{2}\) isospin-\(\frac{1}{2}\) interpolating field which also overlaps with spin-\(\frac{1}{2}\) eigenstates~\cite{Zanotti:2003fx, Lasscock:2007ce}. The spin-flavour structure of this interpolator is very similar to \(\chi_1\), and previous work has demonstrated that inclusion of \(\chi_4\) in the correlation matrix provides little additional information over \(\chi_1\)~\cite{Mahbub:2009nr,Mahbub:2013ala}. Consequently, we focus on \(\chi_1\) and \(\chi_2\) in this work.

In order to expand our variational basis beyond these two interpolators, we exploit the smearing dependence of their overlap with the energy eigenstates~\cite{Mahbub:2009aa, Mahbub:2010jz}. We use five different levels of gauge-invariant Gaussian smearing~\cite{Gusken:1989qx} applied to both \(\chi_1\) and \(\chi_2\), comprised of \(12\), \(24\), \(48\), \(96\), and \(192\) smearing sweeps. This means we use a \(10 \times 10\) correlation matrix, which in turn provides access to ten energy eigenstates in each of the positive- and negative-parity sectors.

The eigenvectors needed to extract the energy eigenstates and corresponding wavefunctions are determined through a correlation-matrix variational analysis~\cite{Michael:1985ne, Luscher:1990ck}. Eigenvector \(n\) is represented as a linear combination of interpolating fields,
\begin{equation} \label{eq:phin}
	\bar{\phi}_n = \sum_i \bar{\chi}_i \, u_n^i \,,
\end{equation}
which leads to a recurrence relation satisfied by the correlation matrix,
\begin{equation} \label{eq:generalisedeigenvalueproblem}
	G_{ji}^\pm(\tau_0 + \Delta\tau) \, u_n^i = e^{-m_n \Delta\tau} G_{ji}^\pm(\tau_0) \, u_n^i \,,
\end{equation}
for variational parameters \(\tau_0\) and \(\Delta\tau\). This is nothing but a generalised eigenvalue problem (GEVP), and the eigenvectors \(\{u_n\}\) can thus be obtained using standard numerical methods. We choose \(\tau_0 = 2\) to be two slices after the source and \(\Delta\tau = 2\), which coincides with Refs.~\cite{Roberts:2013ipa, Roberts:2013oea, Mahbub:2010rm}. An equivalent procedure at the sink leads to left eigenvectors \(v_n\). By adopting the same basis of interpolating fields at the source and sink, our correlation matrix is Hermitian. This implies the left and right eigenvectors are equal.

To be precise in our methods, we normalise each element of the correlation matrix \(G_{ji}(\tau)\) by its diagonal elements two time slices after the source via
\begin{equation} \label{eq:normalisation}
	G_{ji}(\tau) \to \frac{G_{ji}(\tau)}{\sqrt{G_{ii}(2)} \sqrt{G_{jj}(2)}} \,.
\end{equation}
This ensures all matrix elements are \(\sim\order{1}\). The normalisation point is chosen to match \(\tau_0\) in the variational analysis. Additionally, in taking the ensemble average we utilise both \(\{U\} + \{U^*\}\) configurations, which have equal weight in the QCD partition function. This provides an improved unbiased estimator of the two-point functions that is purely real~\cite{Draper:1988xv}. As a result, our correlation matrix is symmetric, which we enforce through explicit symmetrisation of \(G\).

\subsection{Mass spectrum} \label{subsec:massspectrum}
With the left and right eigenvectors in hand, our first consideration is to examine the nucleon mass spectrum. From the eigenvectors, one obtains eigenstate-projected correlation functions,
\begin{equation} \label{eq:eigenstateprojected}
	G_n^\pm(\tau) = v_n^j \, G_{ji}^\pm(\tau) \, u_n^i \,.
\end{equation}
The effective mass for state \(n\) with parity \(\pm\) is then calculated as
\begin{equation} \label{eq:effectivemass}
	m_n^\pm(\tau) = \frac{1}{2} \log \left(\frac{G_n^\pm(\tau)}{G_n^\pm(\tau + 2)}\right) \,.
\end{equation}
The use of slices two time steps apart in the ratio provides an improved determination of the effective mass compared to adjacent slices through a more accurate estimate of the local slope.

Our formalism is designed to excite and isolate the single-particle dominated states of the nucleon spectrum. As multi-particle scattering states also form part of the spectrum, their contributions will appear as contaminants in our projected correlation functions. In practice, the contributions of these states can be suppressed through Euclidean time evolution, ensuring that the time dependence of the fitted correlator is consistent with that of a single state. For the PACS-CS ensemble selected for this analysis, the ground state and the first two negative-parity excitations are single-particle dominated states, lying below the energy eigenstates dominated by multi-particle contributions~\cite{Abell:2023nex}.

Upon calculation of the effective mass, constant fits are performed by optimising the full covariance \(\chi^2_\mathrm{cov}/\mathrm{dof}\) value. The Euclidean time region is selected by implementing a strict cutoff of \(\chi^2_\mathrm{cov}/\mathrm{dof} < 1.2\) where possible. The correlators are manually examined to ensure we are fitting at sufficiently large Euclidean times to isolate the state as well as possible, and before signal is lost. The final mass value is ascertained from an uncorrelated fit over the chosen region, with statistical uncertainties obtained using a second-order jackknife analysis.

We focus on the ground state and two lowest-lying excitations for both positive and negative parity. The next two states for both parities are also speculatively examined. For these higher excitations, we are forced to start the fits one time slice earlier and accept slightly worse \(\chi^2\) values to ensure the fits are performed before the loss of signal. In Table~\ref{tab:massspectrum}, we provide the masses and fit details, including the fit region and \(\chi^2\) values, for each considered state. 

\begin{table}
	\caption{\label{tab:massspectrum} The details of our effective-mass fits, including the masses, fit region \([\tau_\mathrm{min}, \tau_\mathrm{max}]\) (relative to the source), correlated \(\chi^2_\mathrm{cov}/\mathrm{dof}\) used to select the fit region, and uncorrelated \(\chi^2/\mathrm{dof}\) used to quote the mass. The lowest-lying five positive-parity and four negative-parity states are examined.}
	\begin{ruledtabular}
		\begin{tabular}{cS[table-format=1.5]cccc}
			State & \multicolumn{1}{c}{\(m\) (GeV)} & \(\tau_\mathrm{min}\) & \(\tau_\mathrm{max}\) & \(\chi^2_\mathrm{cov}/\mathrm{dof}\) & \(\chi^2/\mathrm{dof}\) \\
			\hline
			\(1^+\) & 1.585(4) & 5 & 10 & 0.11 & 0.16 \\
			\(2^+\) & 2.34(5)  & 5 & 10 & 0.81 & 1.07 \\
			\(3^+\) & 2.69(5)  & 5 &  9 & 0.13 & 0.08 \\
			\(4^+\) & 2.85(19) & 4 &  6 & 2.00 & 1.58 \\
			\(5^+\) & 3.27(11) & 4 &  6 & 1.24 & 0.61 \\
			\hline
			\(1^-\) & 2.19(2)  & 5 &  9 & 0.76 & 0.35 \\
			\(2^-\) & 2.22(2)  & 5 &  9 & 0.72 & 0.40 \\
			\(3^-\) & 2.84(6)  & 4 &  8 & 1.04 & 0.38 \\
			\(4^-\) & 2.94(7)  & 4 &  7 & 3.28 & 2.66
		\end{tabular}
	\end{ruledtabular}
\end{table}

\begin{figure}
	\includegraphics[width=\linewidth]{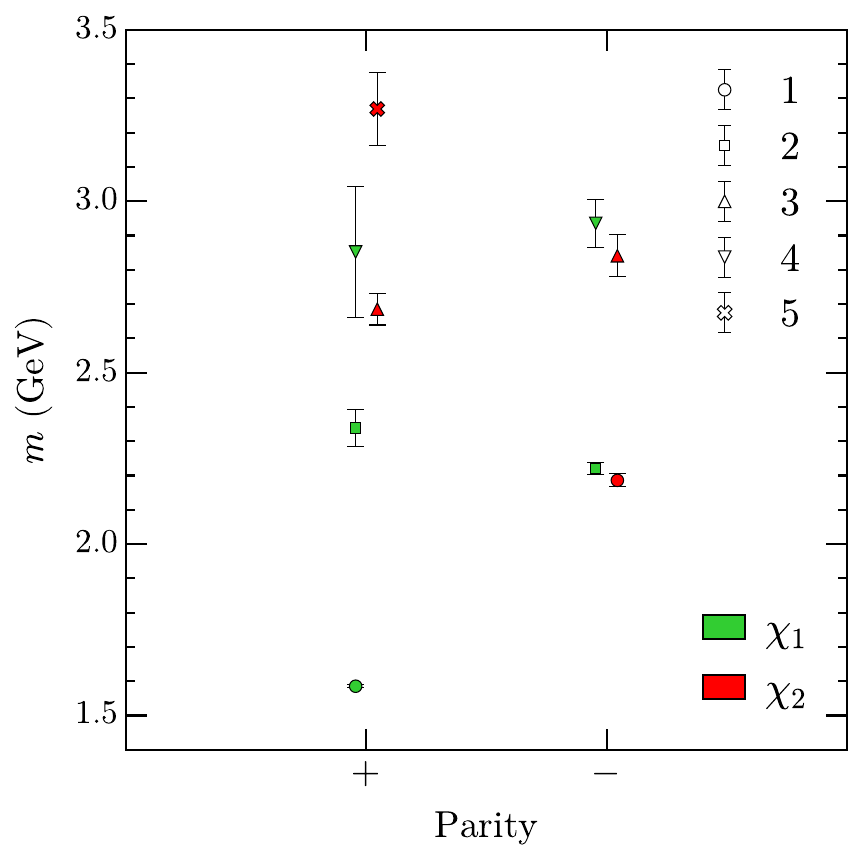}%
	\vspace{-0.5em}
	\caption{\label{fig:massspectrum} The extracted mass spectrum for the ground state and four lowest-lying positive-parity and negative-parity excitations. Each state is labelled by whether it is dominated by \(\chi_1\) (green/light grey) or \(\chi_2\) (red/dark grey) interpolating fields. Beyond the third positive-parity and second negative-parity states, the results become speculative.}
\end{figure}

The complete resulting mass spectrum is displayed in Fig.~\ref{fig:massspectrum}. We label the states as being dominated by one of the \(\chi_1\) or \(\chi_2\) interpolating field. Simply put, this indicates which of \(\chi_1\) or \(\chi_2\) has the strongest overlap with the state, though the precise meaning of this categorisation will become clear in Sec.~\ref{subsec:eigenvectorcomponents}.

The negative-parity states markedly exist in pairs that comprise a \(\chi_1\)- and \(\chi_2\)-dominated state, with the latter sitting slightly lower in each case. Beyond the ground state, the positive-parity excitations also suggestively occur in \(\chi_1/\chi_2\) pairs, though with a larger mass gap between the two states within each pair. In this case, however, it is the \(\chi_1\)-dominated state that is the lower of the pair. This provides our first indication that the \(\chi_1\) and \(\chi_2\) interpolating fields will play opposite roles for positive- and negative-parity states.

\subsection{Wavefunctions} \label{subsec:wavefunctions}
We now turn to calculation of the wavefunction proper. This is achieved by shifting the individual quark fields in the nucleon annihilation operator,
\begin{multline} \label{eq:shiftedchi1}
	\chi_1(x;\mathbf{r}_1,\mathbf{r}_2,\mathbf{r}_3) = \epsilon_{abc}\left(u^T_a(\mathbf{x}+\mathbf{r}_1,\tau) \, C \, \gamma_5 \right. \\ \left. d_b(\mathbf{x}+\mathbf{r}_2,\tau) \vphantom{\textsuperscript{T}}\right) u_c(\mathbf{x}+\mathbf{r}_3,\tau) \,.
\end{multline}
In theory, the spatial position of each quark can vary across the lattice. In practice, the spatial dependence is greatly simplified by taking advantage of rotational and translational symmetry. A near-complete description of the wavefunction can be obtained by separating the two \(u\) quarks along a common axis, and calculating the corresponding \(d\)-quark wavefunction for each such separation. For this work, we leave the two \(u\) quarks fixed at the origin, \(\mathbf{r}_1 = \mathbf{r}_3 = \mathbf{0}\).

The \(d\)-quark wavefunction is then proportional to the overlap of this shifted annihilation operator with the desired energy eigenstate \(\bar{\phi}_n\) (recall \(\mathbf{p} = \mathbf{0}\)),
\begin{equation}
\begin{aligned} \label{eq:wavefunction}
	\psi_n(\mathbf{r};\tau) &= \sum_\mathbf{x} \Braket{\Omega | \mathcal{T}\{\chi_1(x;\mathbf{r}) \, \bar{\phi}_n(0)\} | \Omega} \\
	&= \sum_\mathbf{x} \Braket{\Omega | \mathcal{T}\{\chi_1(x;\mathbf{r}) \, \bar{\chi}_i(0)\} | \Omega} u_n^i \,,
\end{aligned}
\end{equation}
where \(\chi_1(x;\mathbf{r})\) is given by Eq.~(\ref{eq:shiftedchi1}) with \(\mathbf{r}_1 = \mathbf{r}_3 = \mathbf{0}\) and \(\mathbf{r}_2 = \mathbf{r}\). We note that the shifted annihilation operator is no longer smeared. The employment of \(\chi_1\) at the sink follows from the fact that of the two spin-\(\frac{1}{2}\) nucleon interpolators used in this work, only \(\chi_1\) has spin-flavour content that overlaps with the \(\mathrm{SU}(6)\) quark-model wavefunction. Namely, it has a nonrelativistic reduction with a scalar-diquark component, which strongly resembles the ground state~\cite{Leinweber:1994nm, Mahbub:2013ala}. This forces the interesting physics associated with state \(n\) to be contained in the resolved wavefunction.

This definition of the wavefunction is gauge dependent, and as such we fix to Landau gauge. In addition to being a smooth gauge, it has previously been shown that the ground-state wavefunction in Landau gauge is accurately described by the nonrelativistic quark model using standard values for the constituent quark masses and string tension~\cite{Roberts:2013ipa}. This vindicates Landau gauge as providing a strong foundation for probing excited-state wavefunctions. In the future, it would be interesting to perform a comparison with other similar gauges, such as Coulomb gauge and the maximal centre and abelian gauges.

At this point, we emphasise that the wavefunction defined in Eq.~(\ref{eq:wavefunction}) is a \(4\times 4\) Dirac matrix, \(\psi_{\beta\alpha}(\mathbf{r})\), deriving from the spinor components of the source and sink interpolators. Note that although we are computing the \(d\)-quark wavefunction by shifting the spatial position of the \(d\) quark in the \(\chi_1\) interpolator, the source and sink indices \(\alpha\) and \(\beta\) are on the uncontracted \(u\) quark. Importantly, the same Dirac index appears on the nucleon spinor, thus addressing the upper and lower wavefunction components of quark distributions within the nucleon.

Each element of the Dirac matrix provides a wavefunction component to be examined. First, the source index \(\alpha\) identifies the quantum numbers of the wavefunction. Since the proton has total angular momentum \(j = 1/2\), the only allowed orbital angular momenta are \(\ell = 0\) and \(1\). These embody the positive- and negative-parity spectra. The upper (\(\alpha = 1,2\)) components select positive parity, \(\ell = 0\), while the lower (\(\alpha = 3,4\)) components select negative parity, \(\ell = 1\).

Simultaneously, the spin-up (\(\alpha = 1,3\)) and spin-down (\(\alpha = 2,4\)) components naturally isolate the \(m_j = +1/2\) and \(-1/2\) wavefunctions, respectively. Finally, the sink index \(\beta\) then identifies one of the four relativistic components of the resulting spinor wavefunction.

Any parity and spin mismatches between the source and sink are captured by the resolved wavefunction. For example, if one chooses an upper component at the source (\(\alpha = 1,2\)), but a lower component at the sink (\(\beta = 3,4\)), then there is a parity mismatch between them. To account for this, the wavefunction for this component will display a \(p\)-wave structure. Along the same line, if a spin-up component is chosen at the source (\(\alpha = 1,3\)), but spin-down at the sink (\(\beta = 2,4\)), there is a spin mismatch. The resulting wavefunction will have \(m_\ell = +1\) to accommodate this transition.

It is beneficial in this regard to recall the general spin-\(\frac{1}{2}\) solutions to the Dirac equation for a spherically symmetric potential, in which this structure is reflected. These solutions can be written
\begin{equation} \label{eq:diraceqsolution}
	\psi^{\pm}_{jm_j}(\mathbf{r}) = \begin{pmatrix}
		(-i)^\ell \, \psi^{\pm}_{\mathcal{U}}(r) \, \mathcal{Y}_{\ell}^{m_j}(\theta,\phi) \\[0.5em]
		(-i)^{\ell\pm 1} \, \psi^{\pm}_{\mathcal{L}}(r) \, \mathcal{Y}_{\ell \pm 1}^{m_j}(\theta,\phi)
	\end{pmatrix} \,,
\end{equation}
where \(\ell = j \mp 1/2\). For the relevant case of \(j = 1/2\), the choice of sign corresponds to even (\(\ell = 0\)) and odd (\(\ell = 1\)) parity. The factors of \(-i\) are included for convenience, and \(\mathcal{Y}_\ell^{m_j}\) are the spinor spherical harmonics,
\begin{equation} \label{eq:spinorsphericalharmonics}
	\mathcal{Y}_\ell^{m_j} = \frac{1}{\sqrt{2\ell + 1}} \begin{pmatrix}
		\pm \sqrt{\ell \pm m_j + \frac{1}{2}} \, Y_\ell^{m_j - \frac{1}{2}} \\[0.5em]
		\sqrt{\ell \mp m_j + \frac{1}{2}} \, Y_\ell^{m_j + \frac{1}{2}}
	\end{pmatrix} \,.
\end{equation}

Having written these solutions, one can inquire as to the ``effective'' quantum numbers, \(\ell_\mathrm{eff}\) and \(m_\mathrm{eff}\), that specify the spherical harmonics, \(Y_{\ell_\mathrm{eff}}^{m_\mathrm{eff}}(\theta,\phi)\), for each wavefunction component. For positive parity, the upper components of Eq.~(\ref{eq:diraceqsolution}) have \(\ell_\mathrm{eff} = 0\), and the lower components \(\ell_\mathrm{eff} = 1\). This signifies that the upper components are \(s\)-wave in nature, but the lower components are \(p\)-wave, in accord with the prior discussion. The situation is reversed for negative parity, which are \(s\)-wave in the lower components but \(p\)-wave in the upper. The spin-up (first and third) wavefunction components have \(m_\mathrm{eff} = m_j - 1/2\), while the spin-down (second and fourth) components have \(m_\mathrm{eff} = m_j + 1/2\).

This therefore informs the expected shape of the calculated \(d\)-quark wavefunctions. Using the arguments outlined above, we summarise these ``effective'' quantum numbers, \(\ell_\mathrm{eff}\) and \(m_\mathrm{eff}\), for each source and sink combination in Table~\ref{tab:lm_l}. It will be interesting to discover to what extent the angular dependence of our wavefunctions are described by the spherical harmonics with these values of \((\ell_\mathrm{eff},m_\mathrm{eff})\).

\begin{table}
	\caption{\label{tab:lm_l} The ``effective'' quantum numbers, \((\ell_\mathrm{eff}, m_\mathrm{eff})\), that specify the spherical harmonics, \(Y_{\ell_\mathrm{eff}}^{m_\mathrm{eff}}(\theta,\phi)\), for each element of the calculated wavefunctions. Each combination of source and sink indices, \(\alpha\) and \(\beta\), identifies a wavefunction component as described in text. These effective quantum numbers account for any parity and spin mismatches between the source and sink. Any component with \(|m_\mathrm{eff}| > \ell_\mathrm{eff}\) is predicted to vanish.}
	\begin{ruledtabular}
		\begin{tabular}{c|cccc}
			\(\alpha\rightarrow\) & \multirow{2}{*}{1} & \multirow{2}{*}{2} & \multirow{2}{*}{3} & \multirow{2}{*}{4} \\
			\cline{1-1}
			\(\beta\downarrow\) & & & & \\
			\hline
			1 & \((0,0)\) & \((0,-1)\) & \((1,0)\) & \((1,-1)\) \\
			2 & \((0,+1)\) & \((0,0)\) & \((1,+1)\) & \((1,0)\) \\
			3 & \((1,0)\) & \((1,-1)\) & \((0,0)\) & \((0,-1)\) \\
			4 & \((1,+1)\) & \((1,0)\) & \((0,+1)\) & \((0,0)\)
		\end{tabular}
	\end{ruledtabular}
\end{table}

It is important to note that some of the entries in this table recommend quantum number combinations that do not exist. In particular, the \((\ell, m_\ell) = (0,\pm 1)\) combinations do not exist, and therefore these Dirac indices must produce vanishing wavefunctions. Indeed, any component with \(|m_\mathrm{eff}| > \ell_\mathrm{eff}\) is predicted to vanish.

However, it is known from deeply virtual Compton scattering experiments that gluons carry angular momentum, and it has previously been proposed that this could impart a contribution to the quark wavefunctions~\cite{Kamleh:2017lye}. One way this would appear is as a nontrivial structure in these components. We have tested this and found no evidence for gluonic contributions. Without any operators in the correlation functions that explicitly allow sea contributions, including gluons, to acquire angular momentum, no such effects will be present. From a preliminary study of possible autocorrelations, we attribute the result in Ref.~\cite{Kamleh:2017lye} to a small finite-ensemble effect.

\section{Positive-parity states} \label{sec:positiveparity}
Having explained how to access the relativistic wavefunction components, we now proceed to our analysis of the positive- and negative-parity spectra, starting with the former. This will encompass a qualitative examination of wavefunction structure through visualisations in the form of both volume and surface renderings, complemented by a rigorous quantitative analysis of the radial wavefunction dependence.

\subsection{Visualisations} \label{subsec:positiveparityvis}
Our starting point is to produce visualisations that demonstrate our procedure for extracting the relativistic wavefunction components in practice. For this, we prioritise the three lowest-lying positive-parity states. This is sufficient to reveal the salient properties that will form the cornerstone of this analysis.

For each of these states, we visualise the wavefunctions for the \((\beta,\alpha) = (1,1)\), \((3,1)\) and \((4,1)\) Dirac elements. Here, \(\alpha = 1\) at the source selects the positive-parity spin-up wavefunction, while the sink index \(\beta\) corresponds to a given spinor component of said wavefunction. Drawing on Table~\ref{tab:lm_l}, these components have effective quantum numbers \((\ell_\mathrm{eff},m_\mathrm{eff}) = (0,0)\), \((1,0)\) and \((1,+1)\). This covers the three different spherical harmonics relevant for describing angular dependence in this work,
\begin{align}
	Y_0^0(\theta,\phi) &\propto 1 \,, \label{eq:Y00} \\
	Y_1^0(\theta,\phi) &\propto \cos\theta \,, \label{eq:Y10} \\
	Y_1^{\pm 1}(\theta,\phi) &\propto \sin\theta \, e^{\pm i\phi} \,, \label{eq:Y11}
\end{align}
where any normalisation factors are superfluous.

We visualise the wavefunction amplitude, \(|\psi_{\beta\alpha}(\mathbf{r})|\) (i.e.\ the square root of the probability density). Both volume and surface renderings are presented. The former of these renders the full three-dimensional wavefunction above some minimum threshold, with magnitude represented by colour. The wavefunction is cut in half to reveal its interior. This cut is performed along the \(x\)-axis for the \((1,1)\) and \((3,1)\) components, but along the \(z\)-axis for the \((4,1)\) component where a ring structure is expected in the \(xy\)-plane.

For the surface plots, one coordinate is held fixed and the wavefunction is visualised as a function of the remaining two dimensions. To match the volume renders, we show \(x = 0\) for the \((1,1)\) and \((3,1)\) components, and \(z = 0\) for the \((4,1)\) component. In rendering the surface plots, we choose a normalisation that keeps the peak value of the wavefunction constant.

\begin{figure*}
	\includegraphics[width=0.42\linewidth]{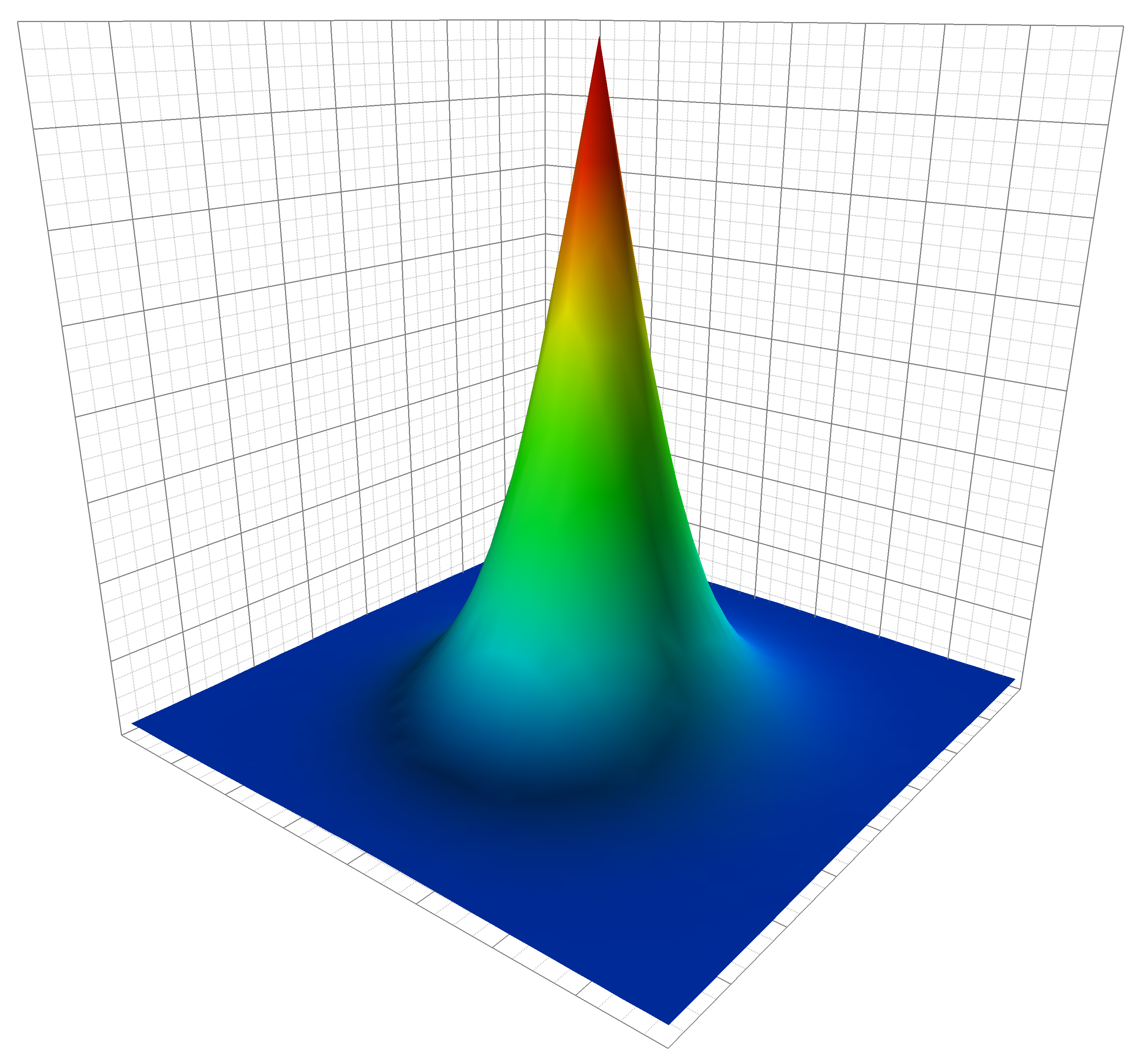}%
	\hspace{5em}
	\includegraphics[width=0.42\linewidth]{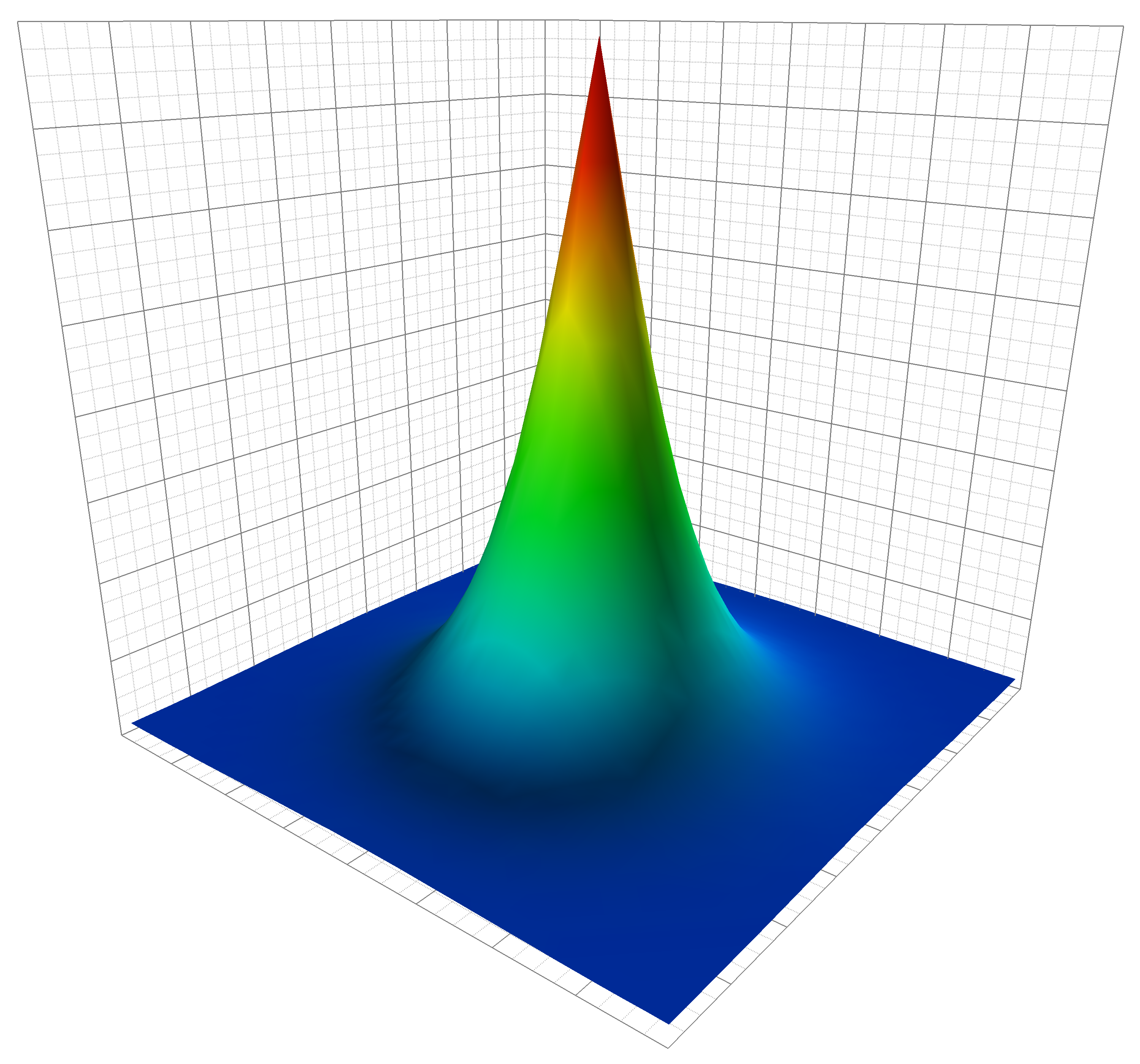}%
	\vspace{0.5em}
	\includegraphics[width=0.42\linewidth]{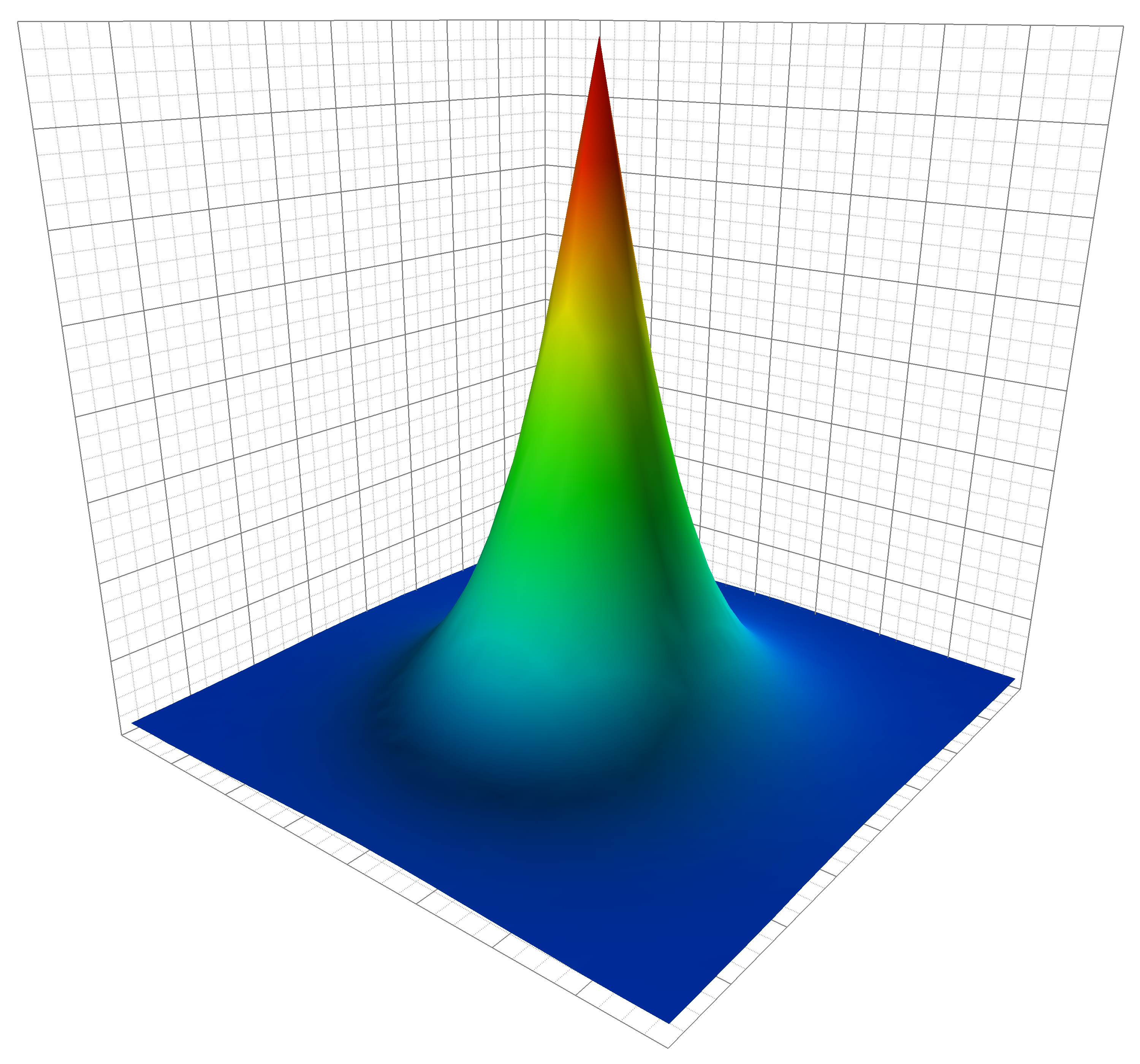}%
	\hspace{5em}
	\includegraphics[width=0.42\linewidth]{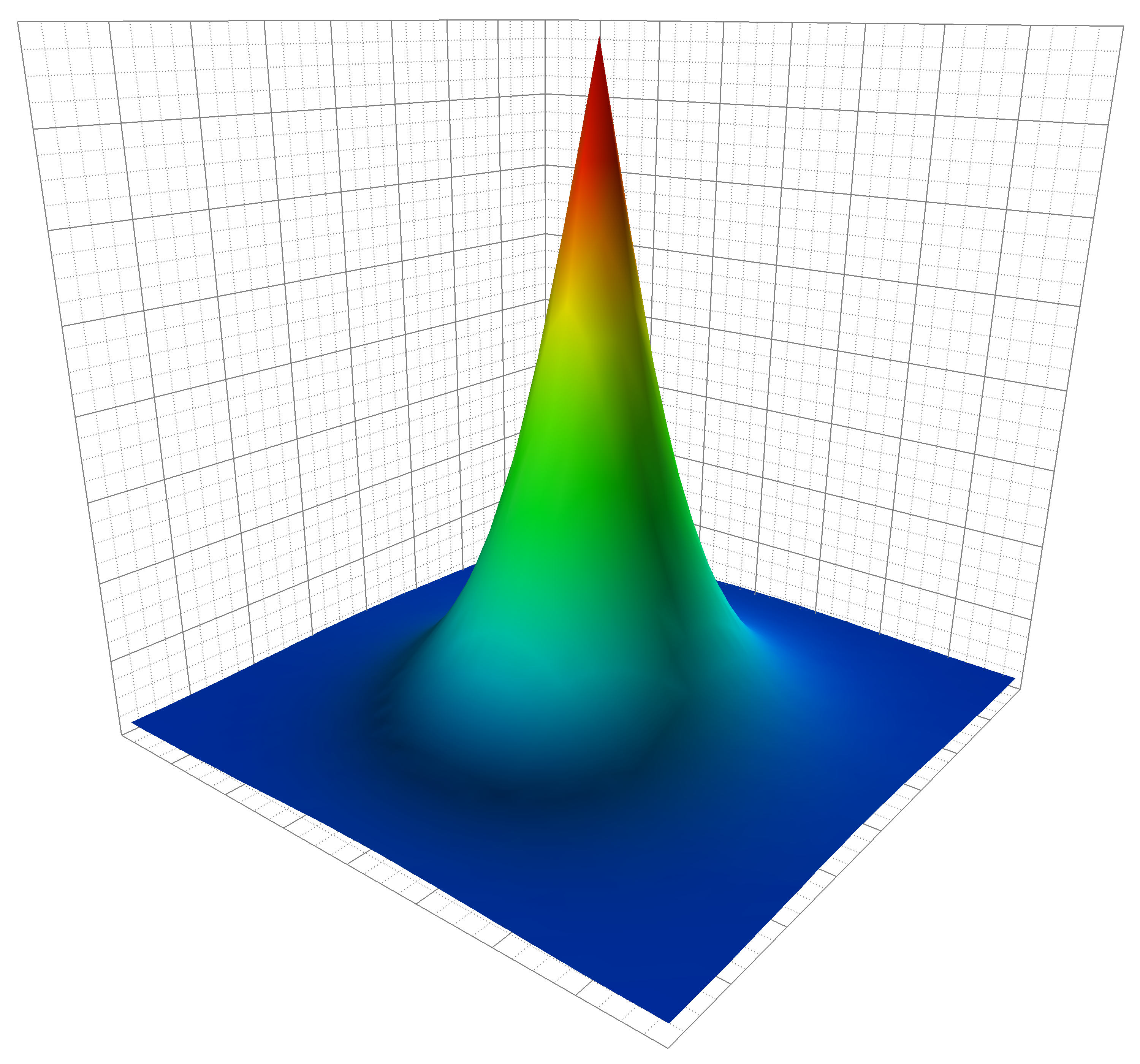}%
	\vspace{0.5em}
	\includegraphics[width=0.42\linewidth]{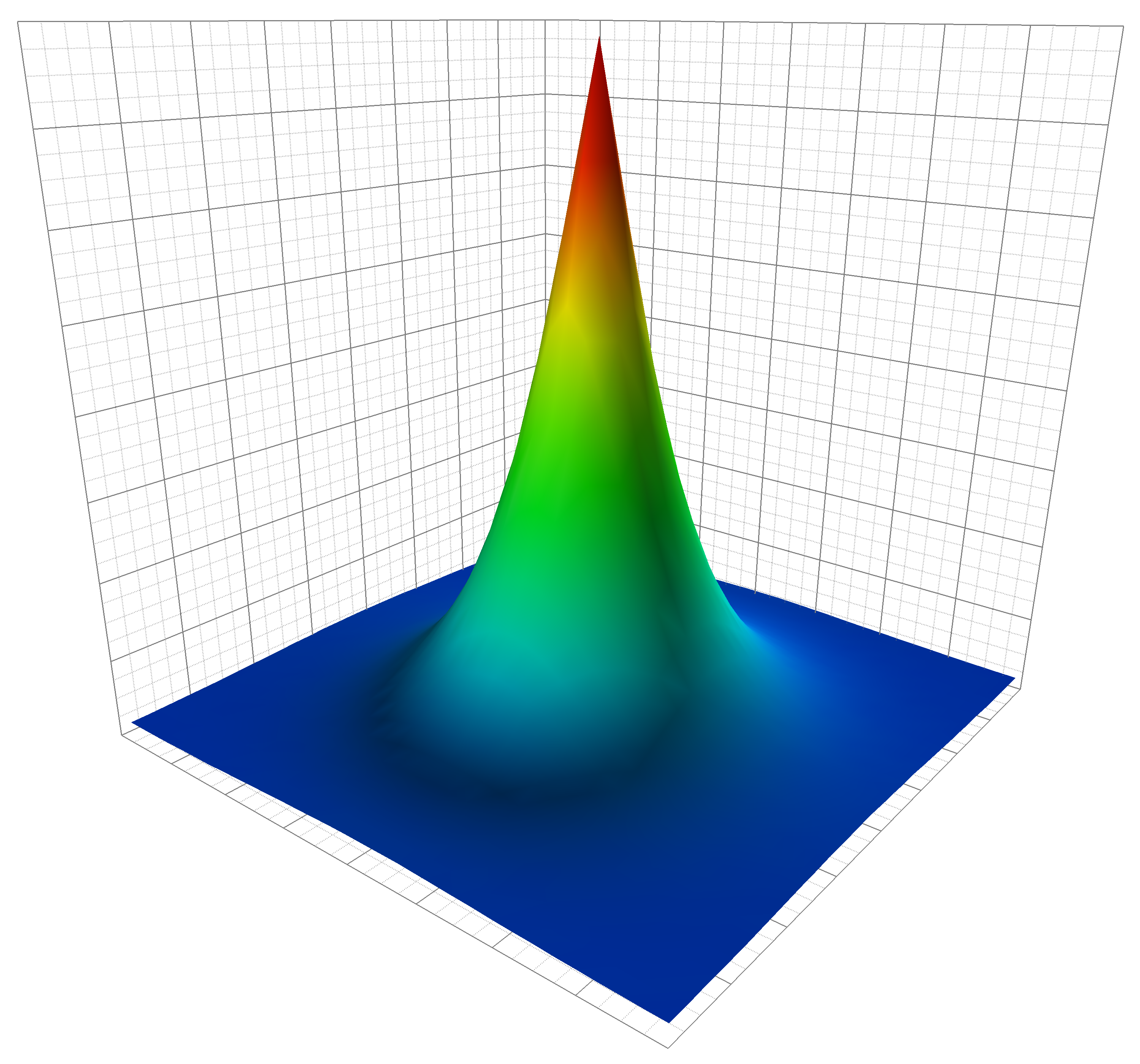}%
	\hspace{5em}
	\includegraphics[width=0.42\linewidth]{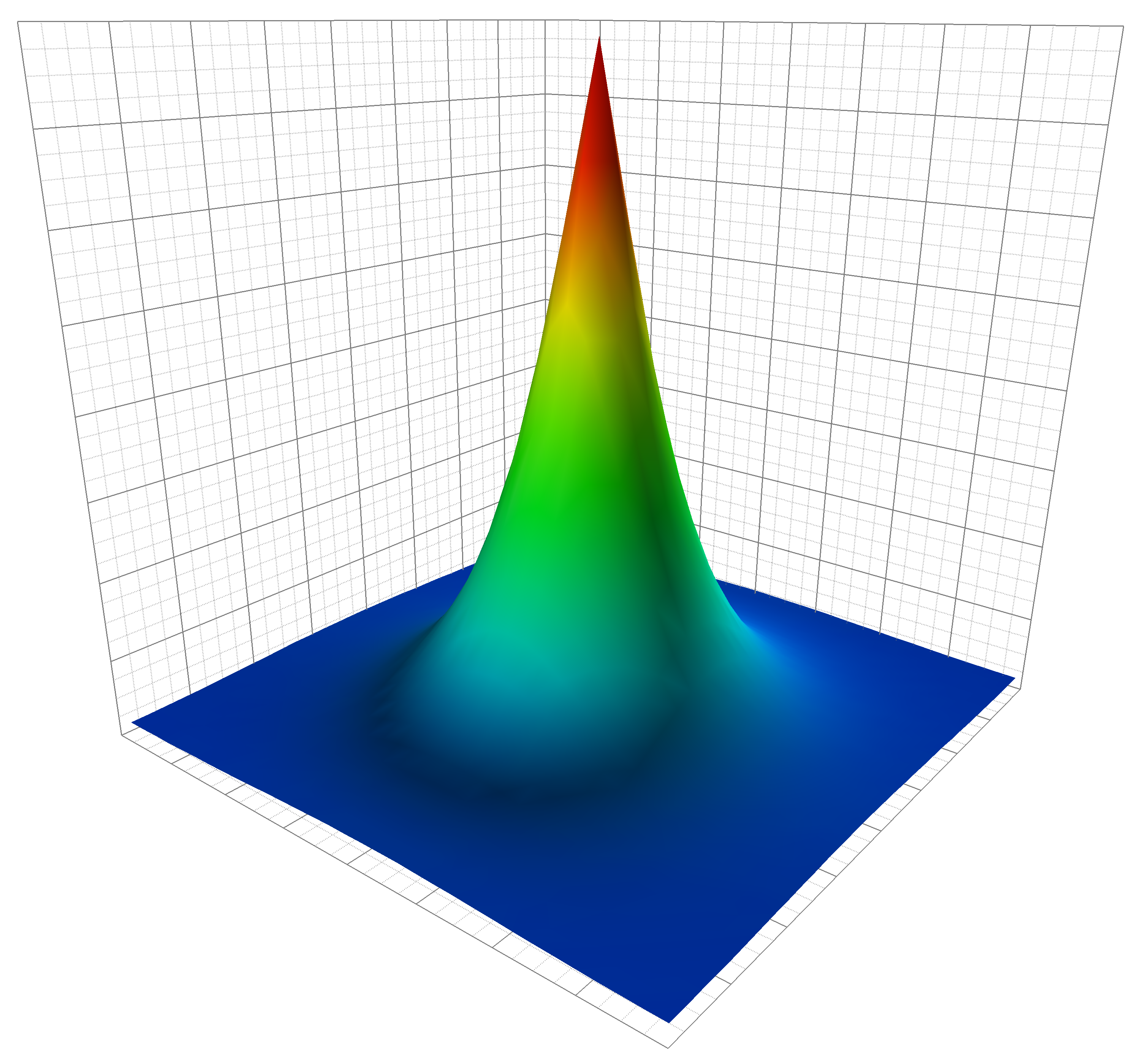}%
	\vspace{-0.5em}
	\caption{\label{fig:euclideantimegroundstate} The Euclidean-time evolution of the \((1,1)\) wavefunction component of the ground state. Visualisations are taken every second time slice after the source, from \(\tau = 2\) to \(12\), displayed from left to right then top to bottom. The wavefunction grows slightly up to Euclidean time \(\tau = 6\) (middle left), but beyond that point is highly stable. This lines up with the effective-mass plateau for the ground state.}
\end{figure*}

Before proceeding to these visualisations, a final but crucial consideration is the Euclidean time at which we examine the wavefunctions. Certainly, we should be sufficiently far after the source to ensure the state in question has been isolated. This suggests a Euclidean time of at least \(\tau_\mathrm{min}\), the starting point of the effective-mass fits in Sec.~\ref{subsec:massspectrum}. To examine this rigorously, we visualise the Euclidean-time evolution of the \((1,1)\) component of the ground state in Fig.~\ref{fig:euclideantimegroundstate}. Surface renderings are shown every second time slice after the source, from \(\tau = 2\) to \(12\). The wavefunction is largely unchanged over the full temporal extent. From a close inspection, the wavefunction is found to become slightly broader up to a Euclidean time of \(\tau \simeq 6\). Beyond that point, it is highly stable. This nicely coincides with the plateau region for the ground-state effective mass of \(\tau \in [5,10]\) (Table~\ref{tab:massspectrum}).

\begin{figure}
	\includegraphics[width=\linewidth]{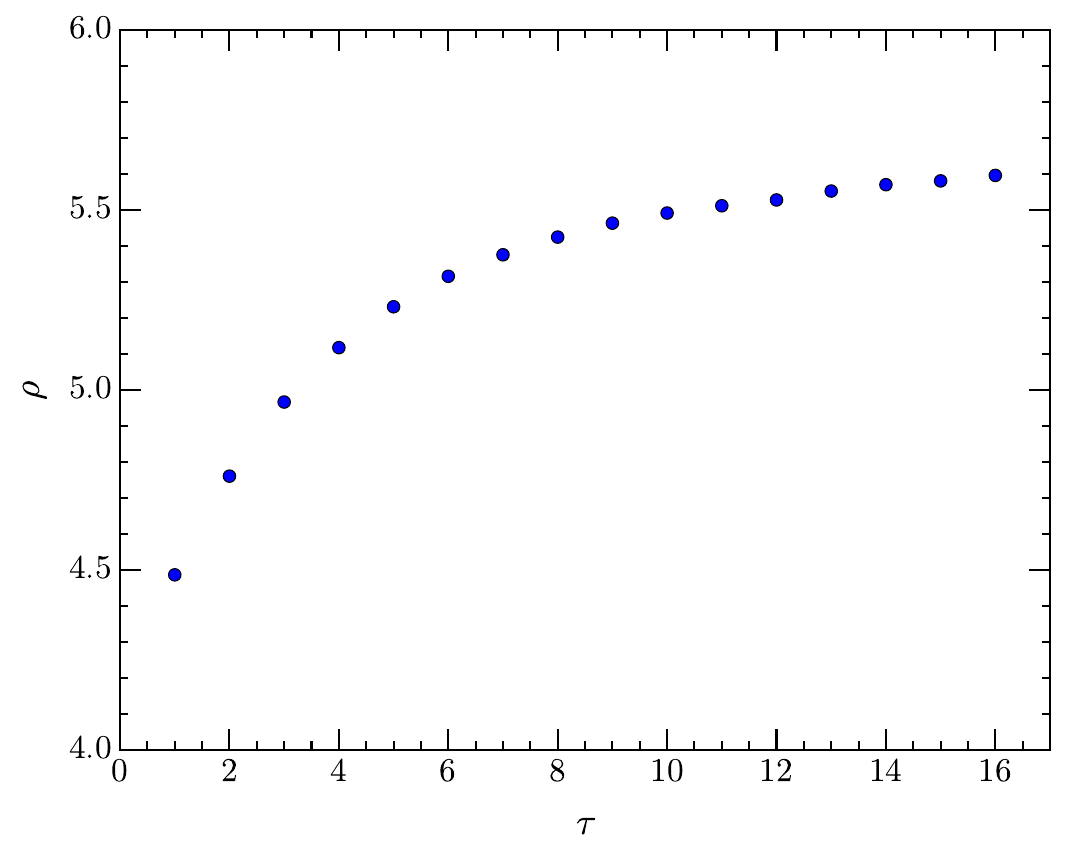}
	\caption{\label{fig:radius} The RMS radius \(\rho\) defined in Eq.~(\ref{eq:radius}) for the ground-state \((1,1)\) wavefunction as a function of Euclidean time. The radius grows considerably over the first few time slices after the source, before converging to a stable value at large Euclidean times.}
\end{figure}

To expand on this, an RMS radius can be calculated for a given wavefunction component in which each point is weighted by its normalised probability density,
\begin{equation} \label{eq:radius}
	\rho = \sqrt{\sum_{\mathbf{r}} \, r^2 \, |\psi(\mathbf{r})|^2} \;.
\end{equation}
We plot the Euclidean-time evolution of \(\rho\) for the ground-state \((1,1)\) wavefunction in Fig.~\ref{fig:radius}. The RMS radius increases rapidly within several time slices of the source, before the state has been properly isolated. The radius is then mostly stable at large Euclidean times, as was captured by the visualisations in Fig.~\ref{fig:euclideantimegroundstate}. It is nonetheless noteworthy that the radius continues to grow over the entire Euclidean time extent, though it increases by only \(\simeq 5\%\) in total beyond \(\tau \geq 6\).

\begin{figure*}
	\includegraphics[width=0.42\linewidth]{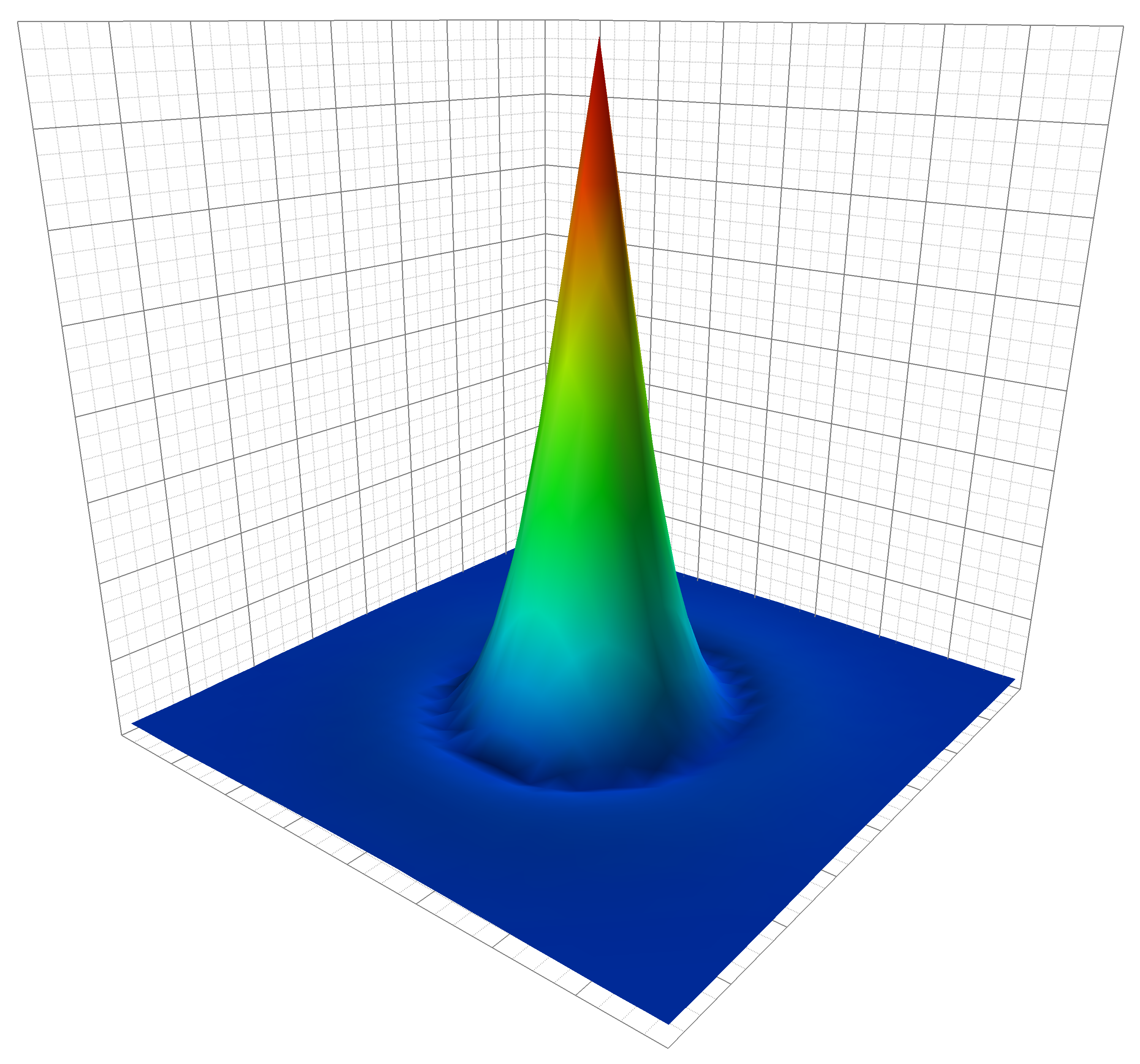}%
	\hspace{5em}
	\includegraphics[width=0.42\linewidth]{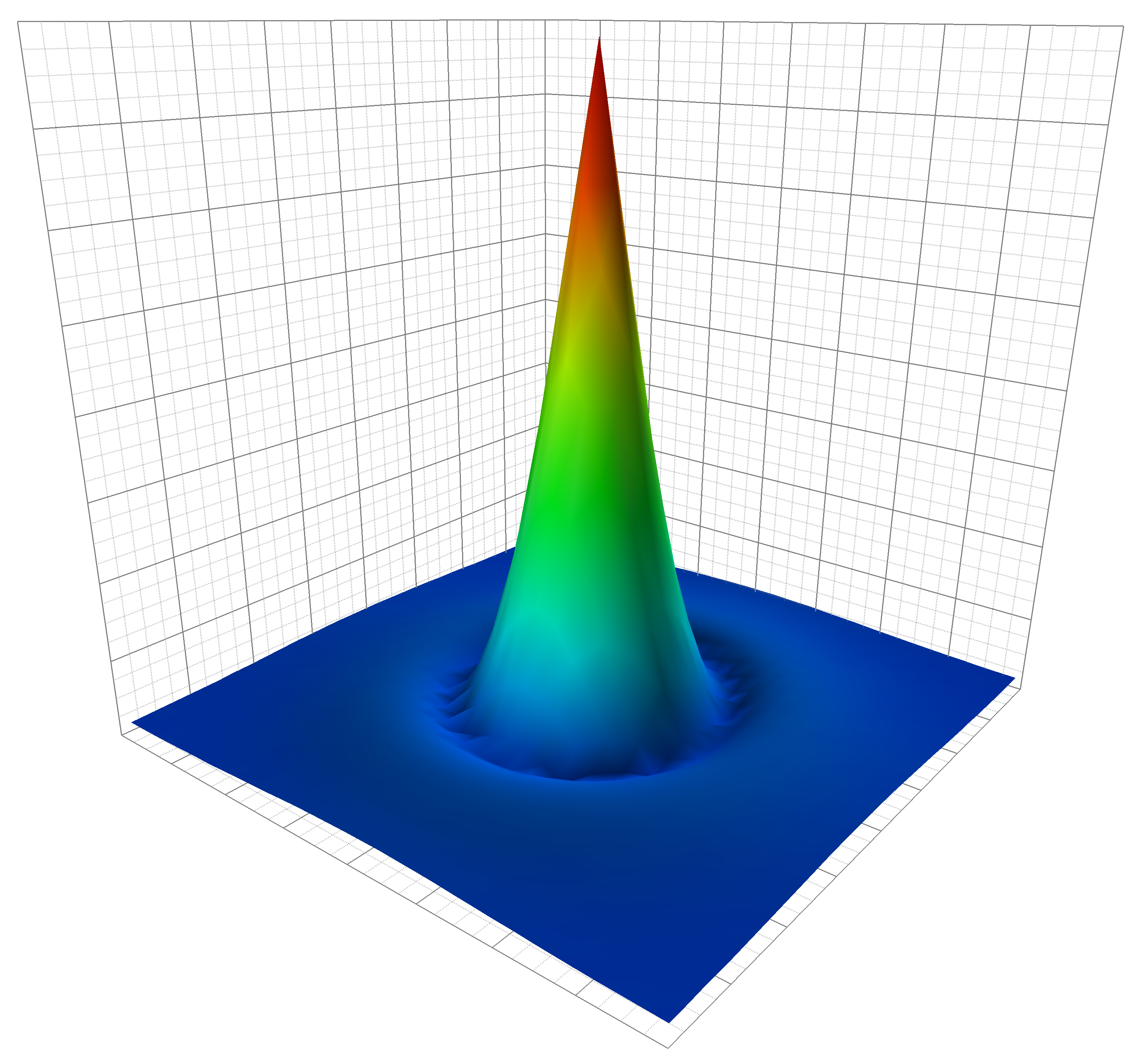}%
	\vspace{0.5em}
	\includegraphics[width=0.42\linewidth]{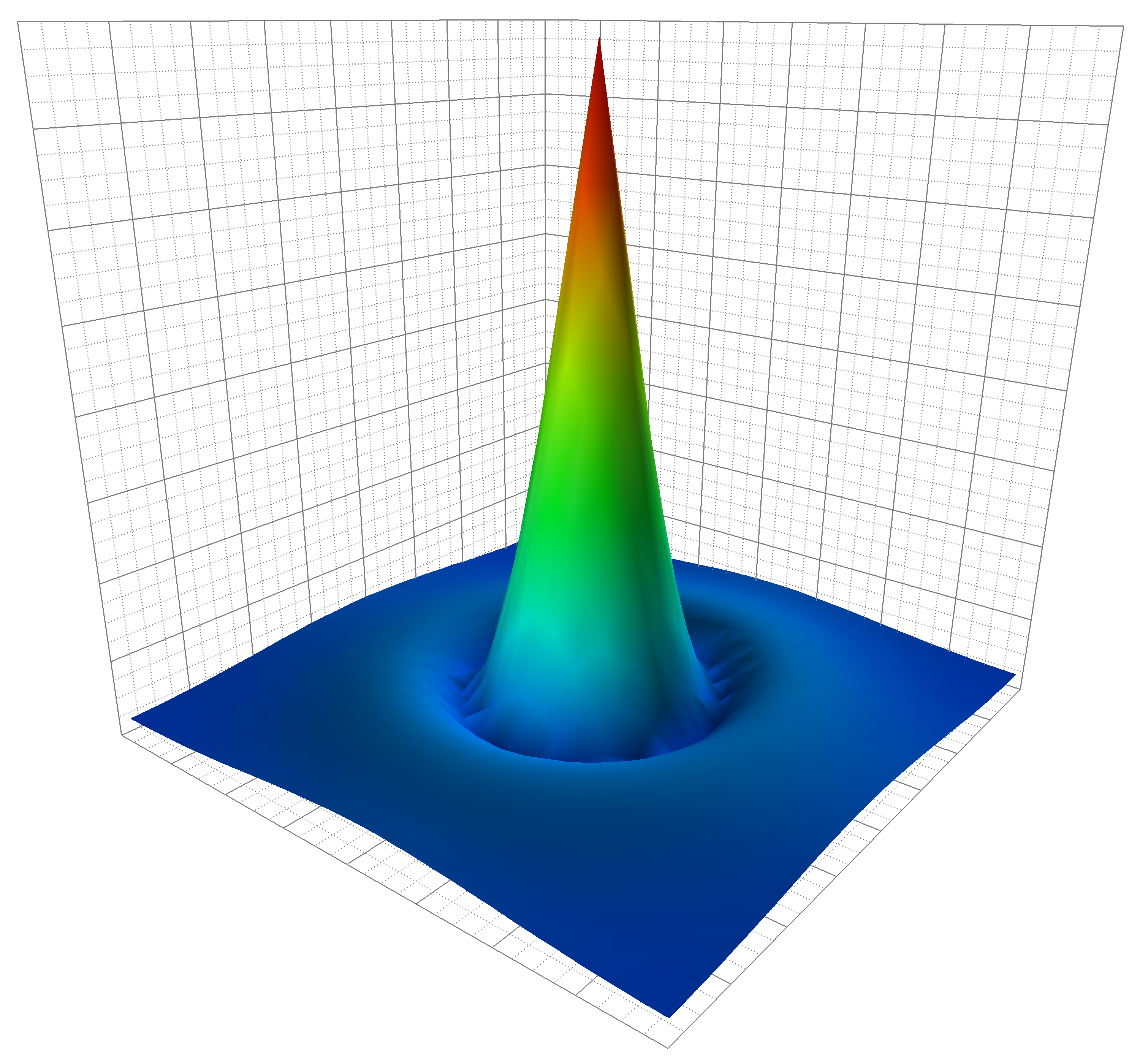}%
	\hspace{5em}
	\includegraphics[width=0.42\linewidth]{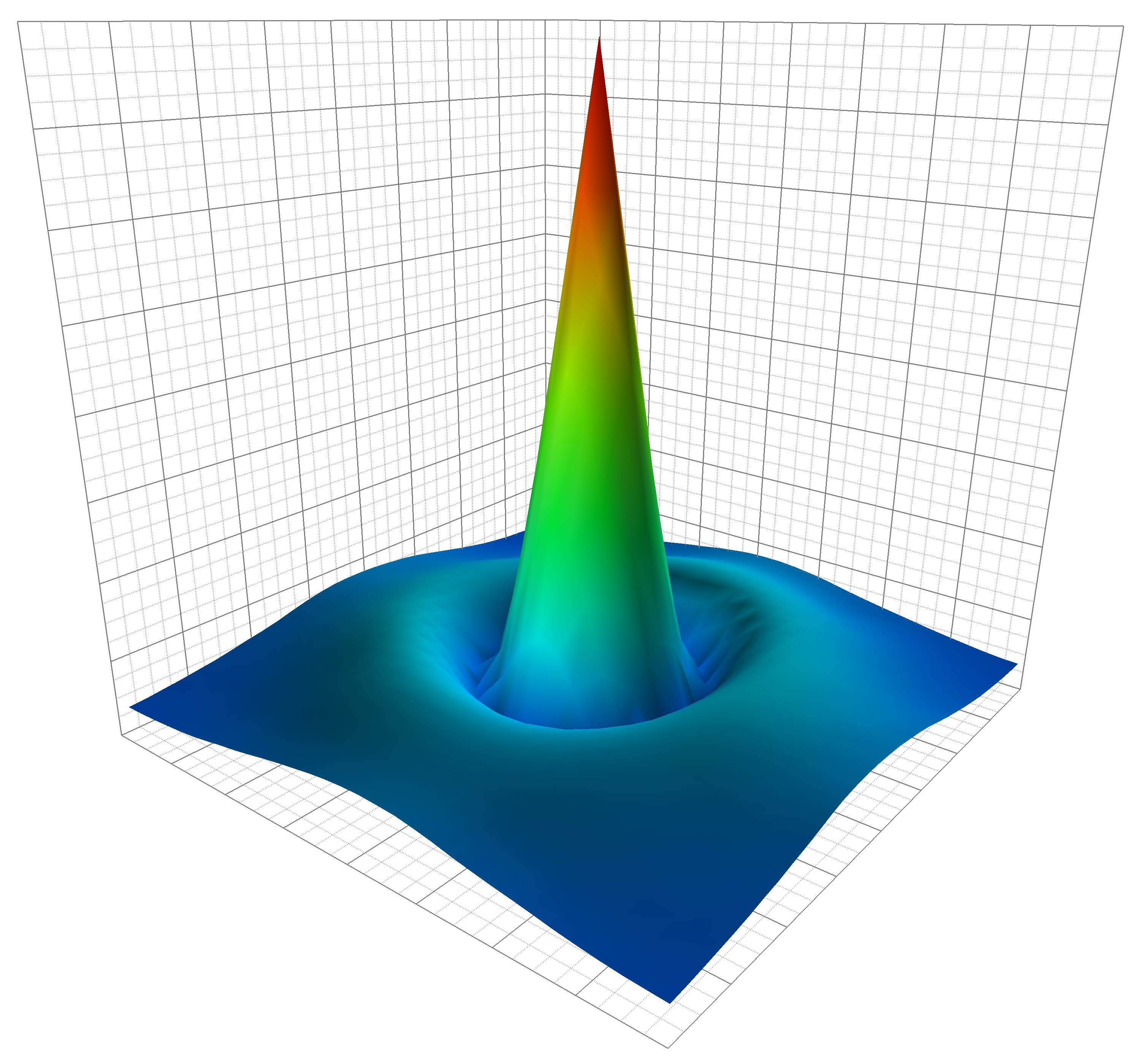}%
	\vspace{0.5em}
	\includegraphics[width=0.42\linewidth]{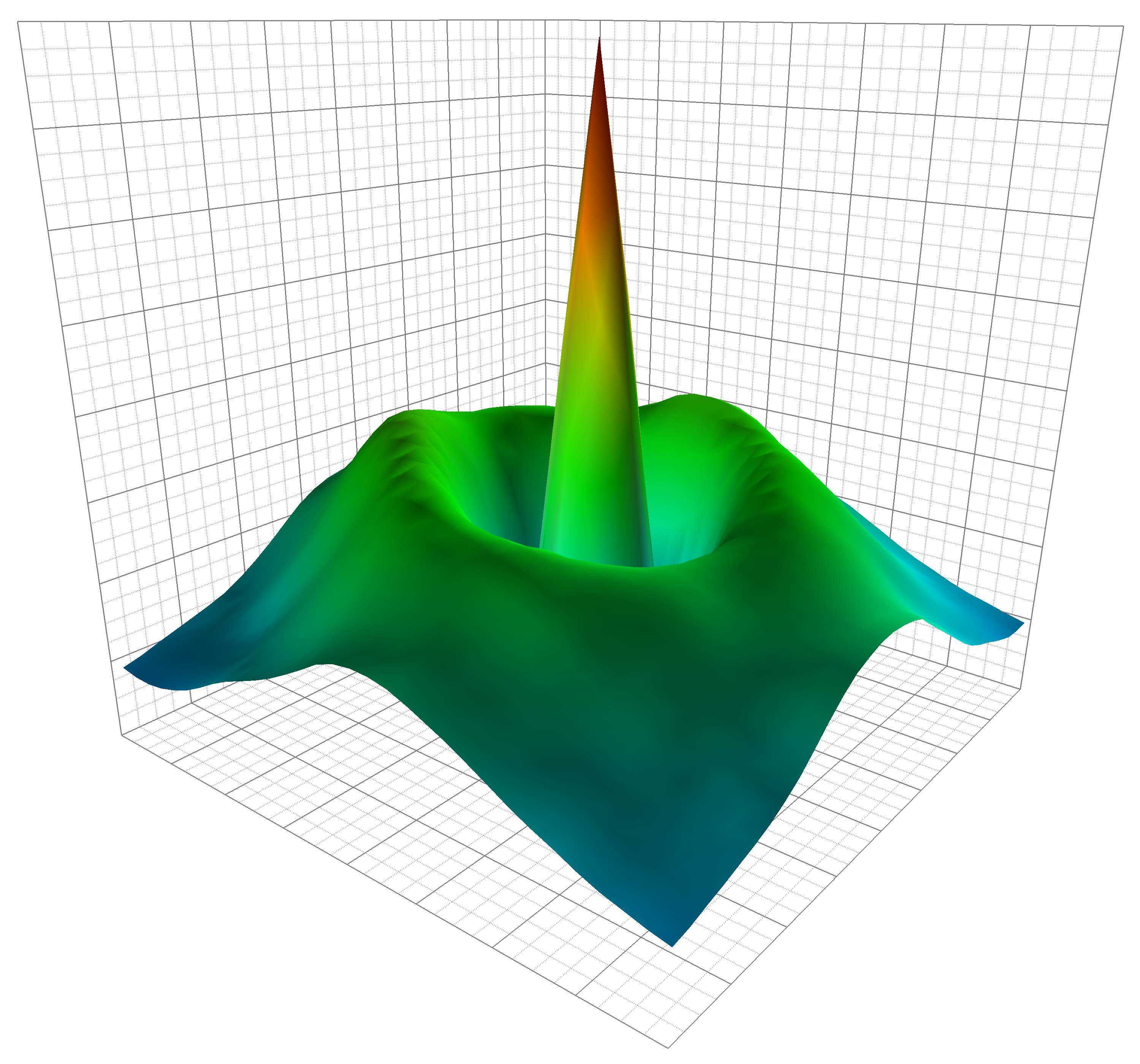}%
	\hspace{5em}
	\includegraphics[width=0.42\linewidth]{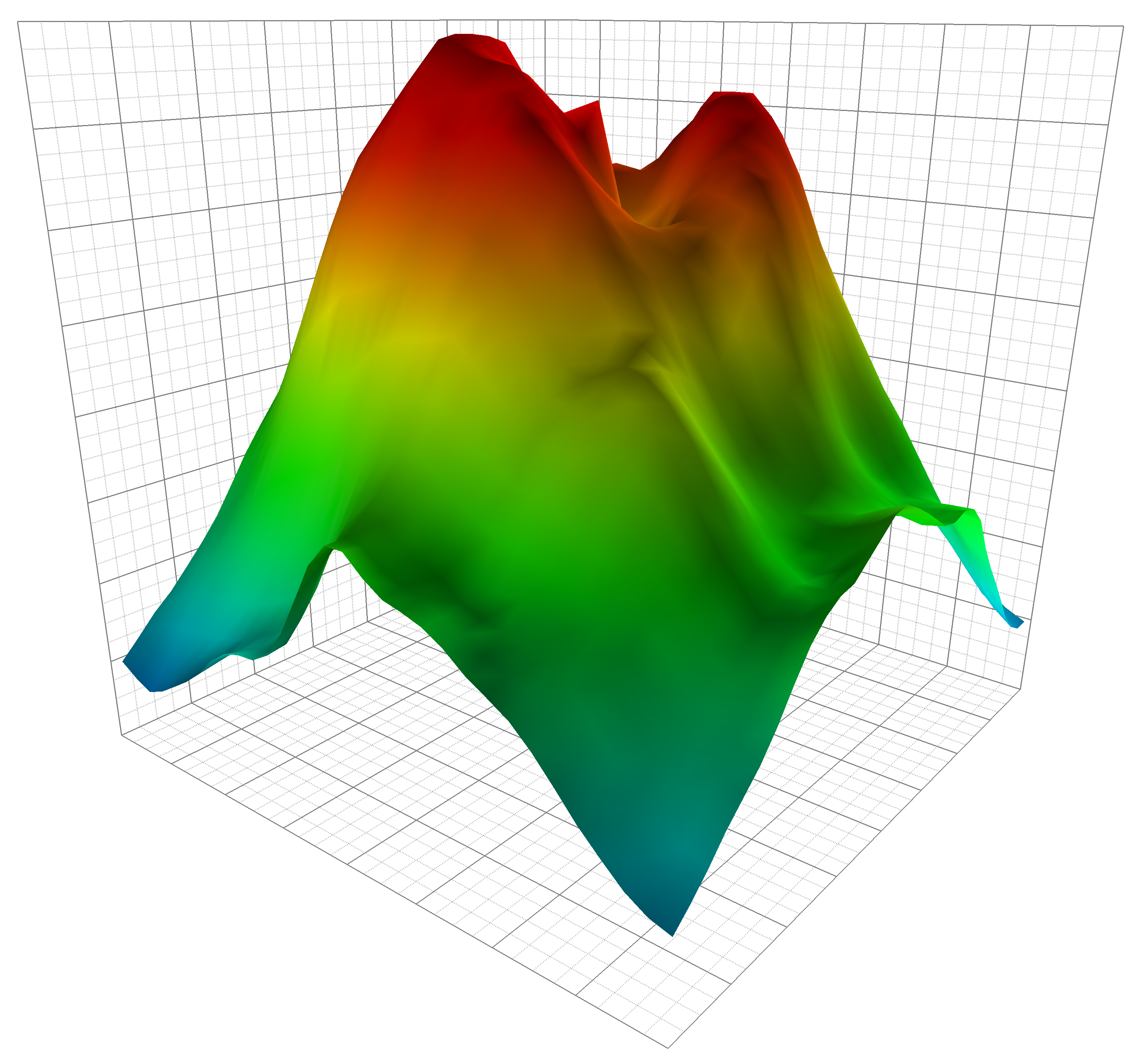}%
	\vspace{-0.5em}
	\caption{\label{fig:euclideantimefirstexcitation} As in Fig.~\ref{fig:euclideantimegroundstate}, but for the first positive-parity excitation. The depression seen in the surface plots is a manifestation of the node found in this state. The wavefunction is less stable than for the ground state, but the node persists over a significant temporal extent, up to and including \(\tau = 10\) (lower left). For later times, the signal is lost.}
\end{figure*}

To accompany the ground state, we also study the Euclidean-time evolution of the first positive-parity excitation. This is presented in Fig.~\ref{fig:euclideantimefirstexcitation}. For this state, a node is now present in the wavefunction. This appears as a depression in the surface plot. Unlike the ground state, the first excited state's wavefunction does not stabilise and instead perpetually broadens. Simultaneously, its node becomes sharper and moves closer to the origin, towards smaller radial distances. The key property, however, is that the node does persist up to (and including) \(\tau = 10\). It is only beyond this point that the signal is lost, and the wavefunction becomes unrecognisable. This again concurs with the effective mass for this state, with our plateau fit over Euclidean times \(\tau \in [5,10]\). For \(\tau > 10\), signal is also lost in the effective mass.

Based on this observation that the wavefunction evolution parallels the effective mass, moving forward we present wavefunctions at Euclidean time \(\tau_\mathrm{min}\) from our effective-mass fits. This ensures we have isolated each state to the best of our ability without losing signal in the wavefunction. For most states this is \(\tau_\mathrm{min} = 5\), though for the higher excitations that are being speculatively examined it is \(\tau_\mathrm{min} = 4\).

\begin{figure*}
	\includegraphics[width=0.42\linewidth]{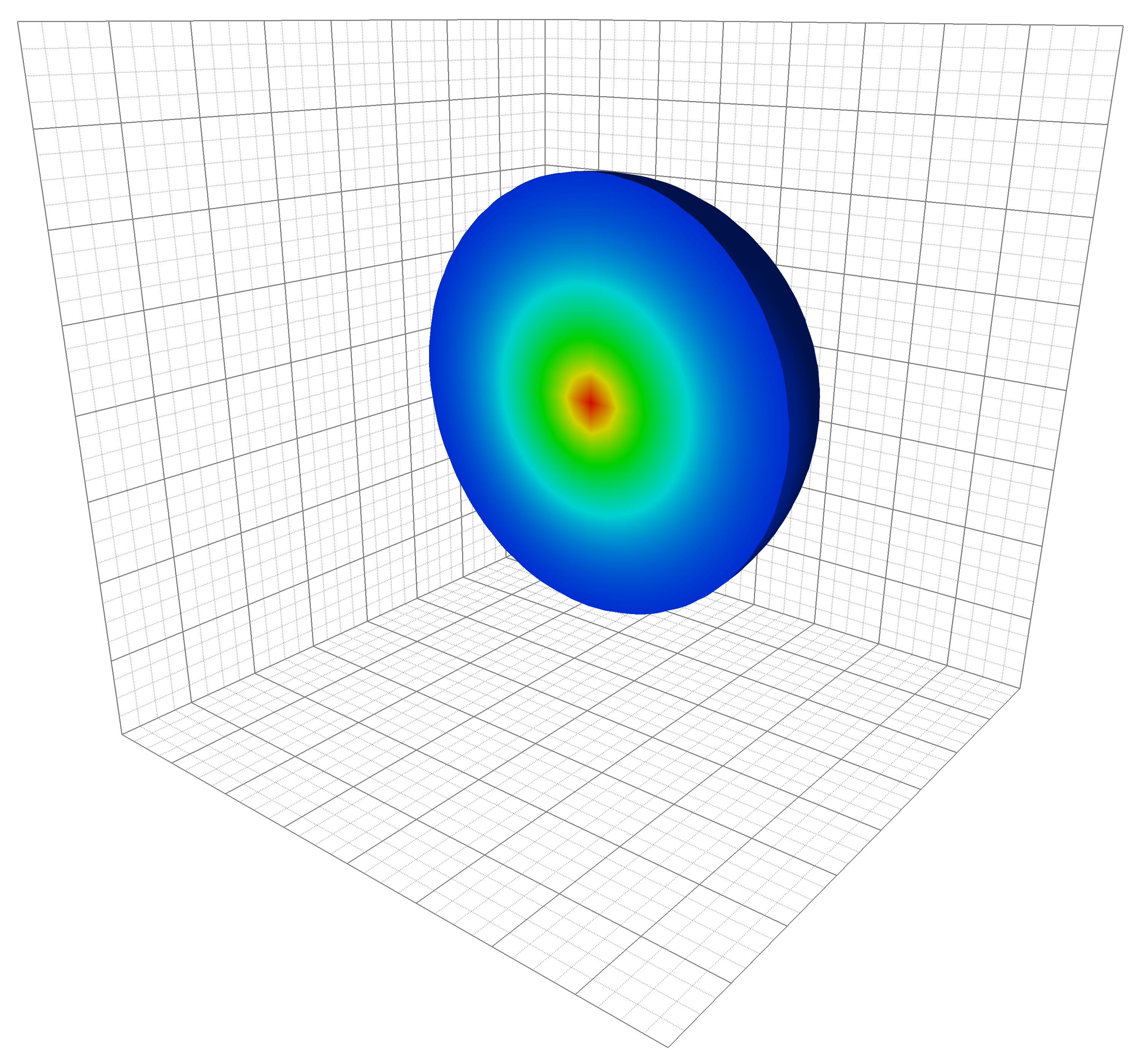}%
	\hspace{5em}
	\includegraphics[width=0.42\linewidth]{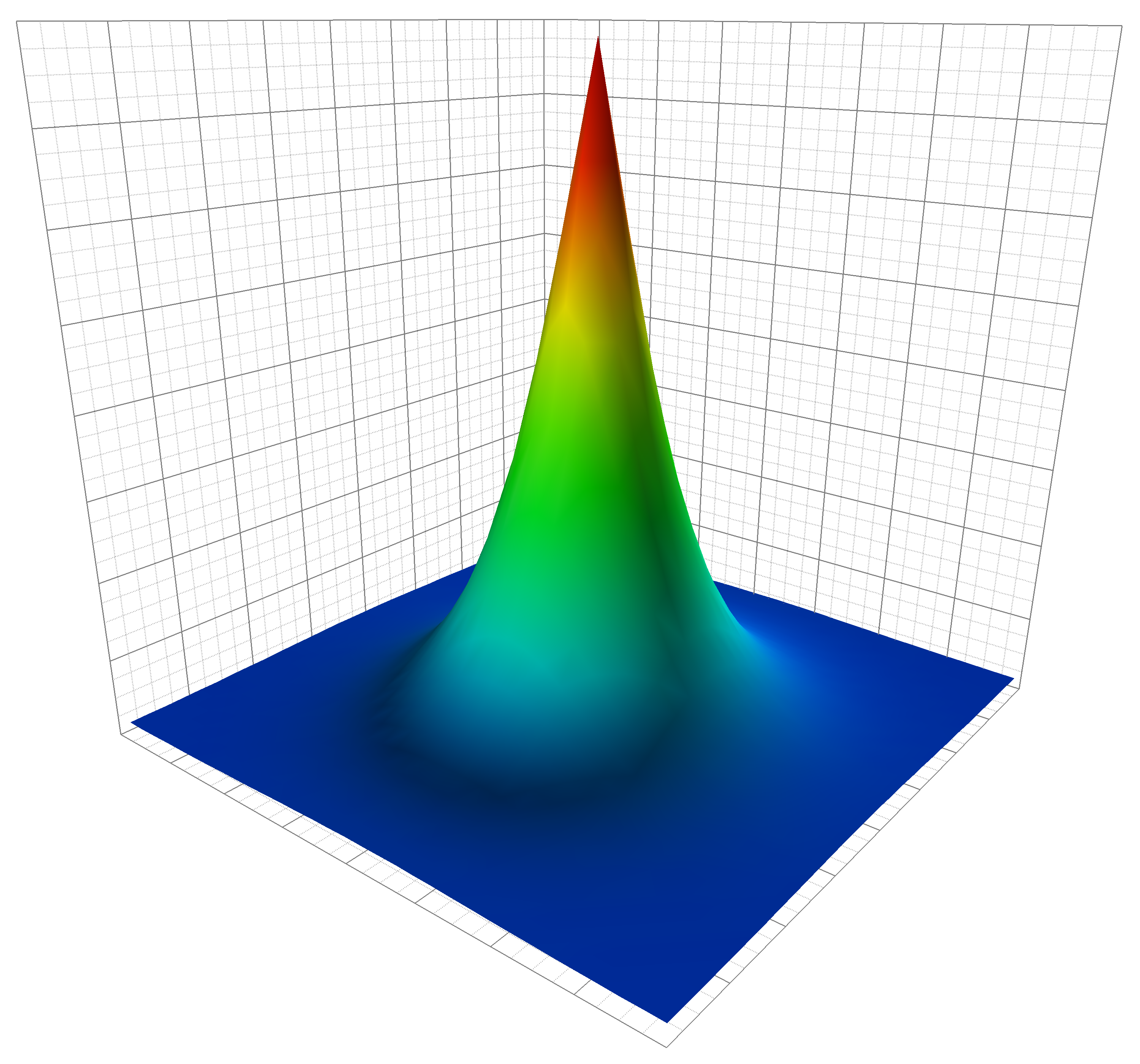}%
	\vspace{0.5em}
	\includegraphics[width=0.42\linewidth]{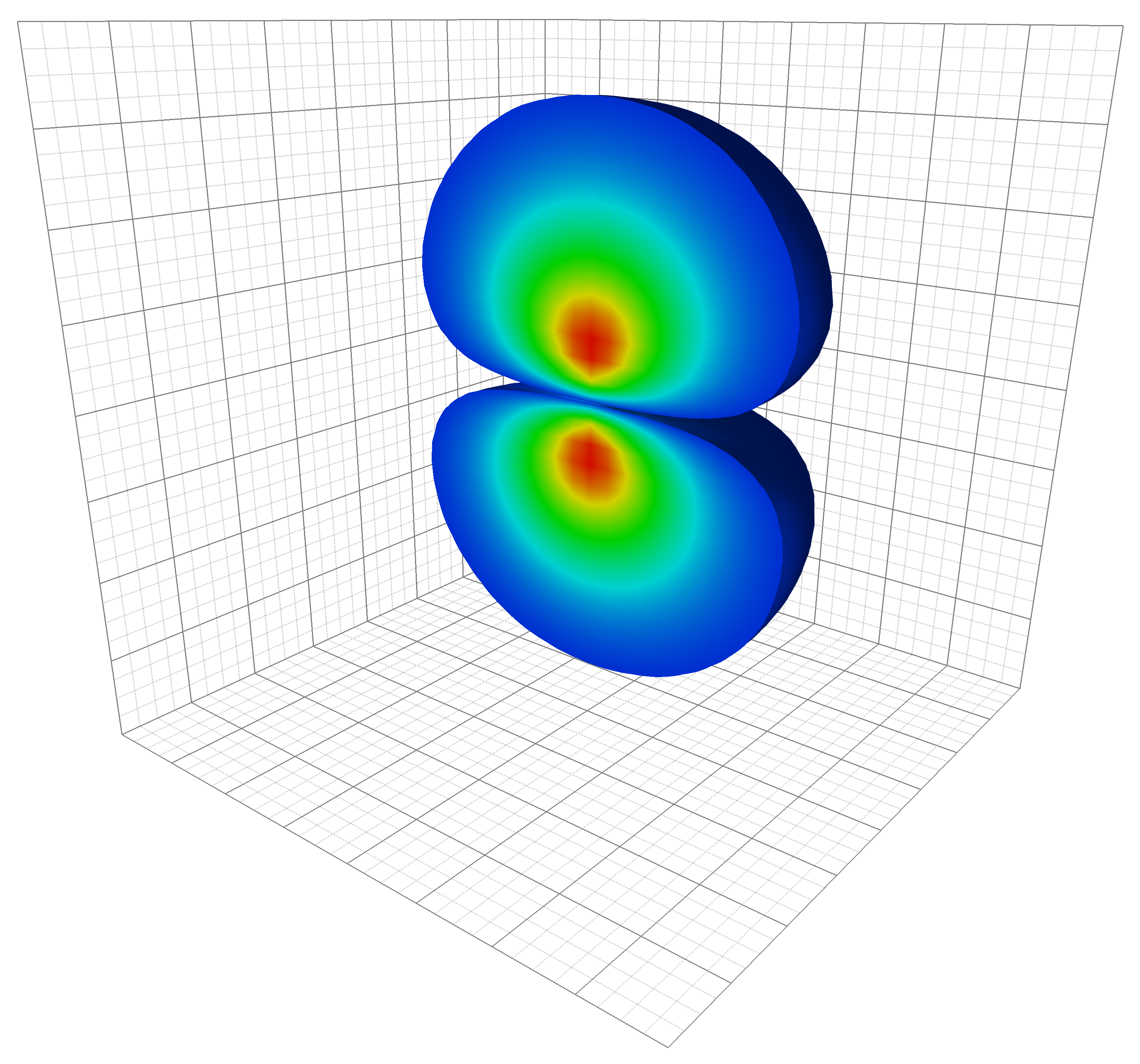}%
	\hspace{5em}
	\includegraphics[width=0.42\linewidth]{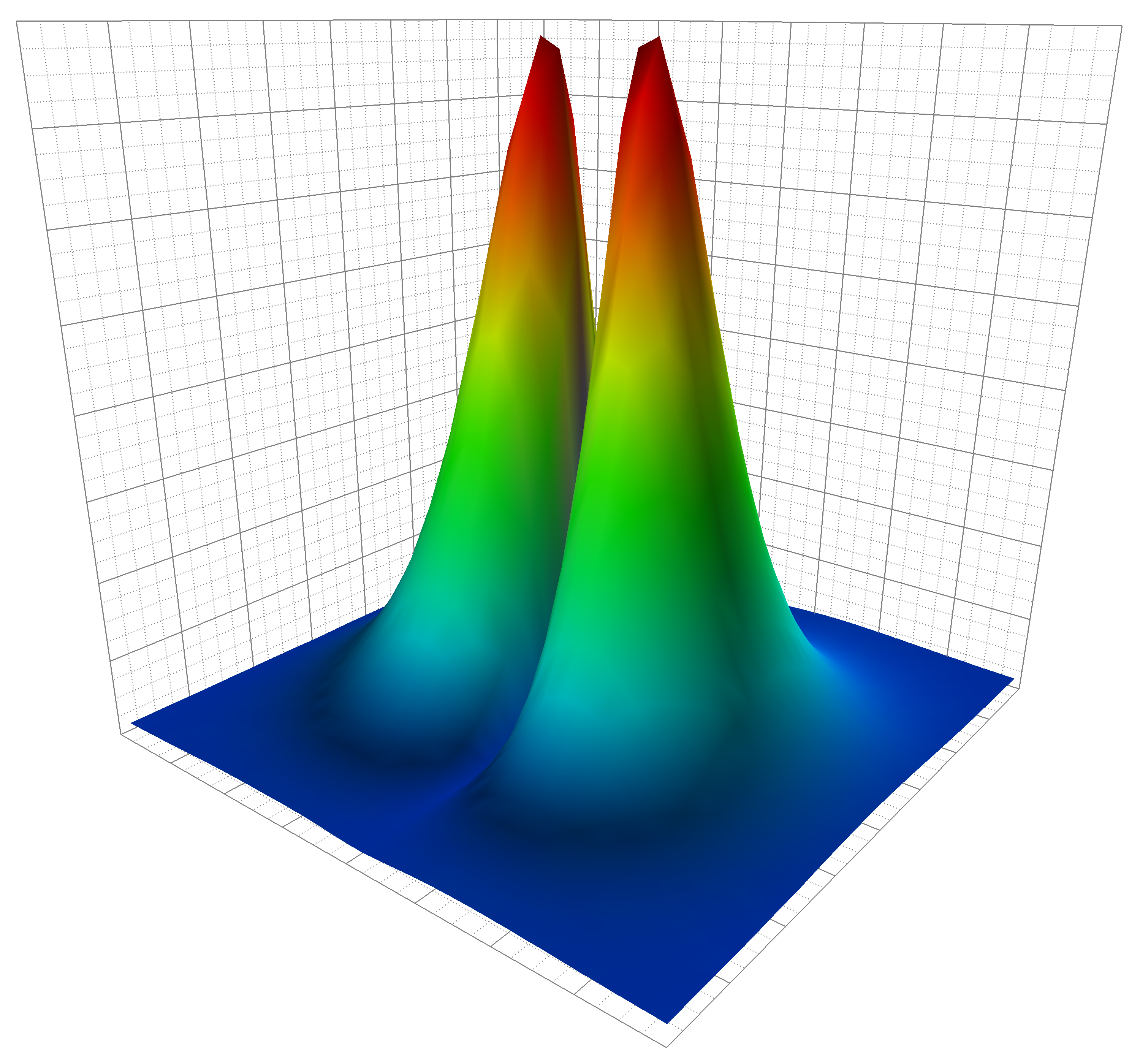}%
	\vspace{0.5em}
	\includegraphics[width=0.42\linewidth]{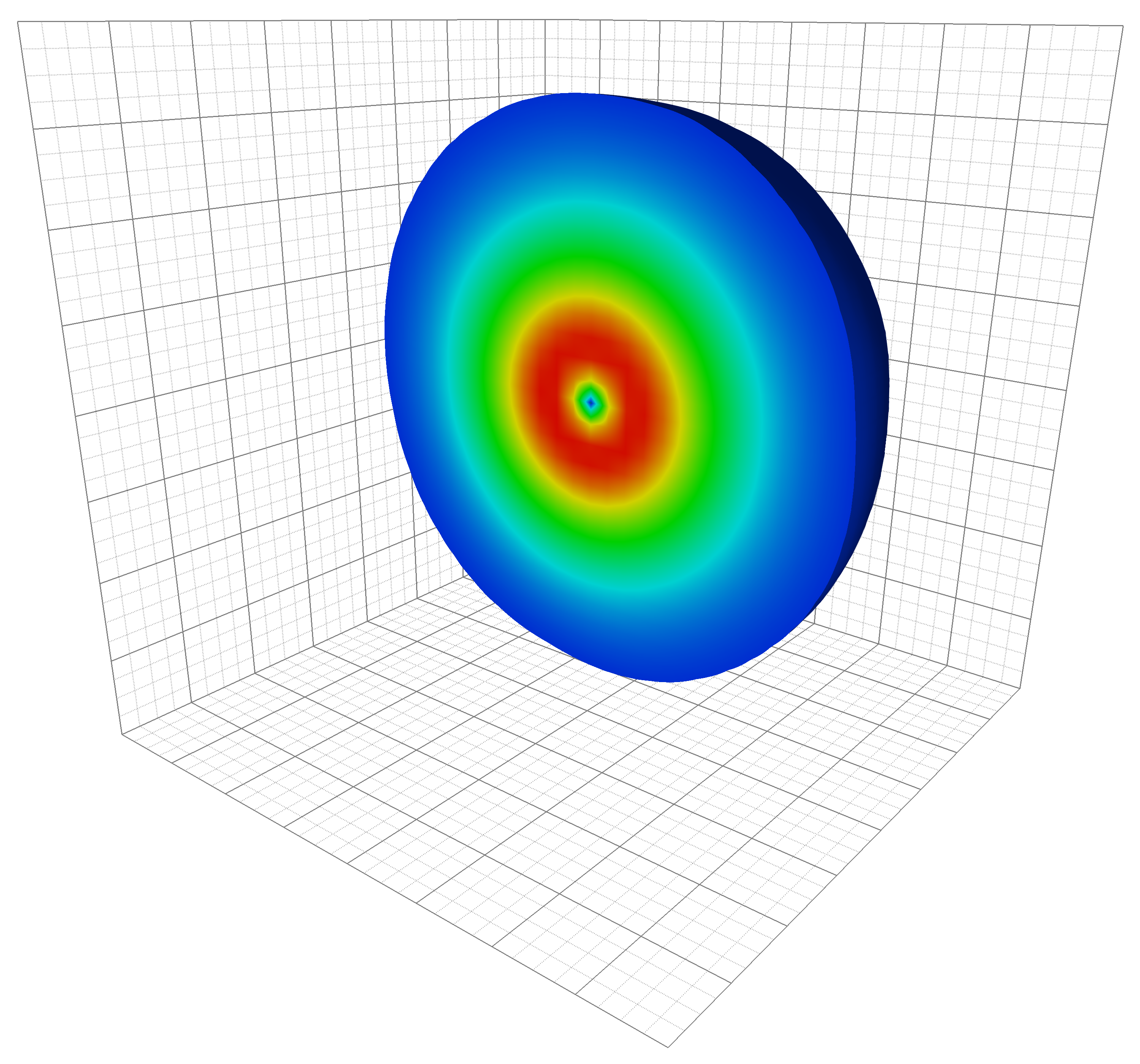}%
	\hspace{5em}
	\includegraphics[width=0.42\linewidth]{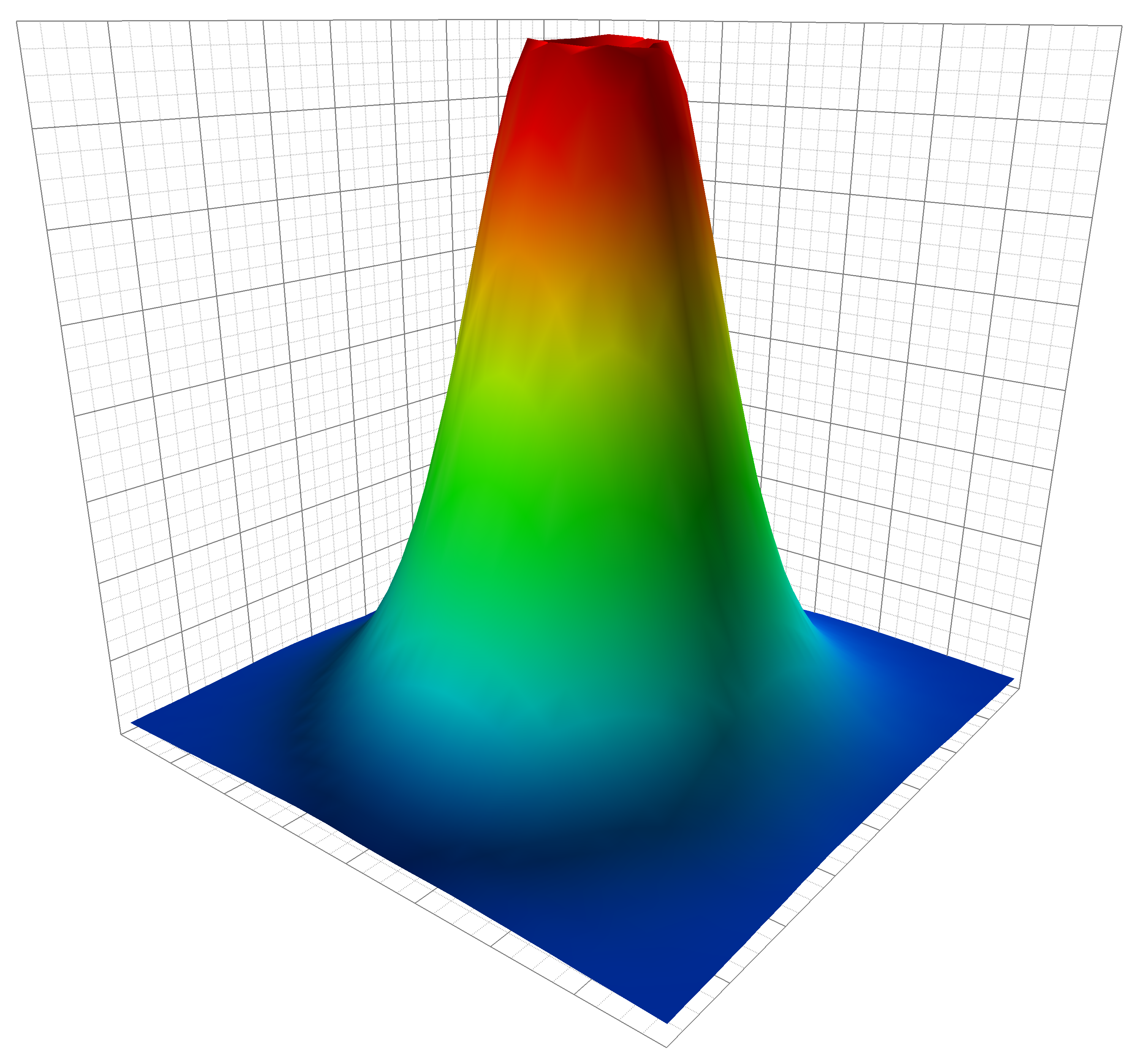}%
	\vspace{-0.5em}
	\caption{\label{fig:groundstatevis} Volume and surface renderings of the \((1,1)\) (\textbf{top}), \((3,1)\) (\textbf{middle}) and \((4,1)\) (\textbf{bottom}) components of the ground-state wavefunction. The \((1,1)\) and \((3,1)\) components are sliced through \(x = 0\), and the \((4,1)\) component is sliced through \(z = 0\). The expected wavefunction shapes are apparent, including the spherically symmetric \((1,1)\) component, the hourglass shape for the \((3,1)\) component and the ring in the \(xy\)-plane for the \((4,1)\) component.}
\end{figure*}

With the ideal Euclidean time determined, we use this to explore the relativistic wavefunction components for the three lowest-lying positive-parity states, starting with the ground state. We display in Fig.~\ref{fig:groundstatevis} the \((1,1)\), \((3,1)\) and \((4,1)\) wavefunction components as both volume and surface renderings, as described at the start of this section. Being the ground state, there are zero nodes in the wavefunction across all components.

Each visualisation reveals the structure expected from the discussion in Sec.~\ref{subsec:wavefunctions}. The upper \((1,1)\) component shows a clear \(s\)-wave structure, peaked at the origin and spherically symmetric. Meanwhile, the lower \((3,1)\) and \((4,1)\) components show a \(p\)-wave structure, starting zero at the origin and rising to a maximum at some nonzero radial distance. By eye, the angular dependence appears to accurately match the appropriate spherical harmonic from Eqs.~(\ref{eq:Y10}) and (\ref{eq:Y11}). This validates using parity and spin mismatches between the correlation-matrix source and sink spinor indices as a means to extract the relativistic wavefunction components. The angular dependence will be quantitatively verified in Sec.~\ref{subsec:positiveparityangular}.

For the \((3,1)\) component, we see an hourglass structure along the \(z\)-axis where the wavefunction amplitude is zero on the \(xy\)-plane and symmetric about the \(z\)-axis. This is typical of angular momentum quantum numbers \((\ell, m_\ell) = (1, 0)\), attributed to the \(\cos\theta\) that identifies the \(Y_1^0\) spherical harmonic. Indeed, \(\cos\theta\) is zero on the \(xy\)-plane (\(\theta = \pi/2\)) and one along the \(z\)-axis (\(\theta = 0\)).

The \((4,1)\) component, which we recall is sliced through the \(z\)-dimension, instead possesses a ring structure in the \(xy\)-plane. This is again typical of the relevant quantum numbers, \((\ell, m_\ell) = (1, \pm 1)\), and associated with the \(\sin\theta\) characterising the \(Y_1^1\) spherical harmonic. The \(\sin\theta\) implies opposite behaviour to \(Y_1^0\), with \(Y_1^1\) maximum on the \(xy\)-plane and zero along the \(z\)-axis. In taking the modulus of this spherical harmonic, as done for the wavefunction, the azimuthal \(e^{i\phi}\) dependence is removed and the ring structure seen in Fig.~\ref{fig:groundstatevis} is formed.

\begin{figure*}
	\includegraphics[width=0.42\linewidth]{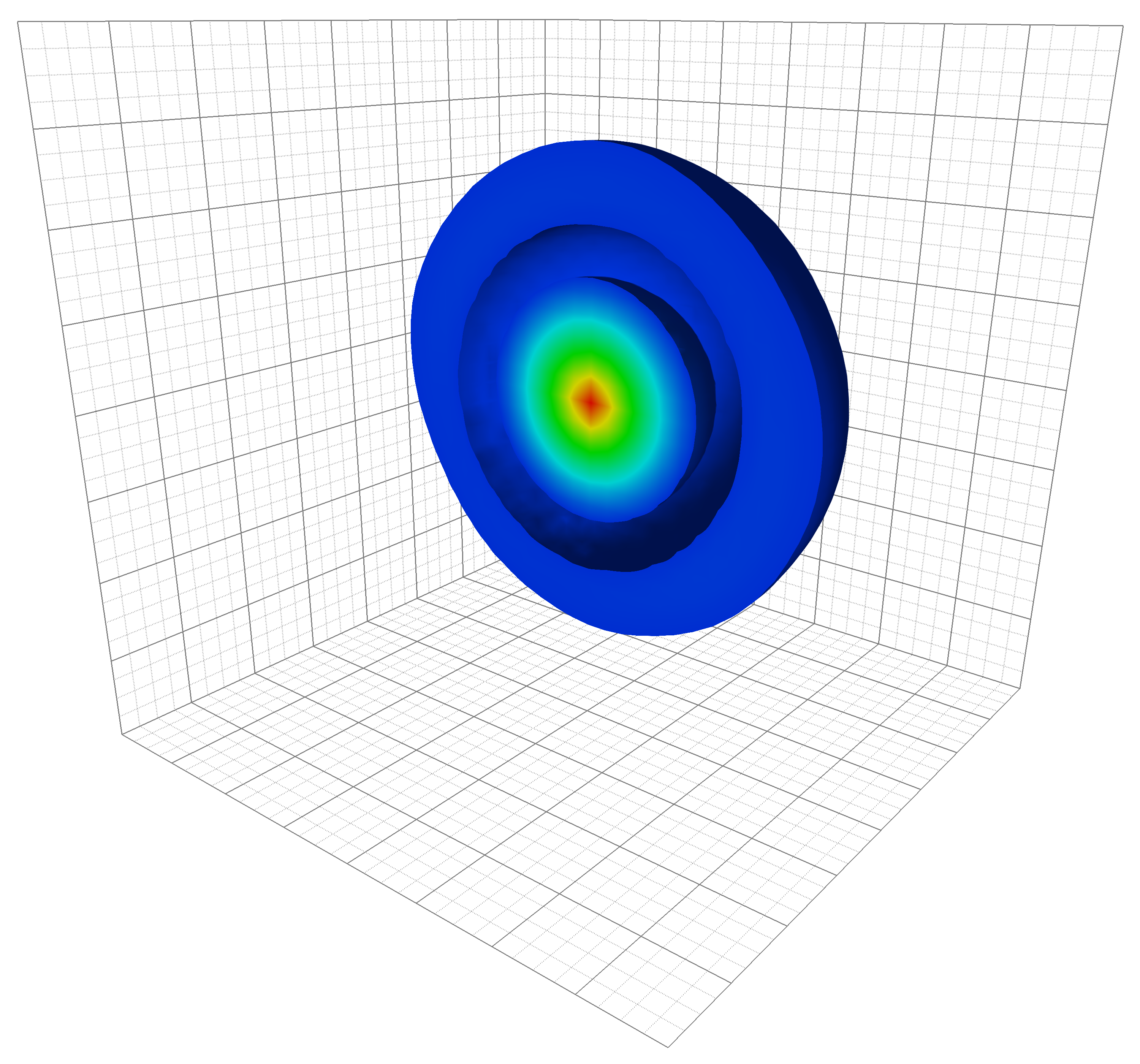}%
	\hspace{5em}
	\includegraphics[width=0.42\linewidth]{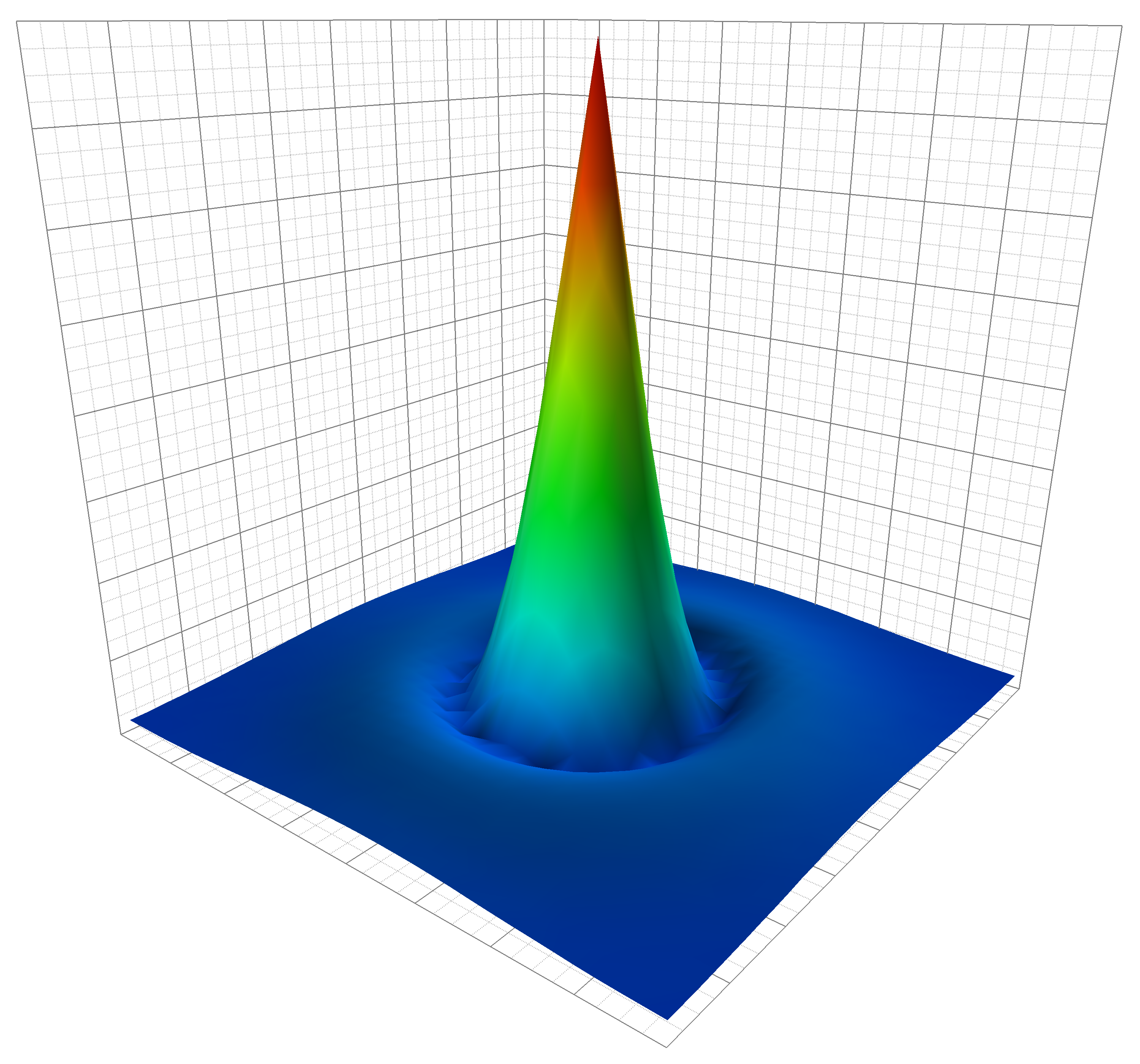}%
	\vspace{0.5em}
	\includegraphics[width=0.42\linewidth]{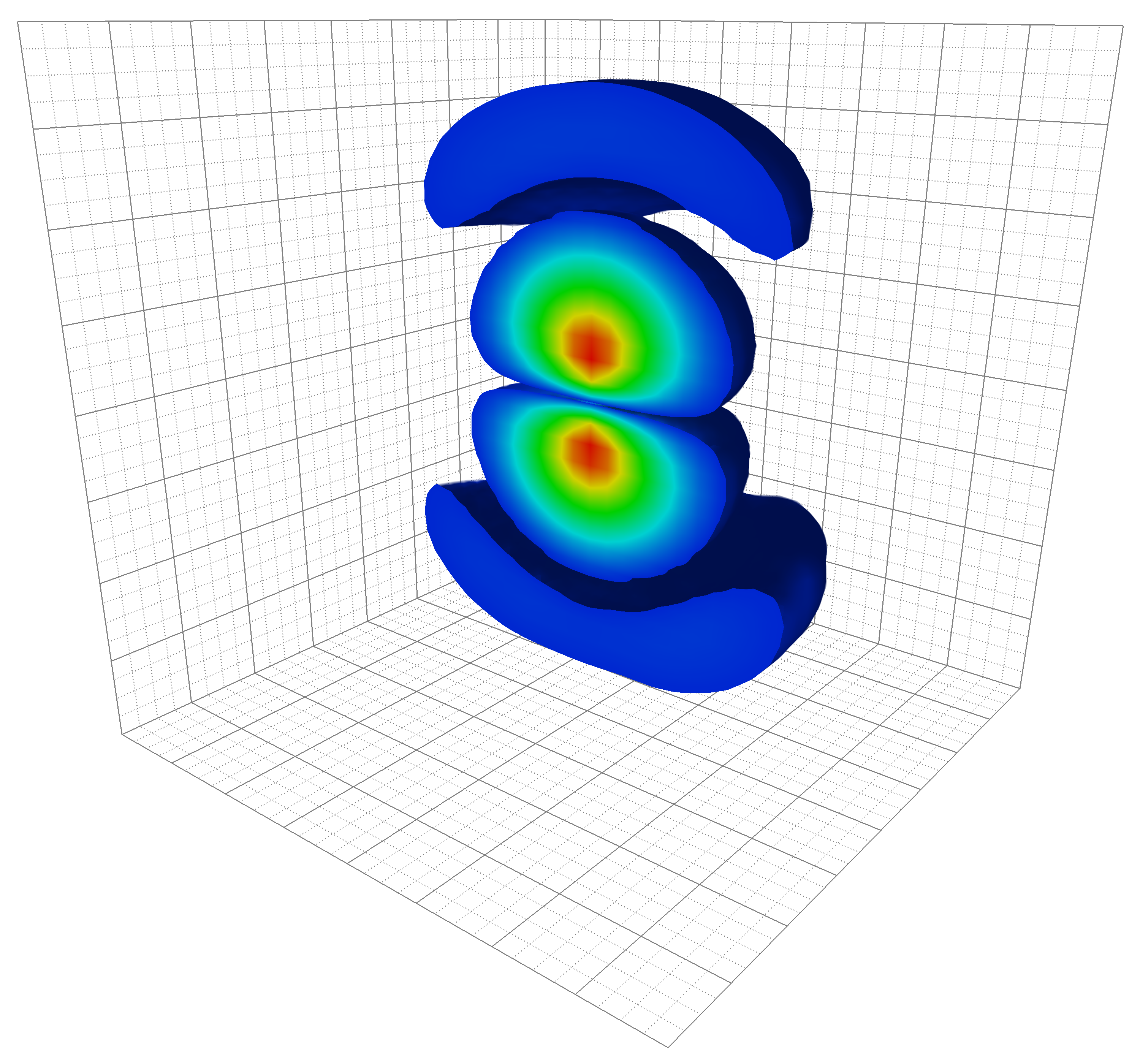}%
	\hspace{5em}
	\includegraphics[width=0.42\linewidth]{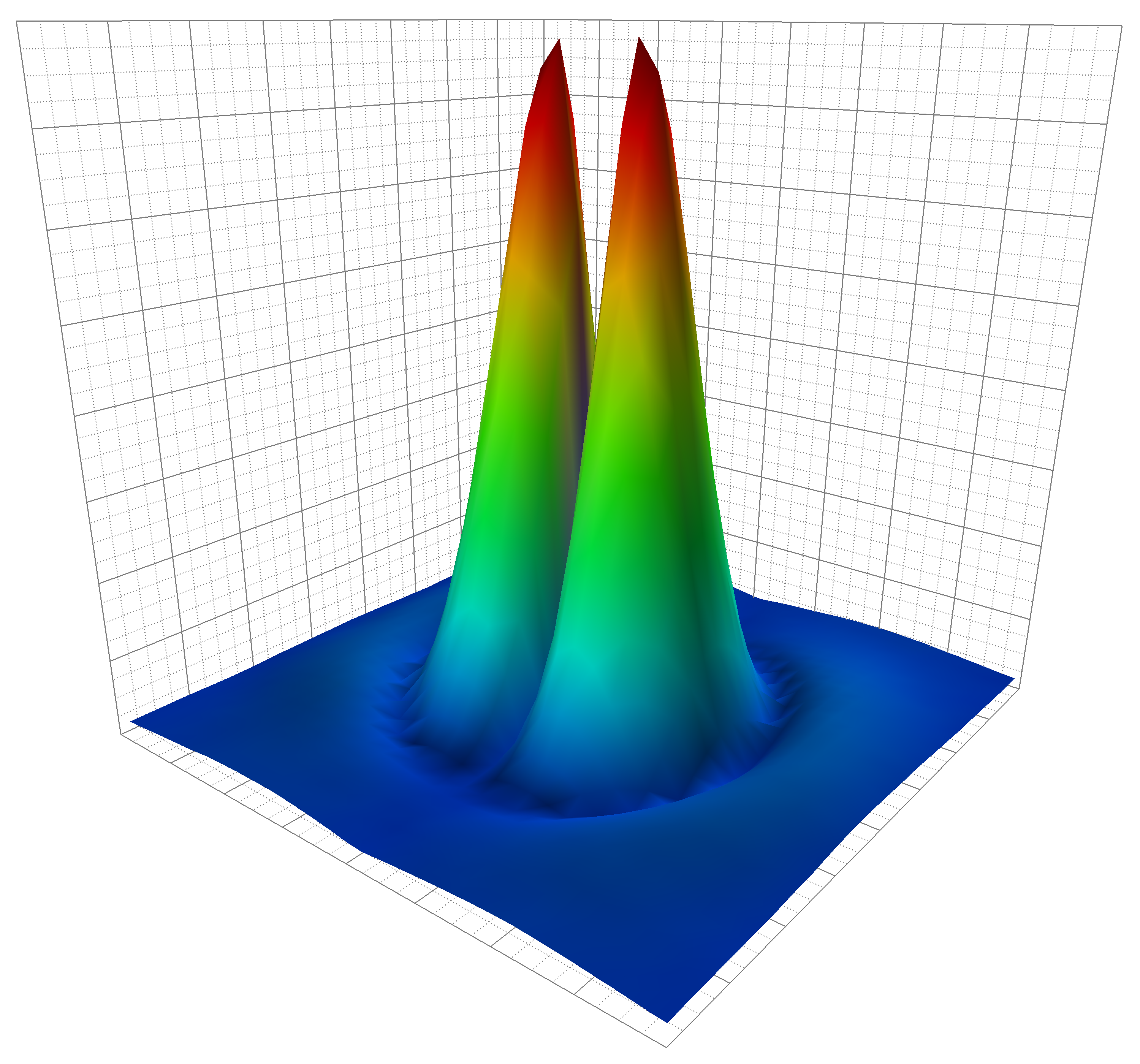}%
	\vspace{0.5em}
	\includegraphics[width=0.42\linewidth]{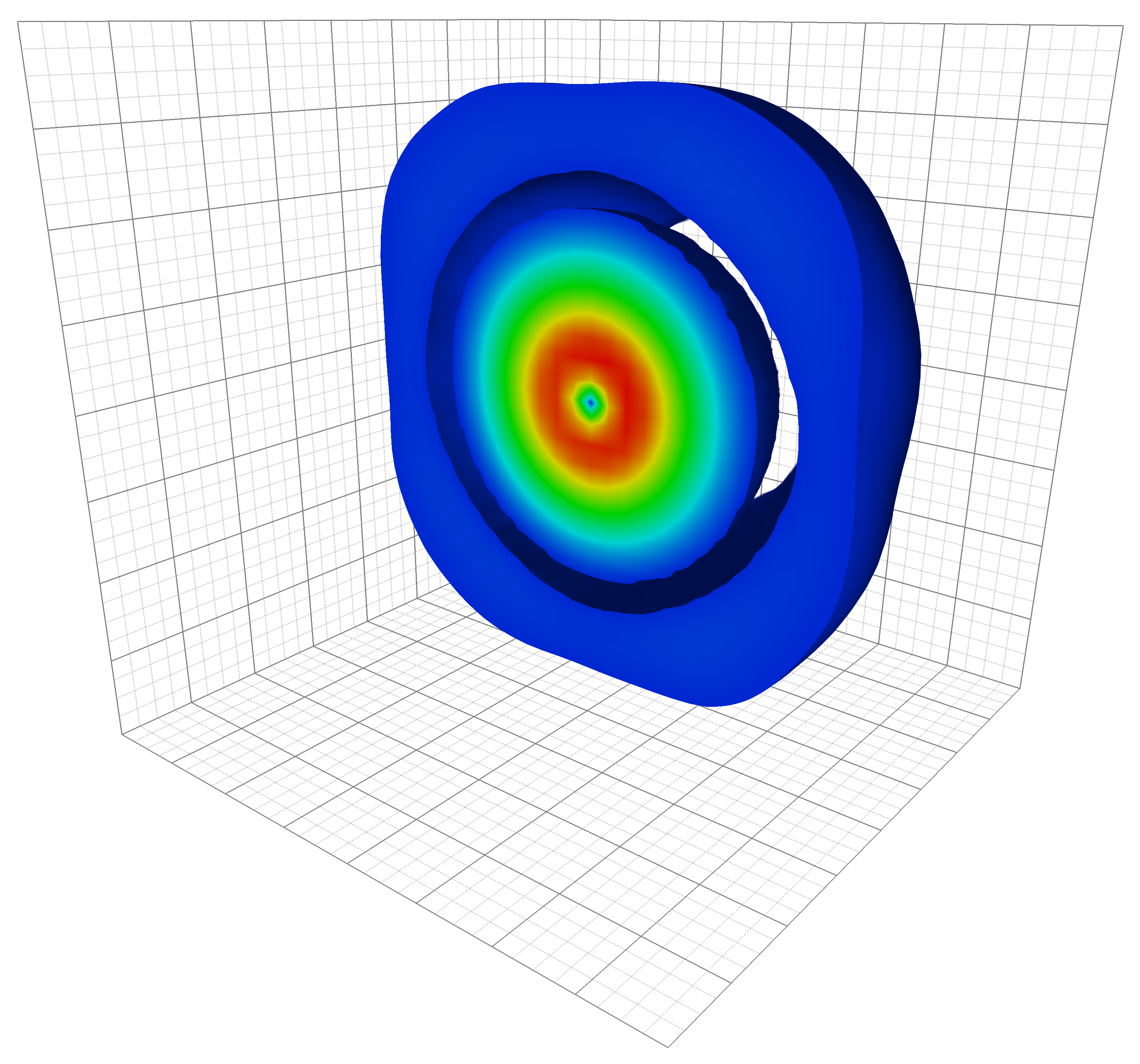}%
	\hspace{5em}
	\includegraphics[width=0.42\linewidth]{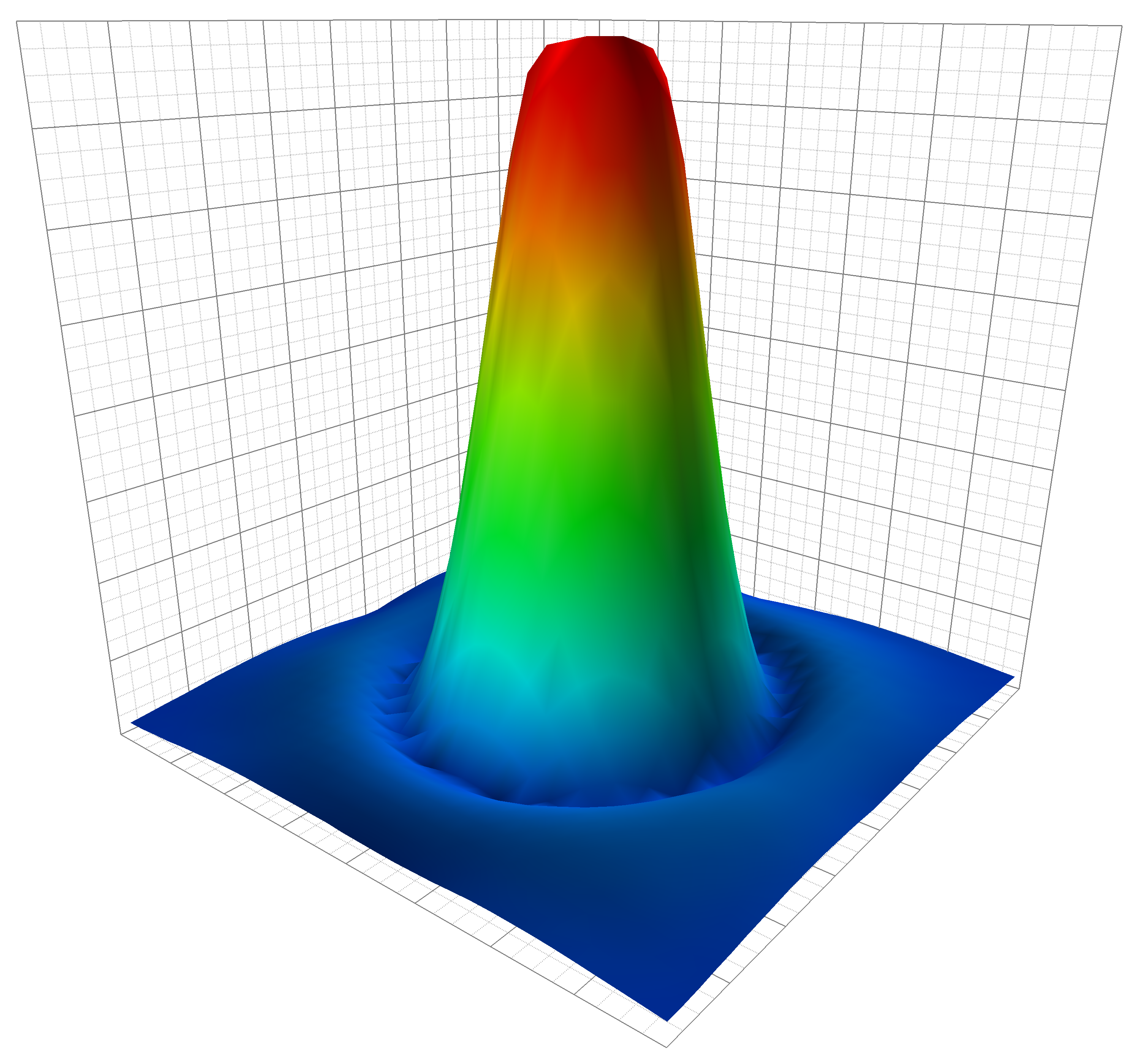}%
	\vspace{-0.5em}
	\caption{\label{fig:firstpositiveparityvis} Volume and surface renderings of the \((1,1)\) (\textbf{top}), \((3,1)\) (\textbf{middle}) and \((4,1)\) (\textbf{bottom}) components of the first positive-parity excited state's wavefunction. The node initially seen in Fig.~\ref{fig:euclideantimefirstexcitation} is found in all three components. It is reflected in the volume renderings by a gap in the wavefunction, where its amplitude temporarily drops below the rendering threshold in the vicinity of a node.}
\end{figure*}

Next, we move to the first positive-parity excitation, and the equivalent visualisations are presented in Fig.~\ref{fig:firstpositiveparityvis}. The same general wavefunction shapes illuminated by the ground state are again observed here, though with one key difference:\ a node is now present in the wavefunction for all spinor components. This was initially observed for the upper component in the surface renders of Fig.~\ref{fig:euclideantimefirstexcitation}, for which the node is identified as a small valley in the render. We now have the first-ever qualitative confirmation that the lower wavefunction components also inherit the node for this state. The nodes are best illustrated in the volume renders, where they appear as a void between the inner and outer shells due to the wavefunction amplitude temporarily falling below the rendering threshold in the region around a node.

\begin{figure*}
	\includegraphics[width=0.42\linewidth]{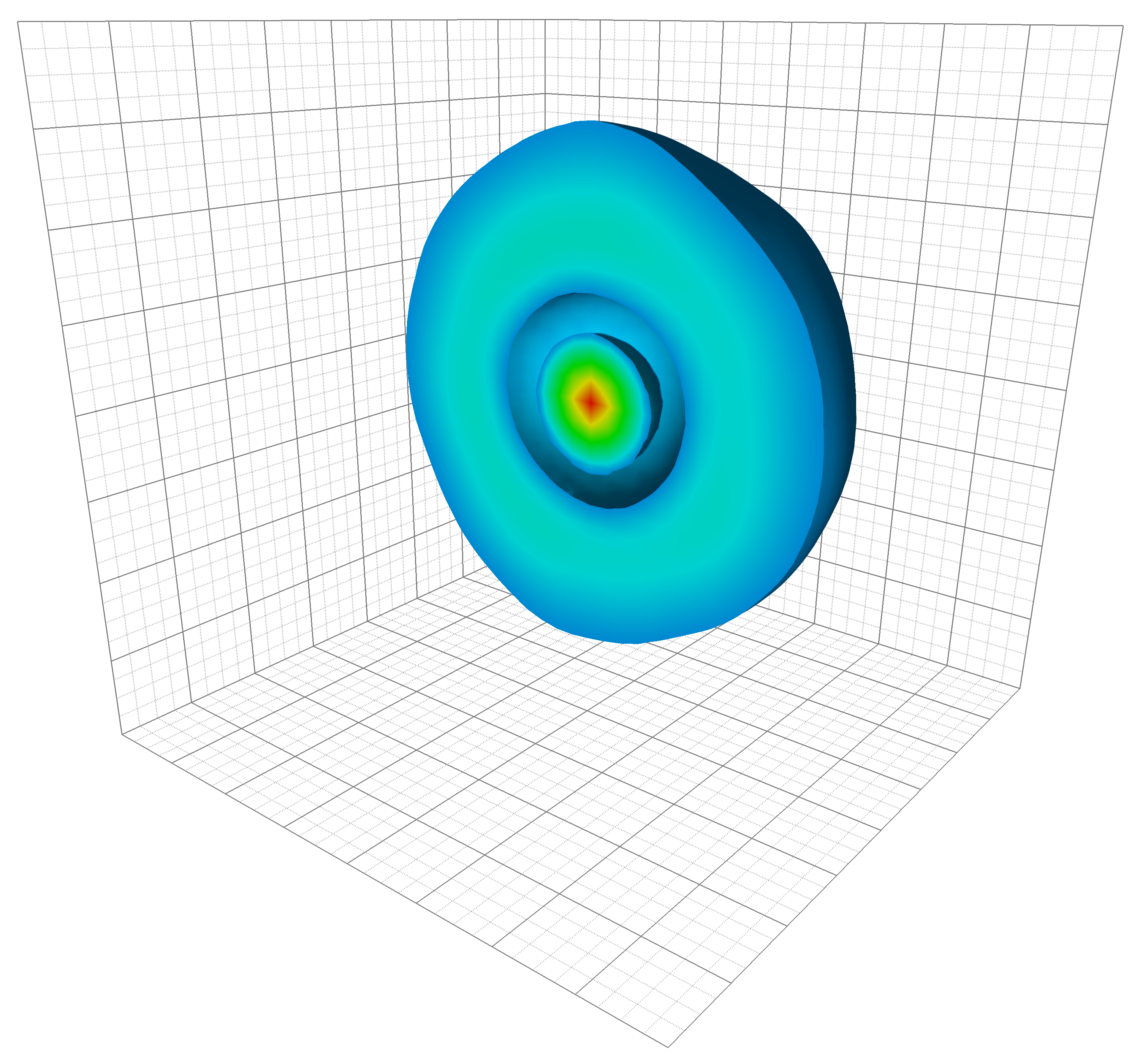}%
	\hspace{5em}
	\includegraphics[width=0.42\linewidth]{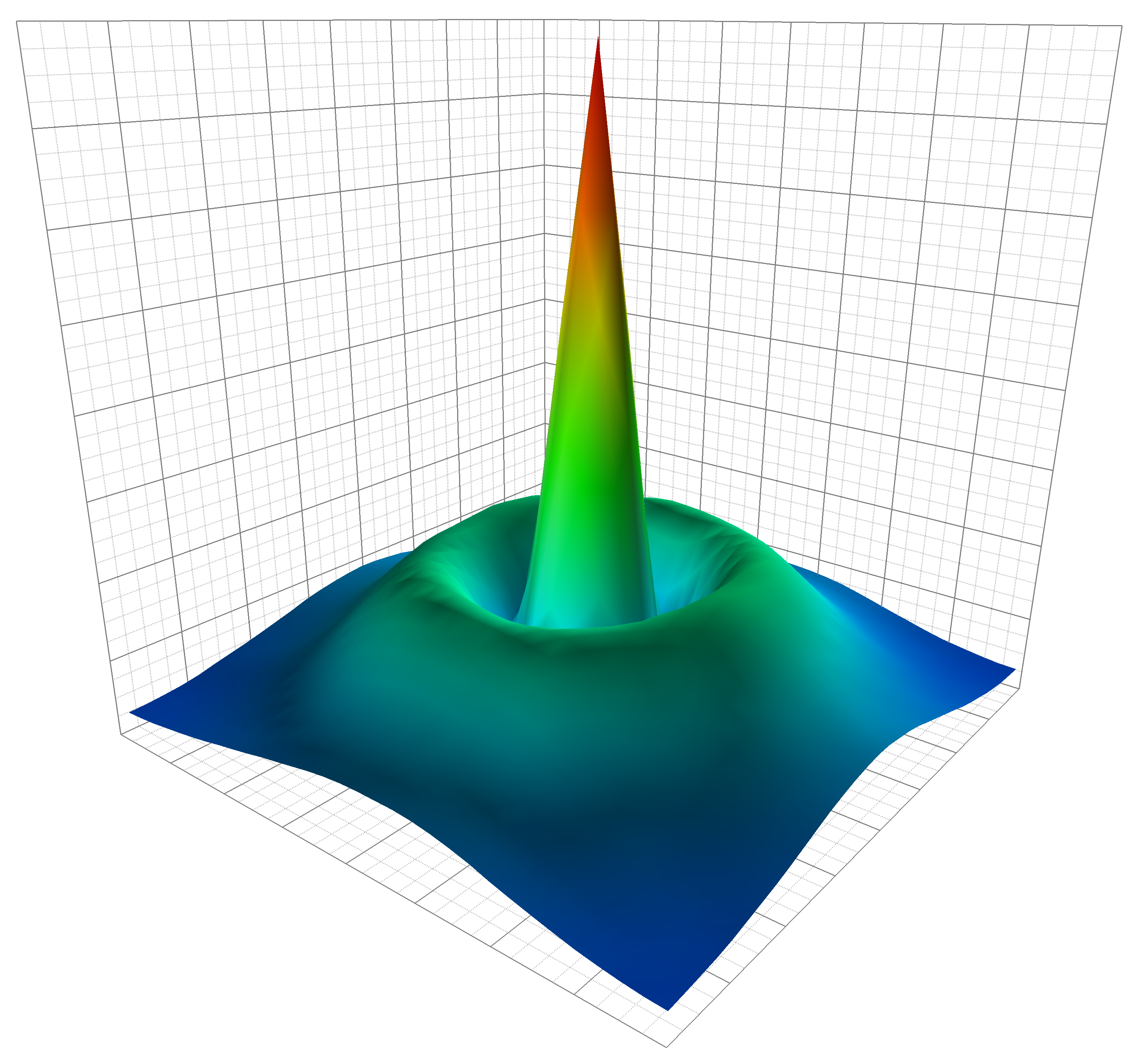}%
	\vspace{0.5em}
	\includegraphics[width=0.42\linewidth]{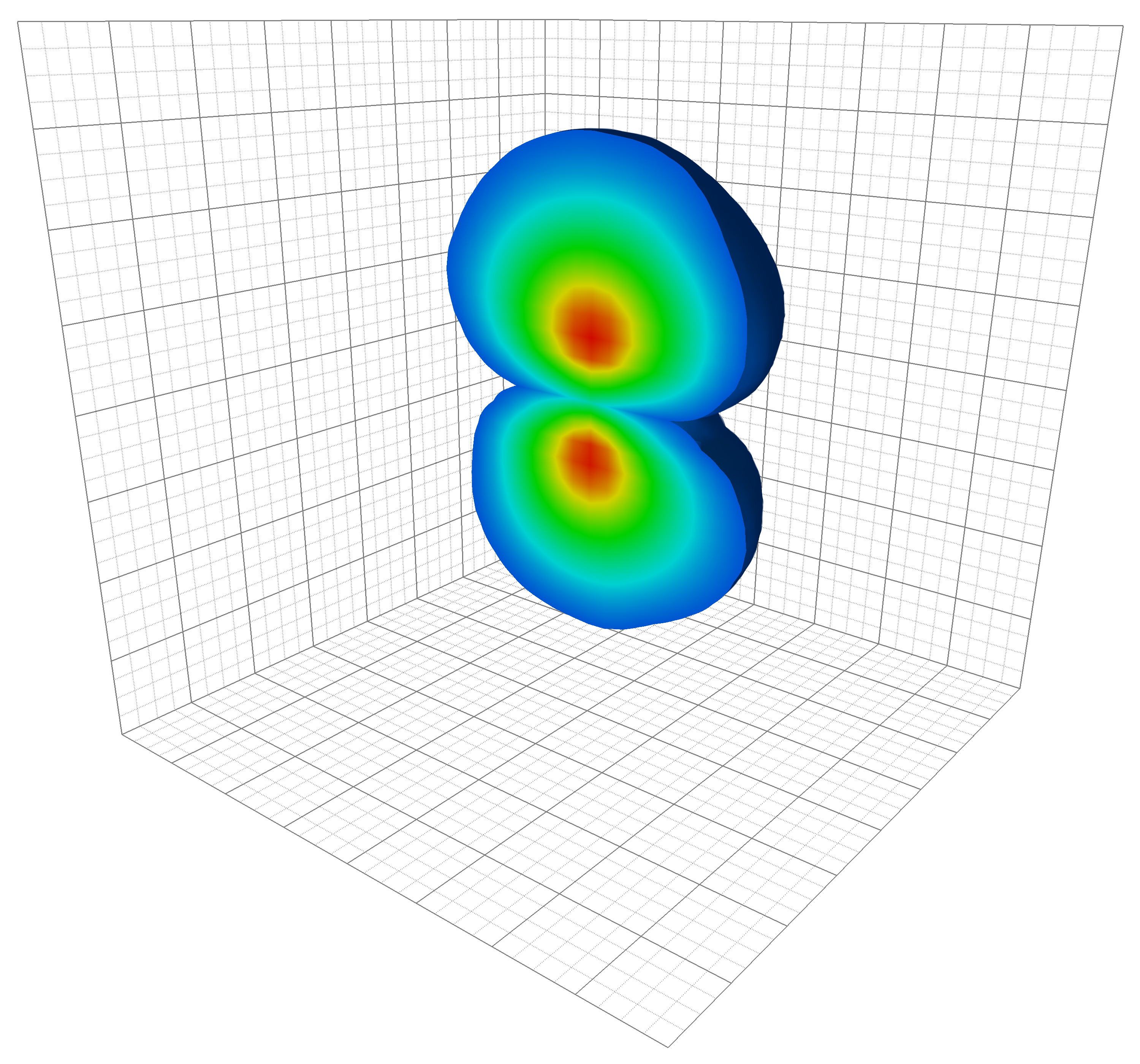}%
	\hspace{5em}
	\includegraphics[width=0.42\linewidth]{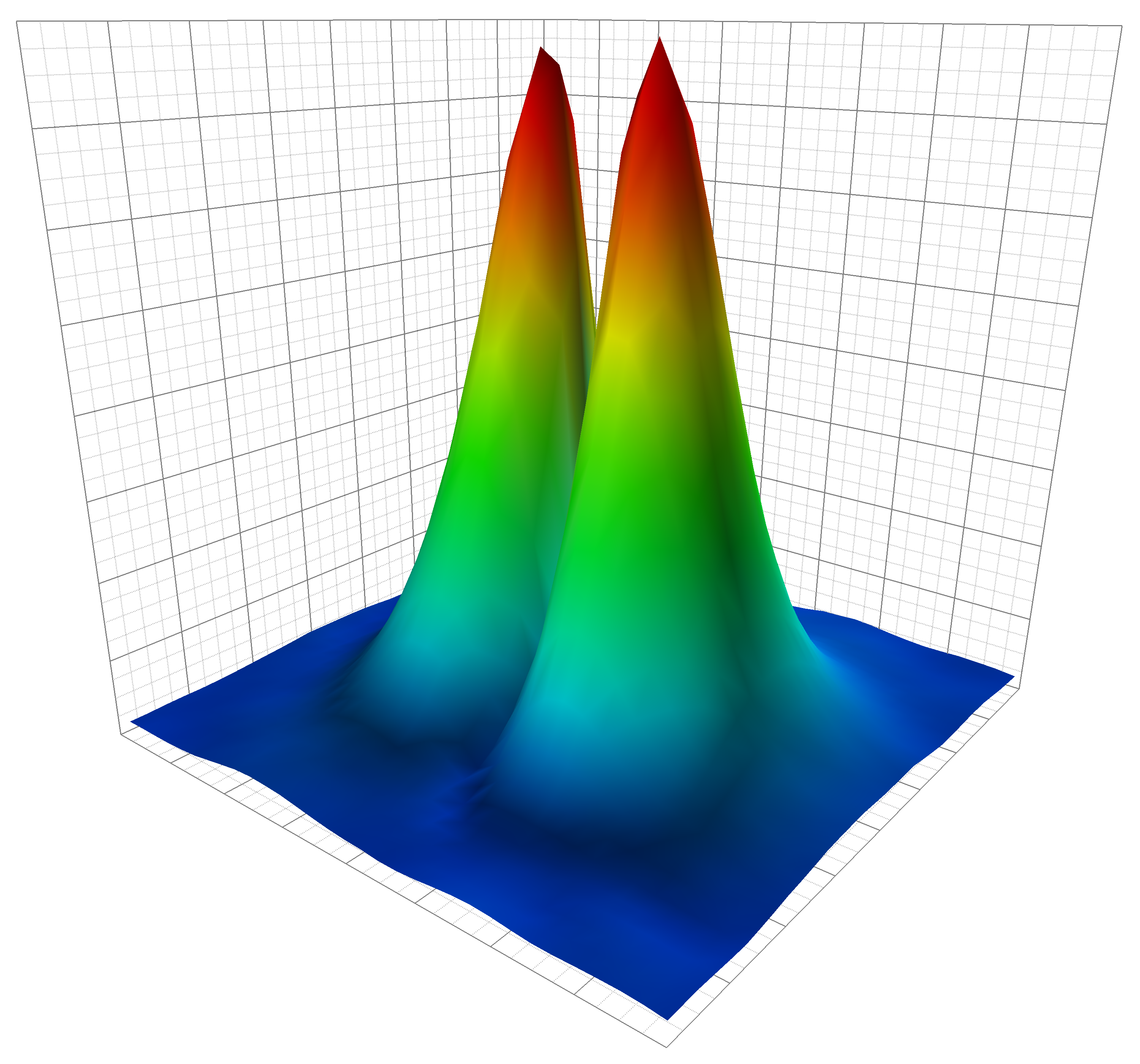}%
	\vspace{0.5em}
	\includegraphics[width=0.42\linewidth]{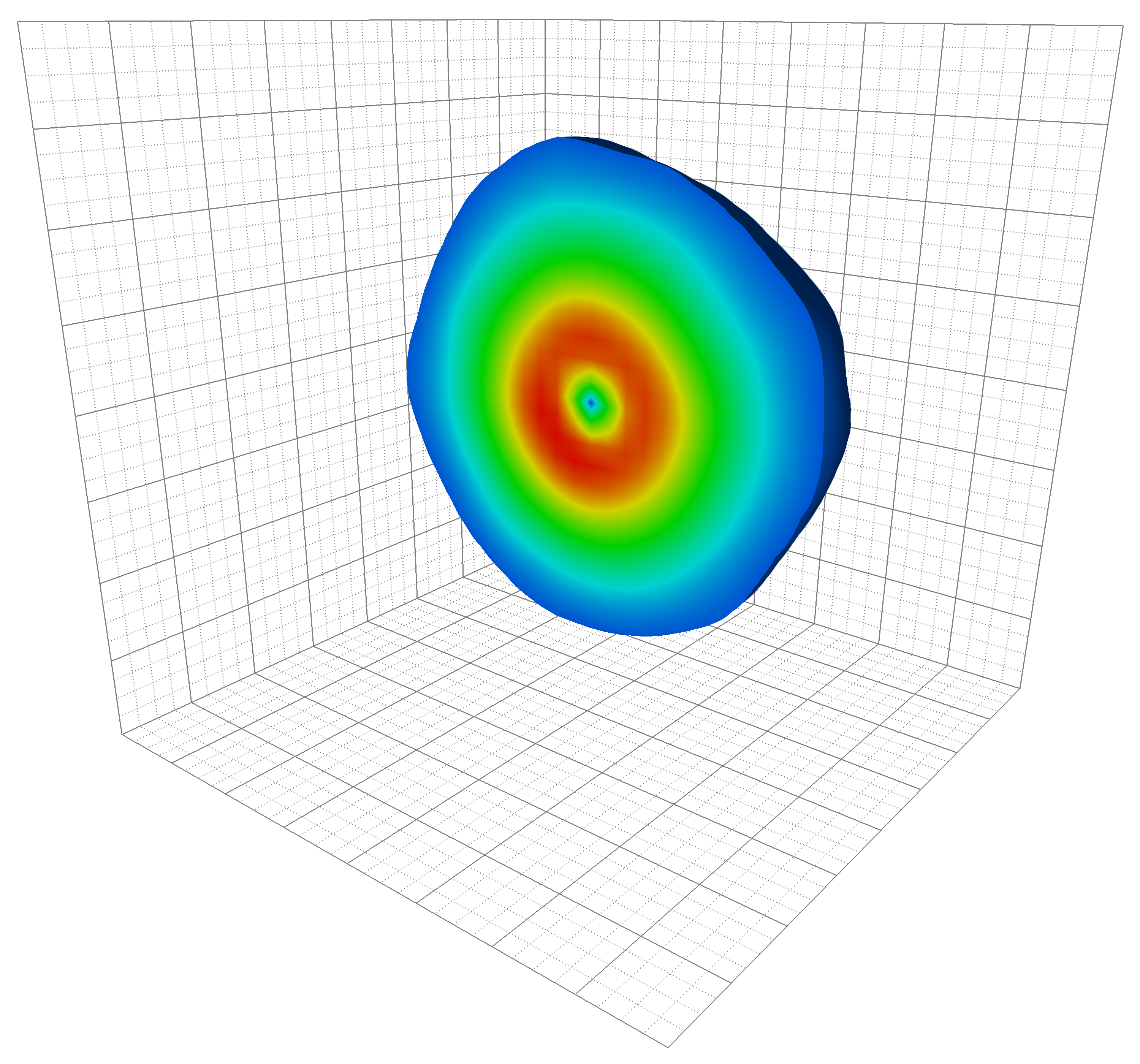}%
	\hspace{5em}
	\includegraphics[width=0.42\linewidth]{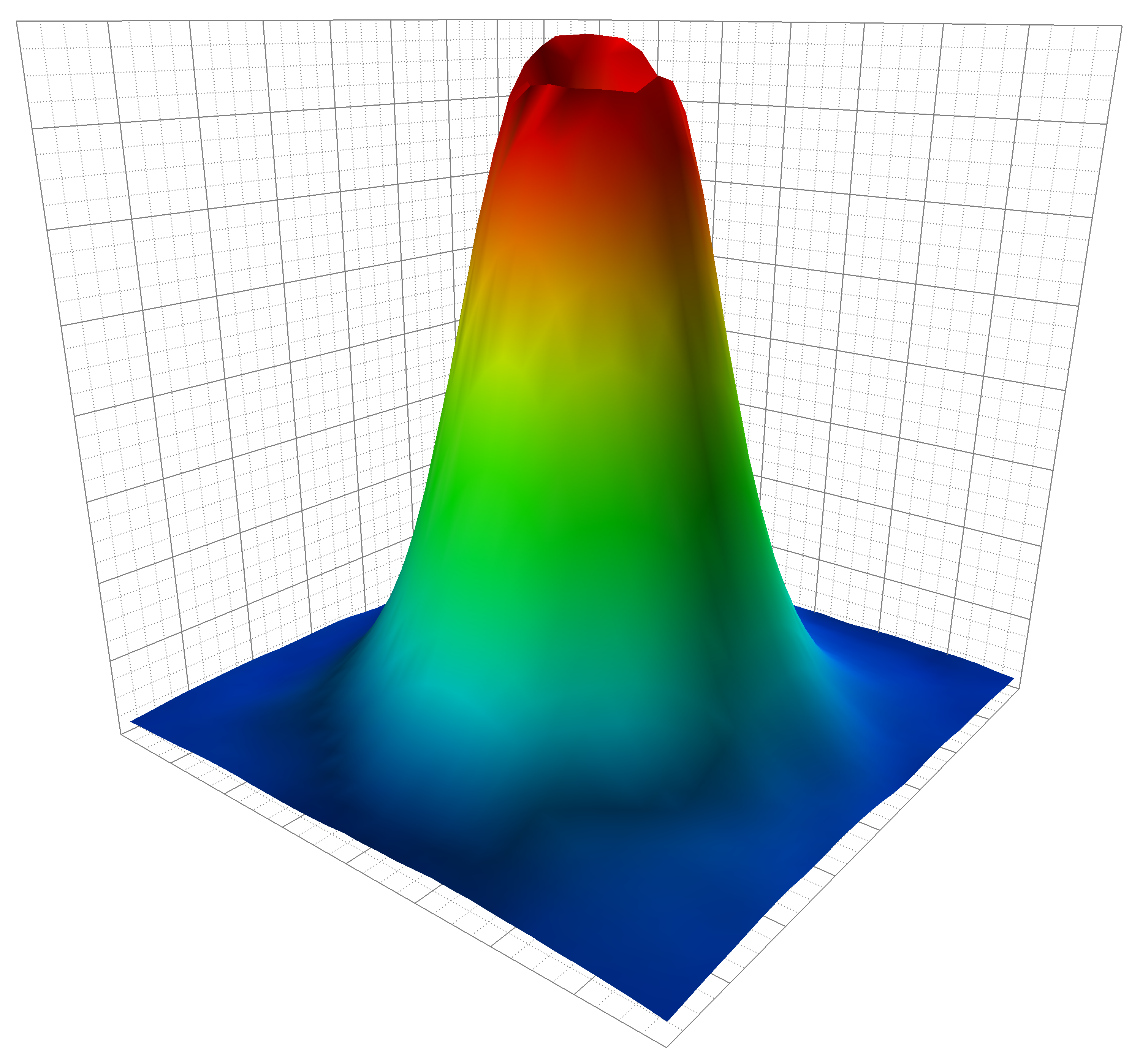}%
	\vspace{-0.5em}
	\caption{\label{fig:secondpositiveparityvis} Volume and surface renderings of the \((1,1)\) (\textbf{top}), \((3,1)\) (\textbf{middle}) and \((4,1)\) (\textbf{bottom}) components for the second positive-parity excitation. In this case, a node is still found in the upper \((1,1)\) component, but it is absent from the lower \((3,1)\) and \((4,1)\) components. This is clearest through the surface renderings, which plot the wavefunction to the edge of the volume. Small distortions are visible in the wavefunctions for this state.}
\end{figure*}

Finally, we take another step up the spectrum to the second positive-parity excitation. The visualisations are provided in Fig.~\ref{fig:secondpositiveparityvis}. Immediately, a salient feature of this state can be perceived. A single node is again found in the upper \((1,1)\) component, but unlike the first excited state, an equivalent node is absent from the lower wavefunction components. This is primarily evident from the surface renderings, which plot the wavefunction amplitude to the edge of the lattice volume. In other words, there is a mismatch in nodes between the upper (\(s\)-wave) and lower (\(p\)-wave) components of the second positive-parity excited state.

This is incredibly fascinating, and suggests that there are two distinct ways in which nodes can form when constructing the wavefunction from the individual interpolating fields (Eq.~(\ref{eq:wavefunction})). This is substantiated by observing that compared to the first excited state, this node in the upper component of the second excited state is considerably sharper and also located closer to the origin. These properties further distinguish the nodes between these states.

Taking a step back, we can note that the same general wavefunction shapes have again been realised, though in this case small distortions are evident. This is most noticeable in the volume renders of the \((3,1)\) and \((4,1)\) components. For the former, there is no gap at the \(xy\)-plane where the spherical harmonic vanishes, and there is a slight bias towards \(z > 0\). For the latter, there are visible deviations from spherical symmetry at the boundary of the rendered region. It is unsurprising that the wavefunctions become noisier as one moves up the spectrum, resulting in these slight imperfections. Naturally, the noise will be most prevalent when displaying only a single wavefunction component, without any averaging. In calculating the radial wavefunctions in Sec.~\ref{subsec:positiveparityradial}, these effects will be mitigated by averaging both over all lattice sites at a given radius and the various wavefunction components predicted to have the same radial dependence.

\subsection{Angular dependence} \label{subsec:positiveparityangular}
With the key features of the low-lying positive-parity wavefunctions identified through visualisation, our next endeavour is to extract the radial wavefunction dependence. To achieve this, it is first necessary to remove the angular dependence. Ideally, one would simply divide by the appropriate spherical harmonic from the Dirac equation for each wavefunction component, as listed in Table~\ref{tab:lm_l}. Hence, it is crucial to establish whether the angular dependence is accurately described by these spherical harmonics. We have seen in the visualisations of Sec.~\ref{subsec:positiveparityvis} that this appears to apply near the origin, though it is prudent to check for any subtle deviations that may arise at large radial distances.

To this end, after dividing by the predicted spherical harmonic value at each lattice site, a scatter plot is produced that displays all points at a given radius individually. If the angular dependence has been accurately accounted for, these points should cluster on top of each other to within a high degree of precision. After verifying that this holds, all points with the same radius can be averaged to obtain an improved estimate of the radial wavefunction.

\begin{figure}
	\includegraphics[width=\linewidth]{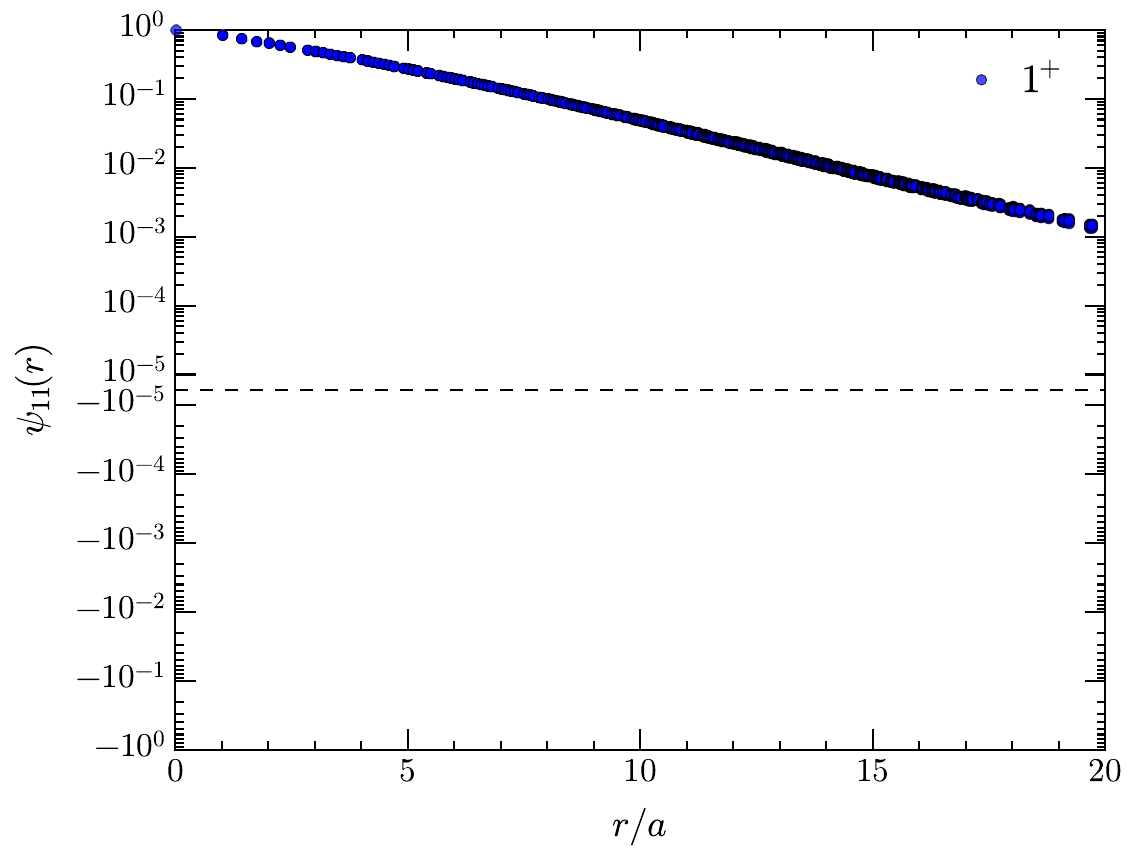}%
	\vspace{0.5em}
	\includegraphics[width=\linewidth]{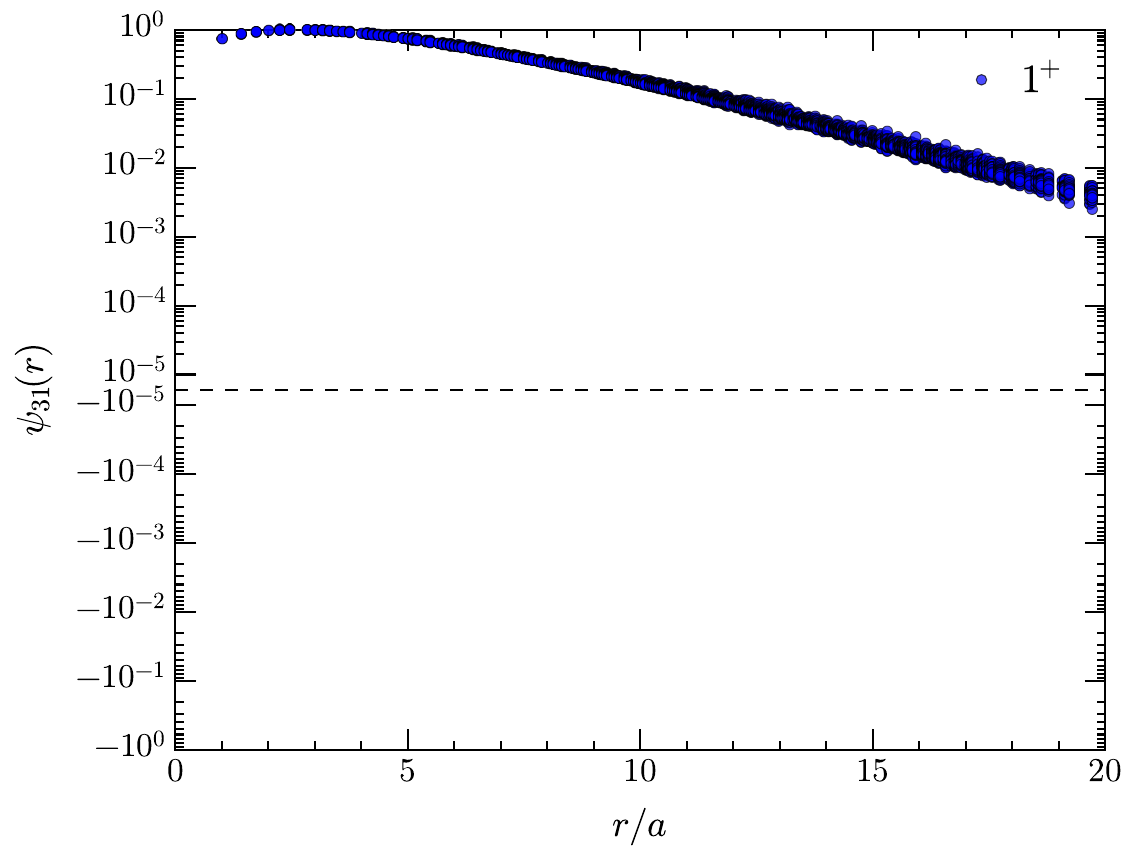}%
	\vspace{0.5em}
	\includegraphics[width=\linewidth]{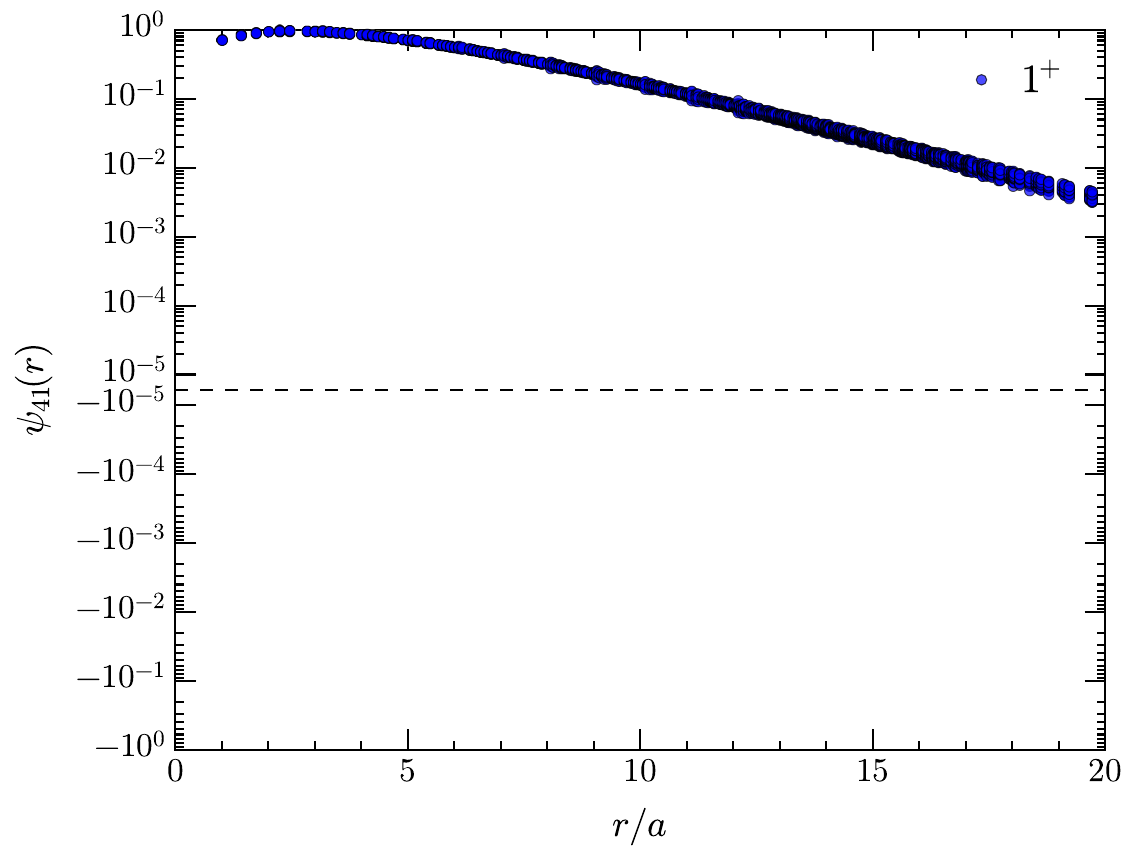}%
	\vspace{-0.5em}
	\caption{\label{fig:groundstateangular} Scatter plots of the \((\beta,\alpha) = (1,1)\) (\textbf{top}), \((3,1)\) (\textbf{middle}) and \((4,1)\) (\textbf{bottom}) wavefunction components for \(n^p = 1^+\) as a function of radial distance, with the appropriate spherical harmonic divided out at each lattice site. Only a small spread in values at a fixed radius is observed, and even then only at large radial distances. This indicates the angular dependence of the calculated wavefunctions is consistent with the spherical harmonics predicted by the Dirac equation.}
\end{figure}

As with the visualisations in Sec.~\ref{subsec:positiveparityvis}, we produce these scatter plots for the \((\beta,\alpha) = (1,1)\), \((3,1)\) and \((4,1)\) wavefunction components of the ground state. This again covers the three different relevant spherical harmonics. A symmetric log scale is used for the vertical axis of these plots. This is optimal for displaying nodes, but necessitates a linear region around zero to accommodate the transition from positive to negative values. This is implemented between \(\pm 10^{-5}\), which is sufficiently narrow to have negligible impact on the plots. The plots are shown in Fig.~\ref{fig:groundstateangular}.

Looking at the \((1,1)\) component, which under this procedure is unmodified (\(Y_0^0(\theta,\phi) = 1\)), we find effectively no deviation from a single trend line across all lattice sites. This strongly indicates that the resolved wavefunction is spherically symmetric, with no suggestion of nontrivial angular dependence.

Turning now to the \((3,1)\) component, a small but visible spread does develop at large radial distances. This is perhaps unsurprising. In dividing out by small values of \(\cos\theta\) near the \(xy\)-plane, minor statistical fluctuations in the wavefunction will be magnified. The \(xy\)-plane itself, on which \(\cos\theta = 0\), is excluded from these calculations. Up to radial distances of \(r/a \simeq 12\), the trend line is near perfect. Based on this, there is no evidence in Fig.~\ref{fig:groundstateangular} of statistically significant departure from the angular dependence of the \(Y_1^0\) spherical harmonic.

Finally, the \((4,1)\) component shows a similar spread at very large radial distances. The \(z\)-axis, along which \(\sin\theta = 0\) and \(\phi\) is undefined, plays a similar role to the \(xy\)-plane for the \((3,1)\) component. That is, dividing out by small values of \(\sin\theta\) will amplify any statistical fluctuations. However, since only a single axis is problematic, as opposed to an entire plane, the effect is not as severe. We again conclude that, within the observed precision, the angular dependence is entirely accounted for by the \(Y_1^{+1}\) spherical harmonic. Note also that the radial dependence of the \((3,1)\) and \((4,1)\) components are consistent with each other, as expected.

\begin{figure}
	\includegraphics[width=\linewidth]{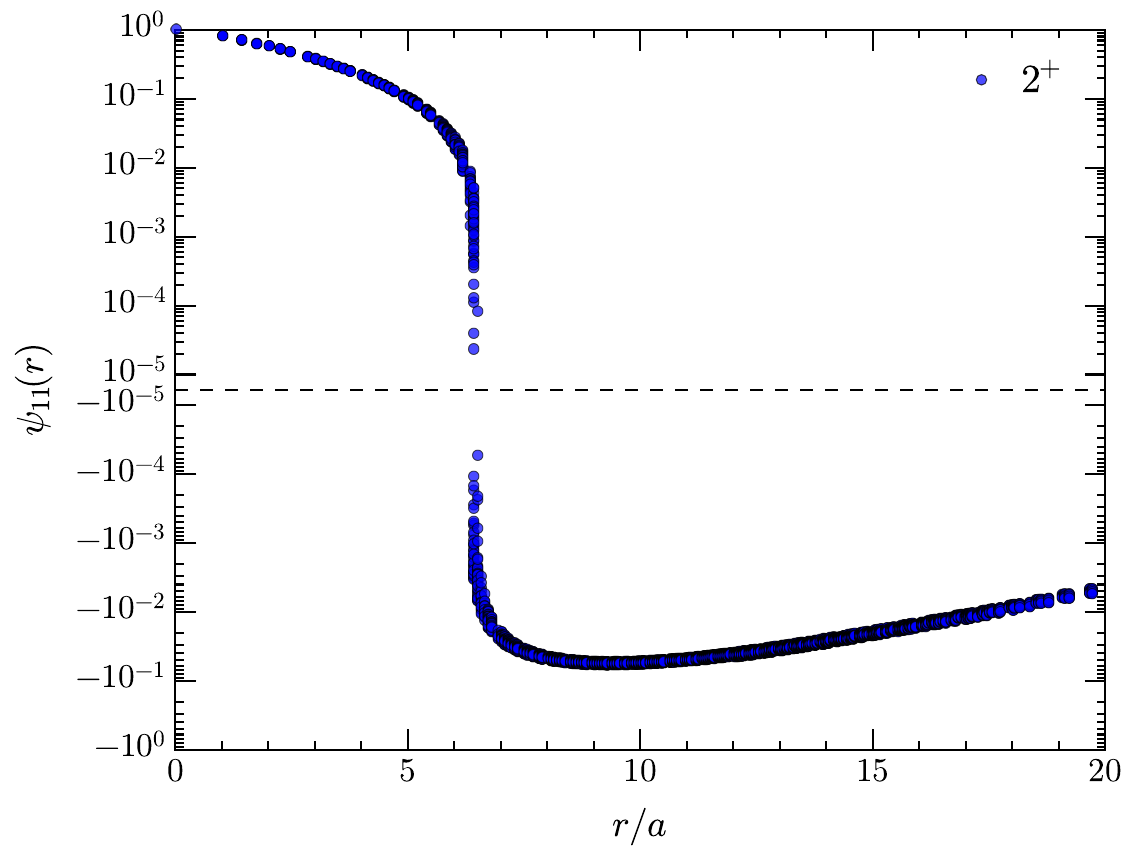}%
	\vspace{0.5em}
	\includegraphics[width=\linewidth]{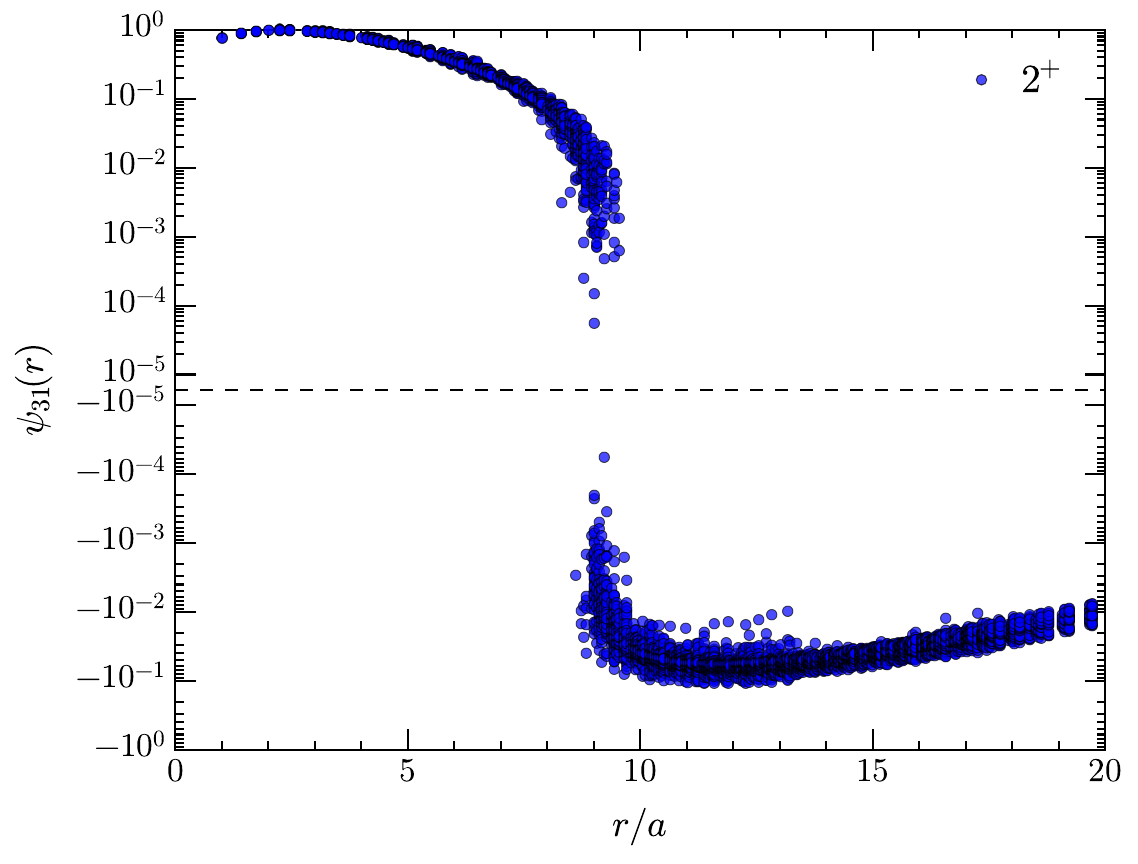}%
	\vspace{-0.5em}
	\caption{\label{fig:firstpositiveparityangular} Scatter plots of the \((\beta,\alpha) = (1,1)\) (\textbf{top}) and \((3,1)\) (\textbf{bottom}) wavefunction components for \(n^p = 2^+\) as a function of radial distance, with the appropriate spherical harmonic divided out at each lattice site. A more pronounced spread of values does occur in the vicinity of a node, especially in the small (lower) components. Nevertheless, we have verified that the set of points at a fixed radial distance is consistent within statistical uncertainty.}
\end{figure}

With the emphasis on node structure in this work, it is pertinent to also assess the angular dependence of wavefunction components with a node. Our scatter plots of the \((1,1)\) and \((3,1)\) components for the first even-parity excited state, which are known from Fig.~\ref{fig:firstpositiveparityvis} to both carry a node, are shown in Fig.~\ref{fig:firstpositiveparityangular}. A similar plot is obtained for the \((4,1)\) component. Away from the node, these resemble the equivalent ground-state scatter plots, though we see a wider spread of values at each radial distance near the node. This is especially notable in the \((3,1)\) component, where there is an apparent broadening of the node. It is unsurprising that the small (lower) wavefunction components will be more sensitive to noise, and there are several points between radial distances of \(r/a = \numrange{11}{13}\) that at first glance could be considered outliers. We have computed the statistical uncertainty on each individual point and verified that the set of points at a fixed radial distance overlaps within statistical uncertainty. This gives us confidence in averaging over each set of points to obtain an improved determination of the node.

\begin{figure}
	\includegraphics[width=\linewidth]{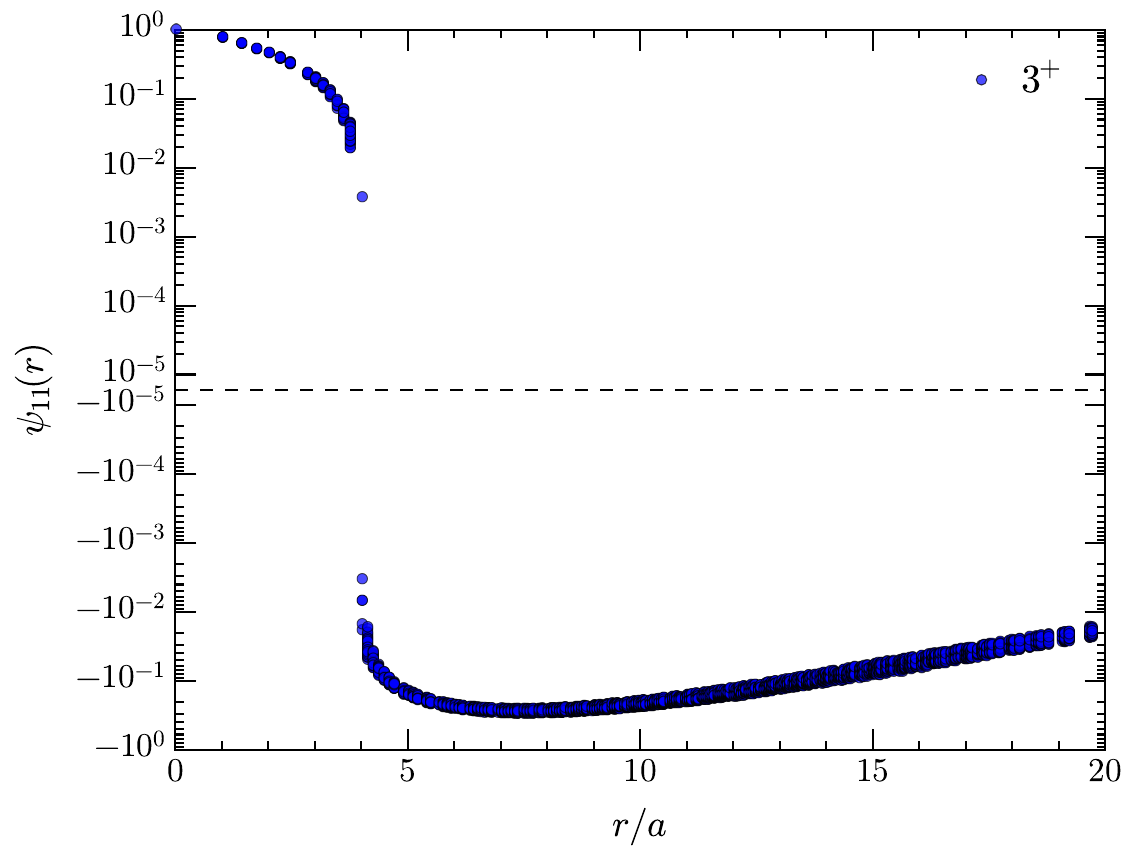}%
	\vspace{-0.5em}
	\caption{\label{fig:secondpositiveparityangular} Scatter plot of the \((\beta,\alpha) = (1,1)\) wavefunction component for \(n^p = 3^+\) as a function of radial distance. The node exhibited by the visualisations in Fig.~\ref{fig:secondpositiveparityvis} is resolved to a very high precision, with minimal spread in values at a fixed radial distance.}
\end{figure}

In addition, our visualisations of the second positive-parity excited state in Fig.~\ref{fig:secondpositiveparityvis} suggest the formation of a novel node in the upper wavefunction components that is absent from its lower components. Therefore, we also carry out the scatter plot analysis for the \((1,1)\) component of this state, which is shown in Fig.~\ref{fig:secondpositiveparityangular}. In this case, we find a remarkably clean signal for the node, with only a very minor spread in values at any given radial distance. This includes in the immediate vicinity of the node, signifying that there is no statistically significant deviation from spherical symmetry attributable to the formation of this novel node. Thus, we can again be confident in averaging over each set of points at a fixed radial distance to obtain an improved determination of the wavefunction.

\subsection{Radial wavefunctions} \label{subsec:positiveparityradial}
Across all wavefunction components, the angular dependence is found to be accurately described by the spherical harmonics predicted by the Dirac equation. Having verified this angular dependence, we now turn to calculation of the radial wavefunctions proper. After removing the angular dependence of each relativistic wavefunction component, as detailed in the previous section, the wavefunction is subsequently averaged over all lattice sites at a given radial distance. We normalise the result to a maximum value of one. The inclusion of diagonal (off-axis) points allows for an unprecedented level of detail in the radial wavefunction, and encompasses distances longer than half the lattice extent in any one spatial dimension.

To minimise finite-volume effects on our \(32^3\) spatial volume, a cut is introduced such that any lattice site with a displacement from the origin greater than 12 in any one direction is excluded. Our observation is that with points any closer to the cubic boundary, small deviations from the trend lines start to appear. By implementing this cut, we are able to reach radial distances of 20 lattice units (using diagonals) while ensuring finite-volume effects are negligible.

In addition, the various components predicted to have the same radial dependence are averaged over. For instance, the \((3,1)\) and \((4,1)\) sink-source combinations correspond to the lower components of the positive-parity wavefunctions. After removing their angular dependence, they can be averaged over to obtain an improved estimate of the lower radial wavefunction, \(\psi^+_\mathcal{L}(r)\).

To take this further, we recall that \(\{U\}\) and \(\{U^*\}\) configurations are equally weighted in the partition function. At zero momentum, there is a clear-cut analytic relationship between the correlation-matrix components for \(U\) and \(U^*\)~\cite{Melnitchouk:2002eg},
\begin{align}
	G[U^*] &= \left(\widetilde{C} \, G[U] \, \widetilde{C}^{-1}\right)^* \,, & \widetilde{C} &= C \gamma_5 \,.
\end{align}
From this, we find the equivalences \(G[U^*]_{11} = G[U]_{22}^*\), \(G[U^*]_{31} = G[U]_{42}^*\), and \(G[U^*]_{41} = -G[U]_{32}^*\). As such, these components can be appropriately combined to obtain an improved estimate without explicitly calculating the wavefunctions for \(U^*\) configurations. In future extension to nonzero momentum, computing wavefunctions for the \(\{U^*\}\) ensemble will provide a genuine improvement.

\begin{figure*}
	\includegraphics[width=0.49\linewidth]{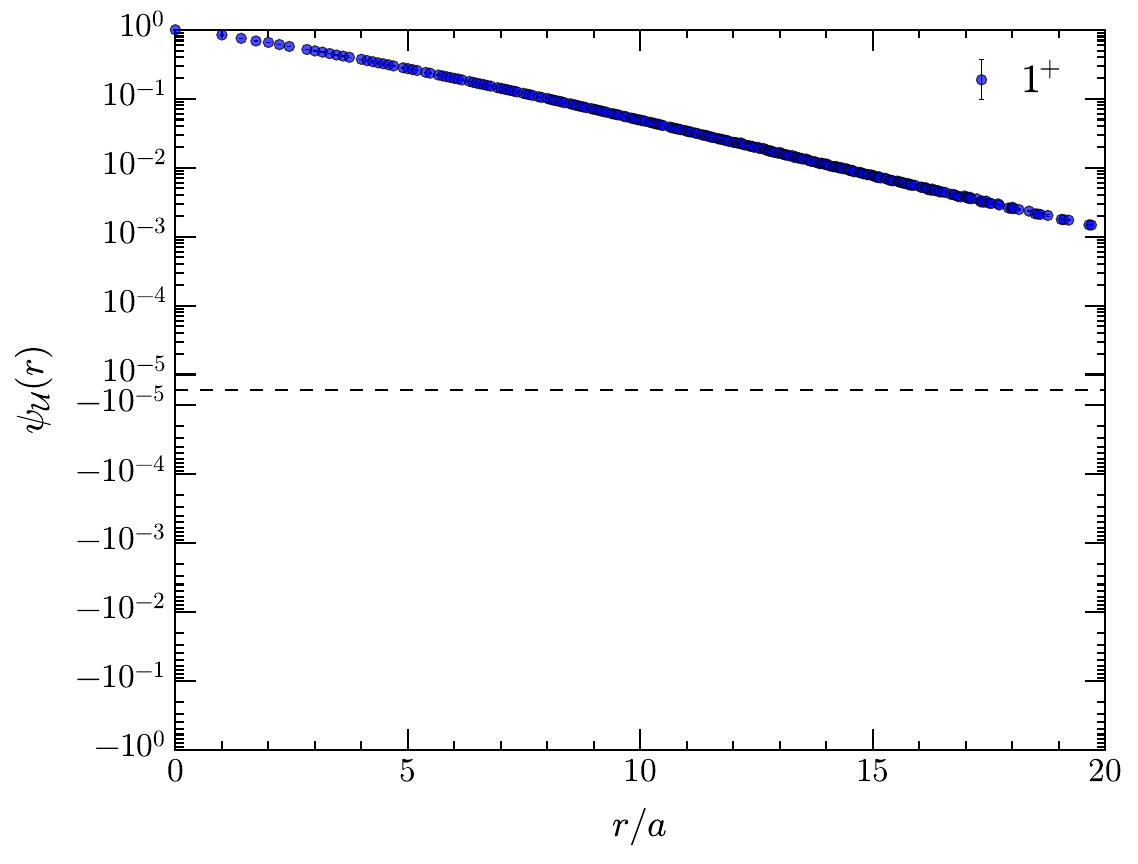}%
	\hfill
	\includegraphics[width=0.49\linewidth]{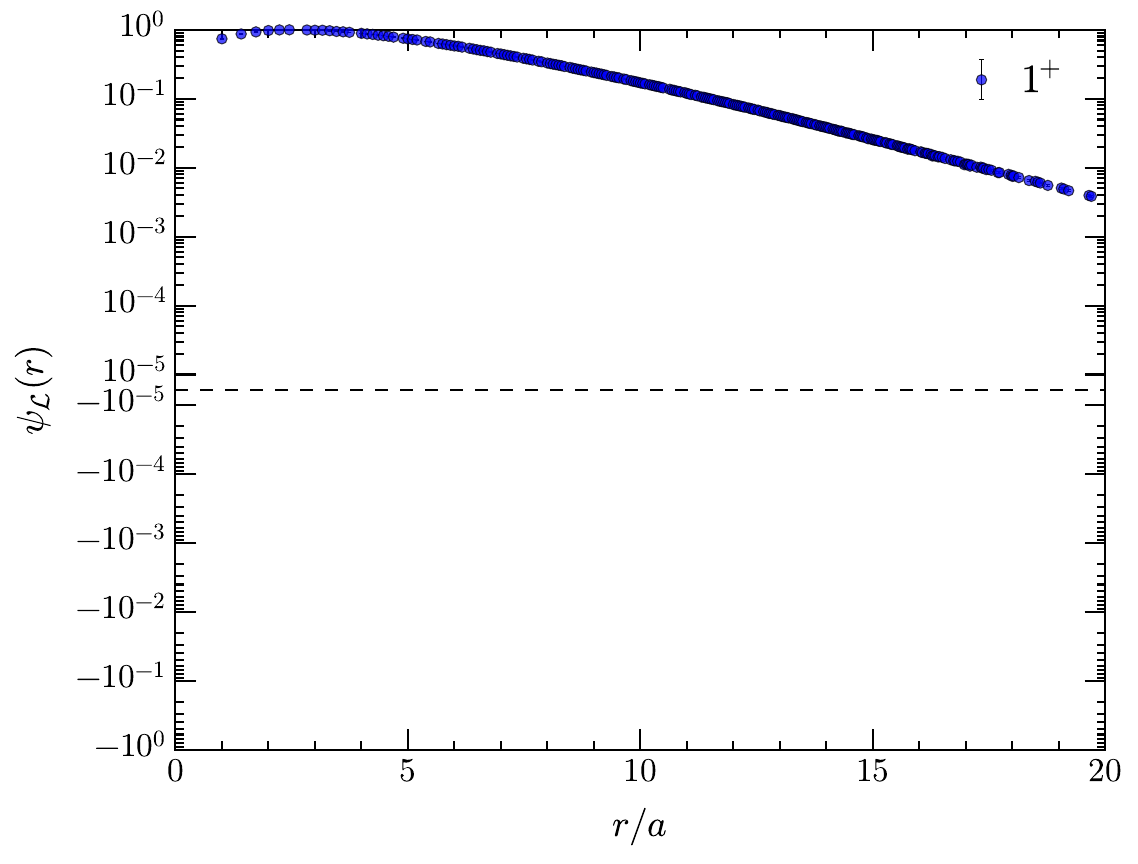}%
	\vspace{0.2em}
	\includegraphics[width=0.49\linewidth]{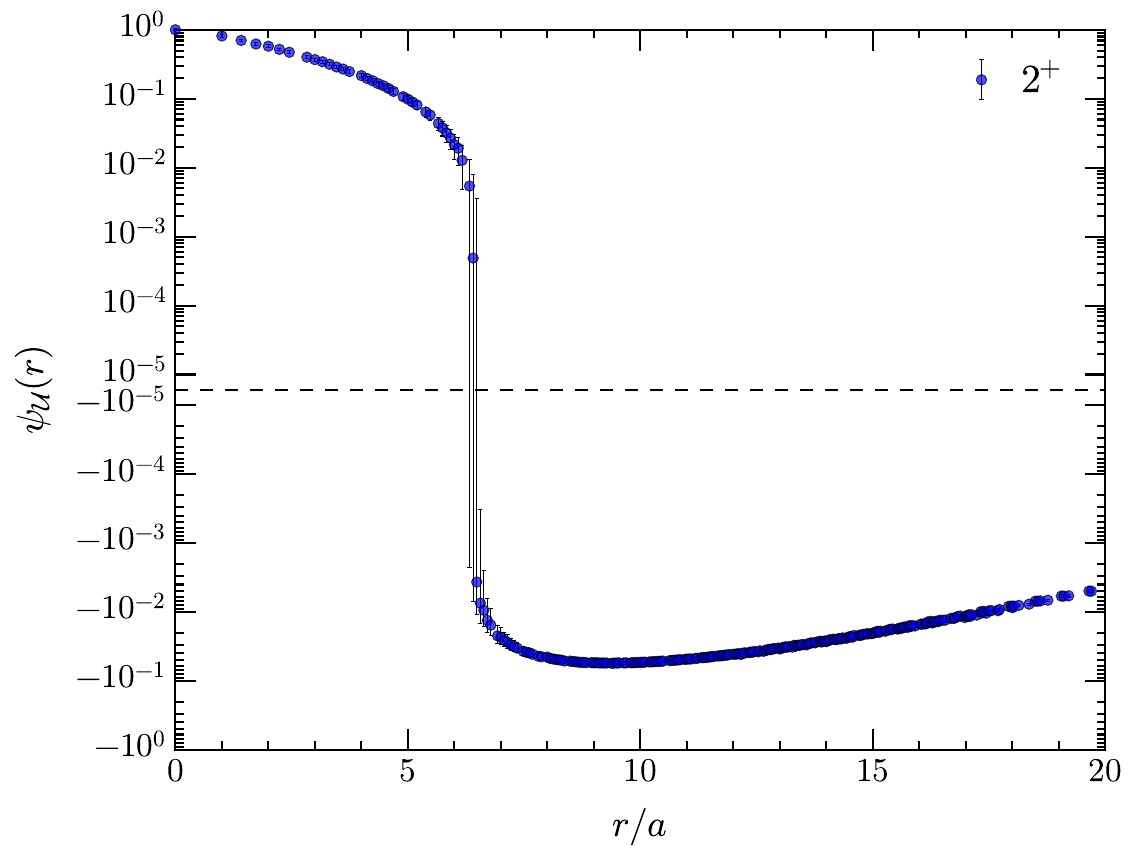}%
	\hfill
	\includegraphics[width=0.49\linewidth]{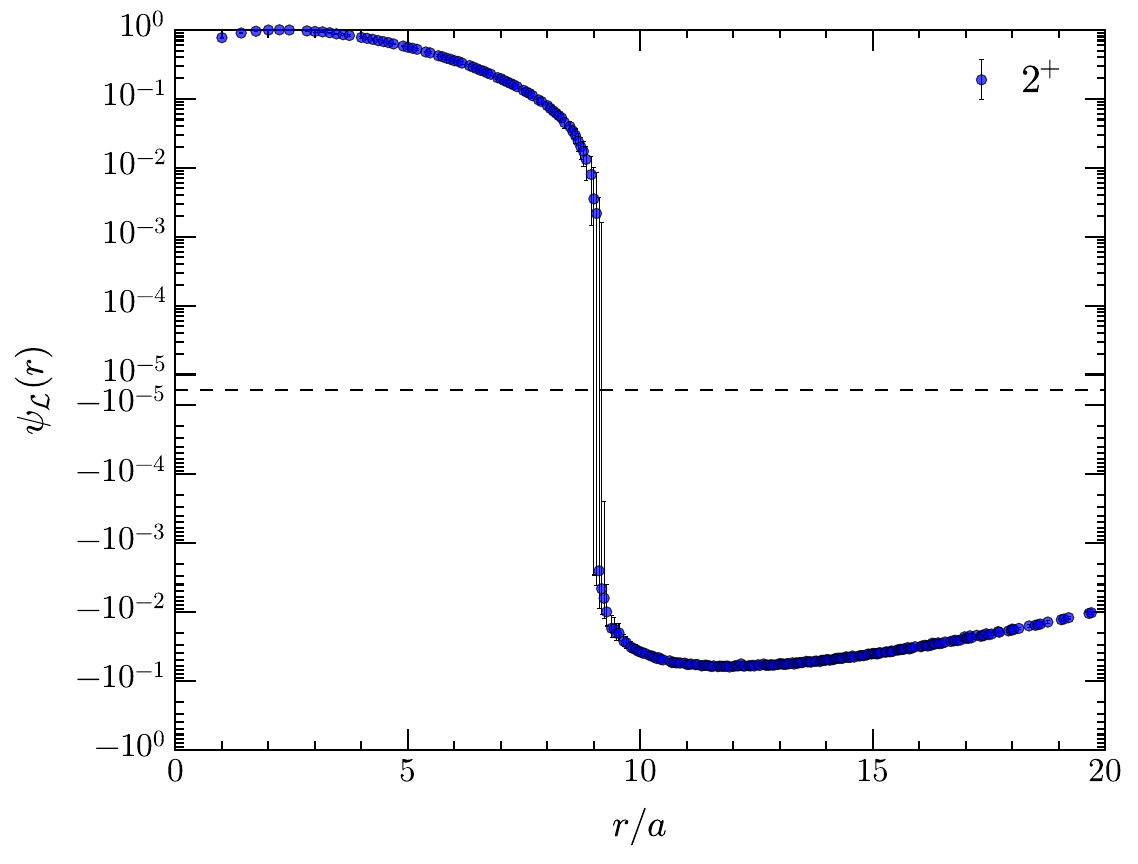}%
	\vspace{0.2em}
	\includegraphics[width=0.49\linewidth]{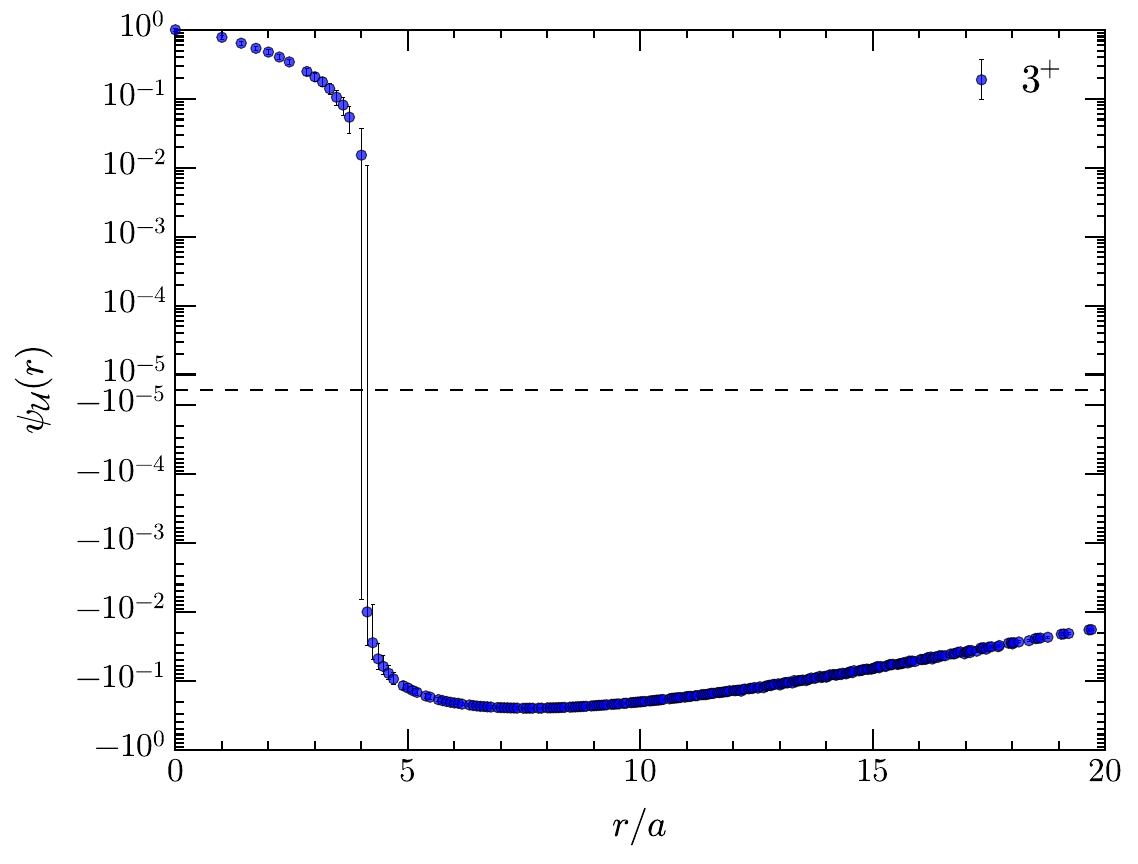}%
	\hfill
	\includegraphics[width=0.49\linewidth]{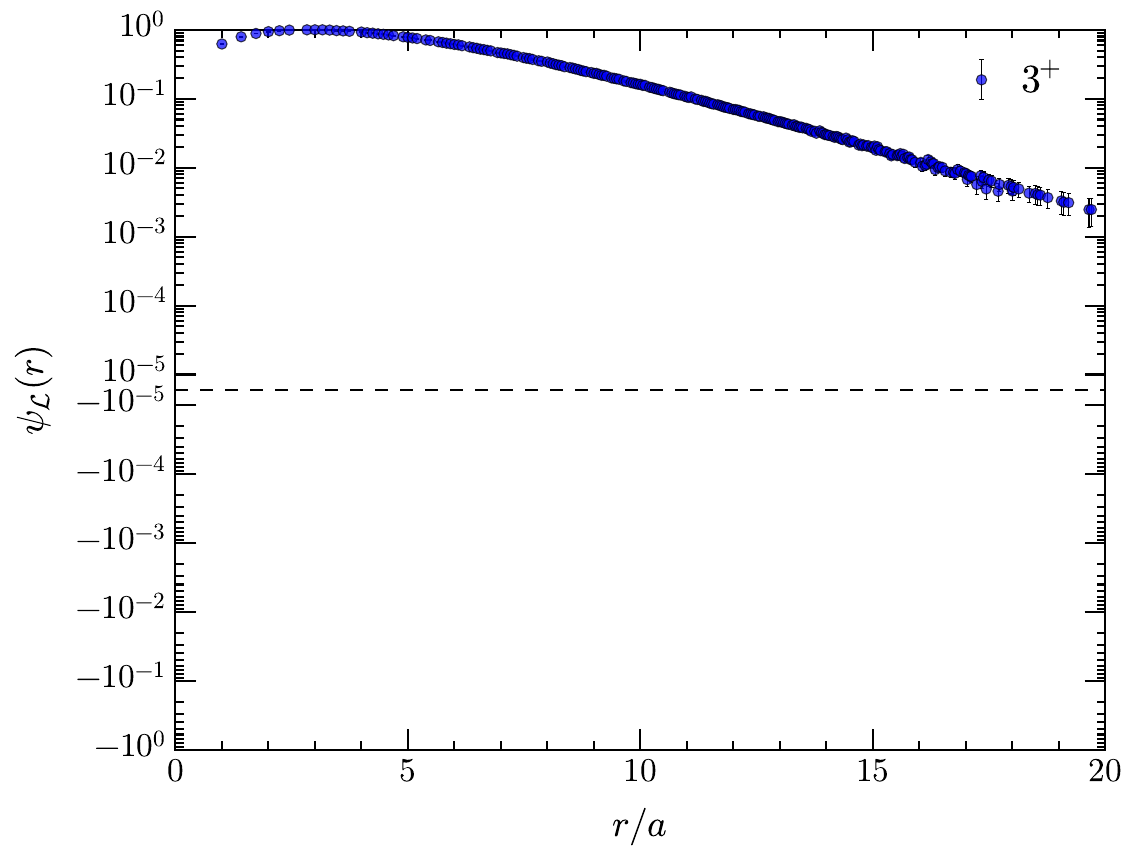}%
	\vspace{-0.5em}
	\caption{\label{fig:positiveparityradial1-3} The upper (\textbf{left}) and lower (\textbf{right}) radial wavefunctions, \(\psi_\mathcal{U}(r)\) and \(\psi_\mathcal{L}(r)\), for the three lowest-lying positive-parity states at Euclidean time \(\tau_\mathrm{min} = 5\) slices after the source. Their node structure agrees with that suggested by the visualisations in Figs.~\ref{fig:groundstatevis}--\ref{fig:secondpositiveparityvis}. Namely, the ground state (\textbf{top}) has zero nodes across all wavefunctions components, the first excited state (\textbf{middle}) has one node in both upper and lower components, and the second excited state (\textbf{bottom}) has a node in the upper (\(s\)-wave) components but not in the lower (\(p\)-wave) components. The statistical error bars are mostly negligible, except for in the region surrounding a node.}
\end{figure*}

With this explained, we now present our plots of the radial wavefunctions for positive-parity states. In line with our analysis of the mass spectrum in Sec.~\ref{subsec:massspectrum}, we focus on the three lowest-lying states. Their upper and lower radial wavefunctions are provided in Fig.~\ref{fig:positiveparityradial1-3}. We include statistical uncertainties from a single-elimination jackknife, performed using an optimised interpolator projected from a GEVP on the ensemble average. As established in Sec.~\ref{subsec:positiveparityvis}, these are presented at Euclidean time \(\tau_\mathrm{min} = 5\) from the effective-mass fits in Table~\ref{tab:massspectrum}.

\begin{figure*}
	\includegraphics[width=0.49\linewidth]{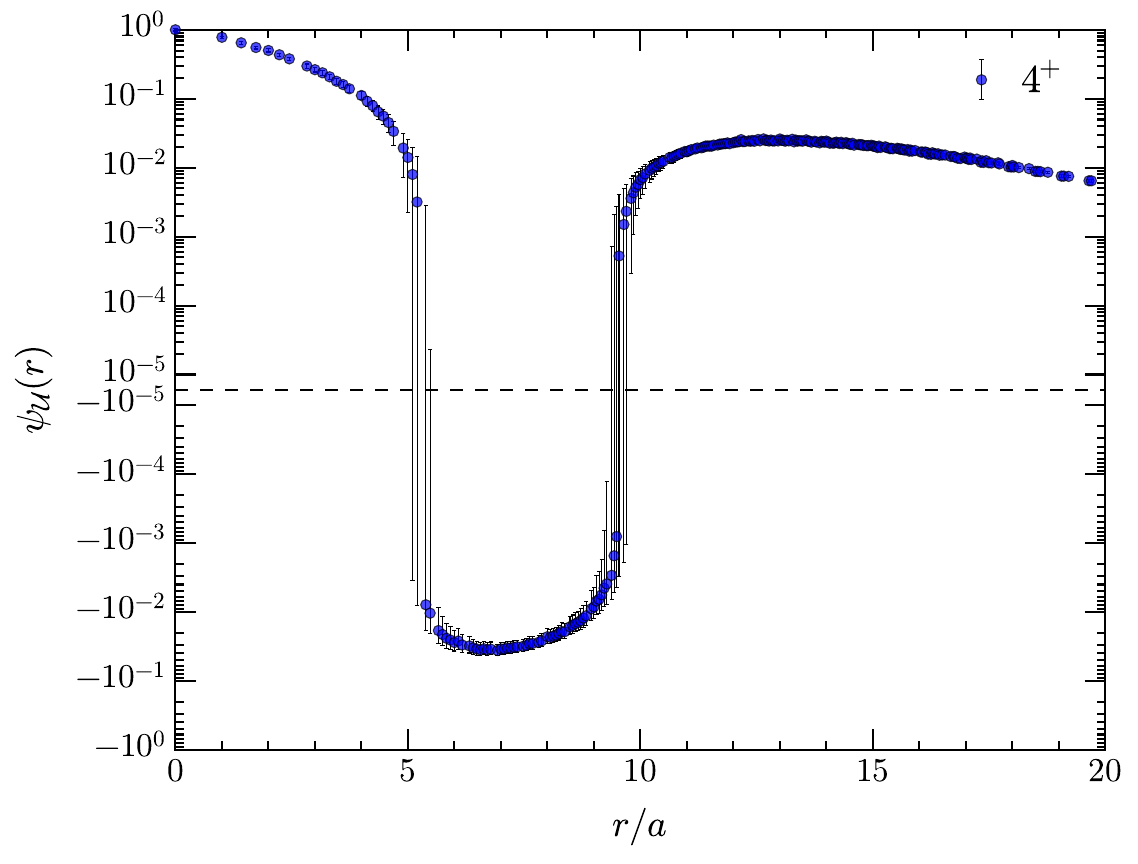}%
	\hfill
	\includegraphics[width=0.49\linewidth]{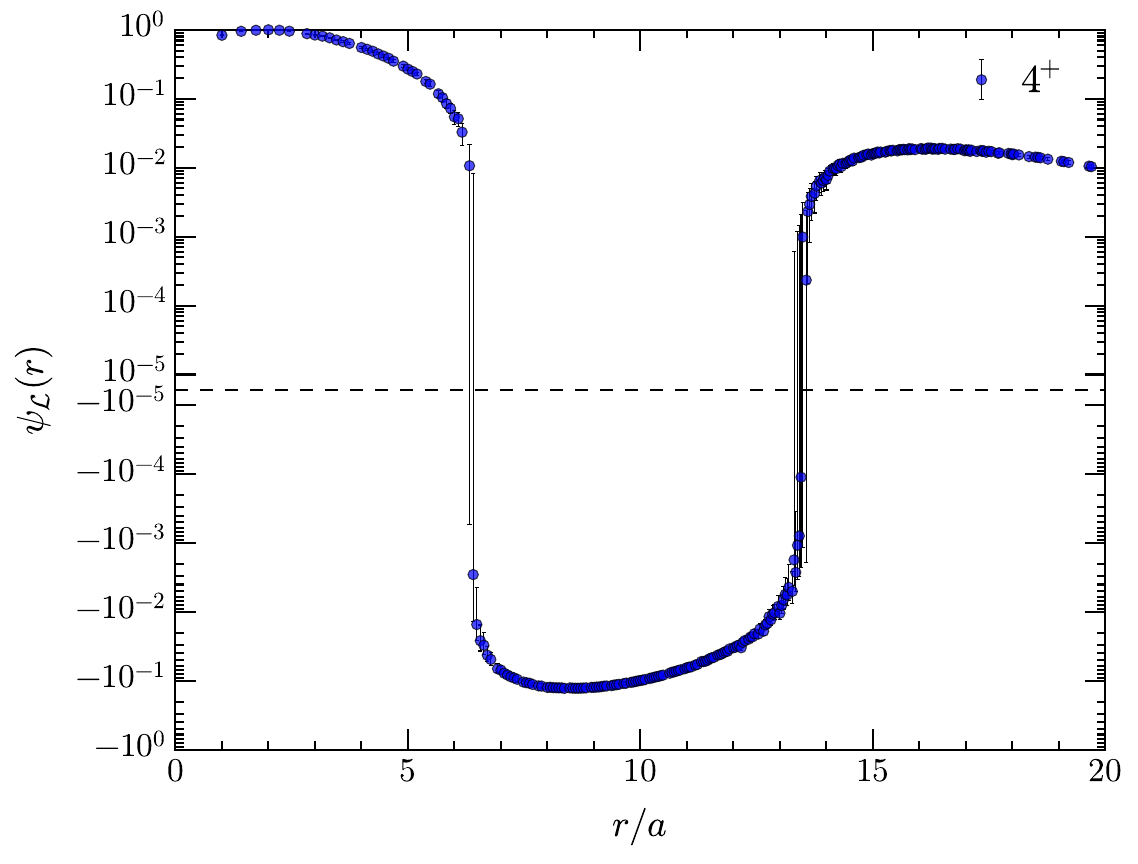}%
	\vspace{0.2em}
	\includegraphics[width=0.49\linewidth]{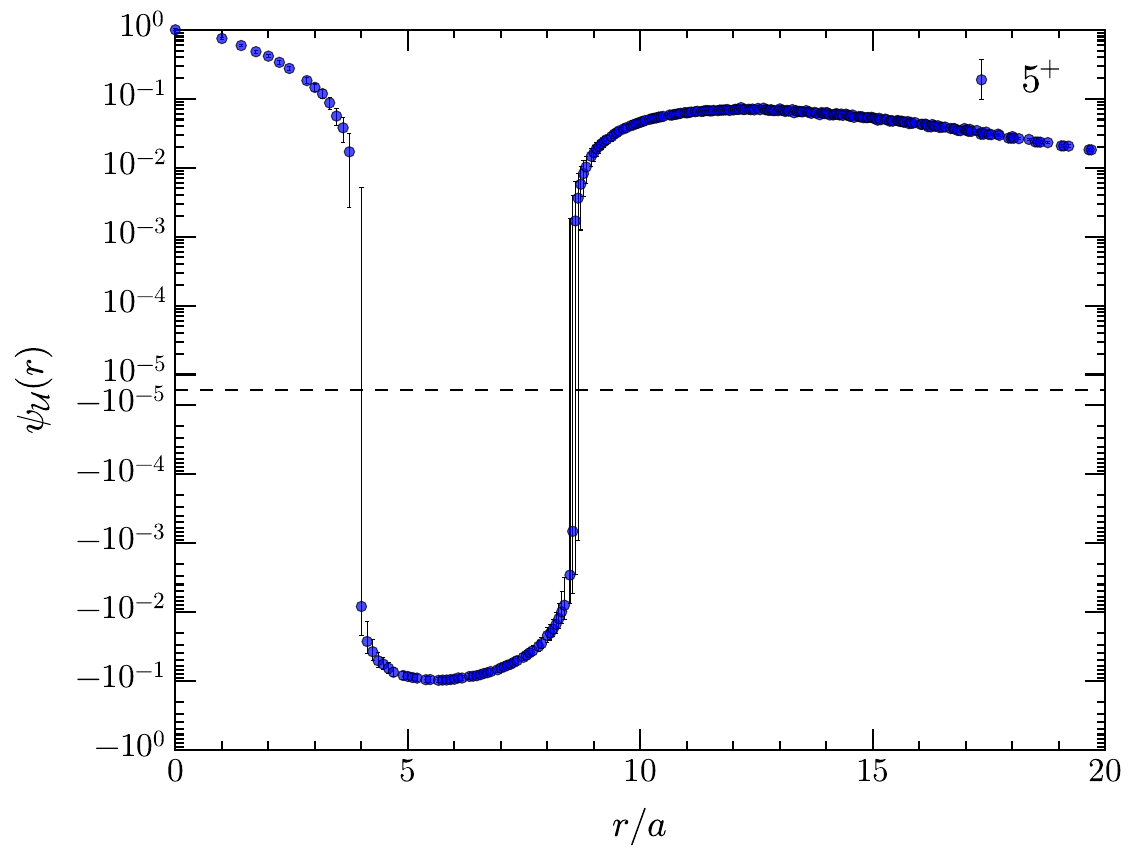}%
	\hfill
	\includegraphics[width=0.49\linewidth]{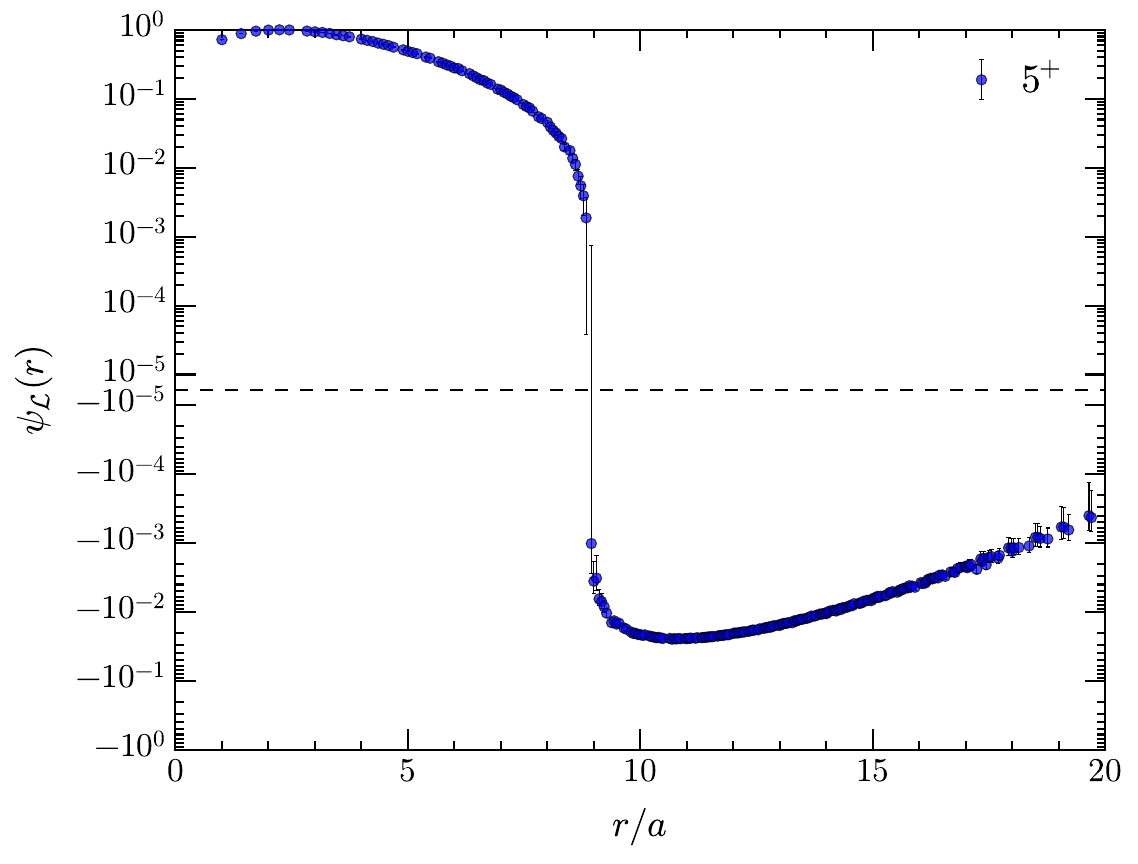}%
	\vspace{-0.5em}
	\caption{\label{fig:positiveparityradial4-5} As in Fig.~\ref{fig:positiveparityradial1-3}, but for the fourth (\textbf{top}) and fifth (\textbf{bottom}) positive-parity states at Euclidean times \(\tau_\mathrm{min} = 4\) and \(3\), respectively. The fourth state contains two nodes in both its upper and lower wavefunction components, while for the fifth state, two nodes are present in its upper (\(s\)-wave) components but one node in its lower (\(p\)-wave) components. This suggests that the node mismatch initially observed for the third state (Fig.~\ref{fig:positiveparityradial1-3}) is a recurring feature of the positive-parity spectrum.}
\end{figure*}

First, we note that the \(s\)-wave structure expected in the upper components, and the \(p\)-wave structure expected in the lower components, have both been revealed. Although the \(p\)-wave should formally vanish at \(r = 0\), we cannot have a data point there because of the inability to divide by \(\cos\theta\) everywhere on the \(xy\)-plane. Note that when counting nodes, we only include nodes for \(r > 0\) (i.e.\ we \emph{exclude} that the \(p\)-wave radial dependence starts zero at \(r = 0\)) as those are the nontrivial ones.

These radial wavefunctions also reflect the node structure indicated by the visualisations in Figs.~\ref{fig:groundstatevis}--\ref{fig:secondpositiveparityvis}. The ground state has zero nodes, the first excited state has one node in both upper and lower wavefunction components, and most importantly, the second excited state has one node in the upper components but zero in the lower. This provides unambiguous confirmation of a node mismatch between the \(s\)-wave and \(p\)-wave components of this state. We are aided in this regard by high statistical precision, with uncertainties largely negligible except for around a node. Compared against the first excited state, the characteristics of this node in the second excited state are also readily discerned, being sharper and located at a smaller radial distance.

This difference in node location naturally explains the mass splitting between the first and second even-parity excited states, as seen in the spectrum of Fig.~\ref{fig:massspectrum}. Because the \(s\)-wave node in the second excitation is closer to the origin, it pushes the probability density out to larger radial distances such that the \(d\) quark is more likely to be found farther from the two \(u\) quarks at the origin. This results in a larger confining potential between the quarks, which in turn generates the higher mass for this state.

We also speculatively examine the next two even-parity excitations, whose radial wavefunctions are shown in Fig.~\ref{fig:positiveparityradial4-5}. The former of these is taken at Euclidean time \(\tau_\mathrm{min} = 4\) from the effective-mass fits in Table~\ref{tab:massspectrum}, though the latter is taken one time step earlier at \(\tau = 3\) due to an observed loss of signal in its wavefunction for \(\tau \geq 4\).

For the third excited state, we return to a consistent node structure across all components of the wavefunction, with two nodes found in both upper and lower components. Then, in the fourth excited state, we find another instance of a mismatch by one extra node in the upper (\(s\)-wave) components of its wavefunction. To be precise, this state possesses two nodes in its upper (\(s\)-wave) components compared to one in its lower (\(p\)-wave) components. The ``additional'' \(s\)-wave node is again sharper and found closer to the origin. As explained for the second excited state, this will generate a higher mass for this state by pushing the probability density out to larger radial distances. This suggests that the node mismatch is a recurring feature of the positive-parity nucleon spectrum, where for each state with \(n > 0\) nodes in all components there is a corresponding state with a mismatch by one fewer nodes in its lower (\(p\)-wave) components.

\section{Negative-parity states} \label{sec:negativeparity}
Our next course of action is to repeat the above analysis, but for the negative-parity spectrum. This is the first study of the wavefunctions for negative-parity excited states. Therefore, our starting point is again to produce visualisations that qualitatively expose the structure of the lowest-lying wavefunctions. This will be followed by calculation of the radial wavefunction dependence, as described for positive parity.

\subsection{Visualisations} \label{subsec:negativeparityvis}
We produce visualisations for the first two negative-parity excitations. These states exist very close to each other in the spectrum (Fig.~\ref{fig:massspectrum}), and thus it will be fascinating to observe any structural differences between the states that underlie this split. For both states, we visualise the wavefunctions for the \((\beta,\alpha) = (1,4)\), \((2,4)\) and \((4,4)\) Dirac elements. Here, \(\alpha = 4\) at the source selects spin-down negative parity. Similar results are obtained with \(\alpha = 3\) (i.e.\ spin-up negative parity). Drawing on Table~\ref{tab:lm_l}, these components have effective quantum numbers \((\ell_\mathrm{eff},m_\mathrm{eff}) = (1,-1)\), \((1,0)\) and \((0,0)\). Similar to the positive-parity visualisations in Sec.~\ref{subsec:positiveparityvis}, this covers the three relevant spherical harmonics.

\begin{figure*}
	\includegraphics[width=0.42\linewidth]{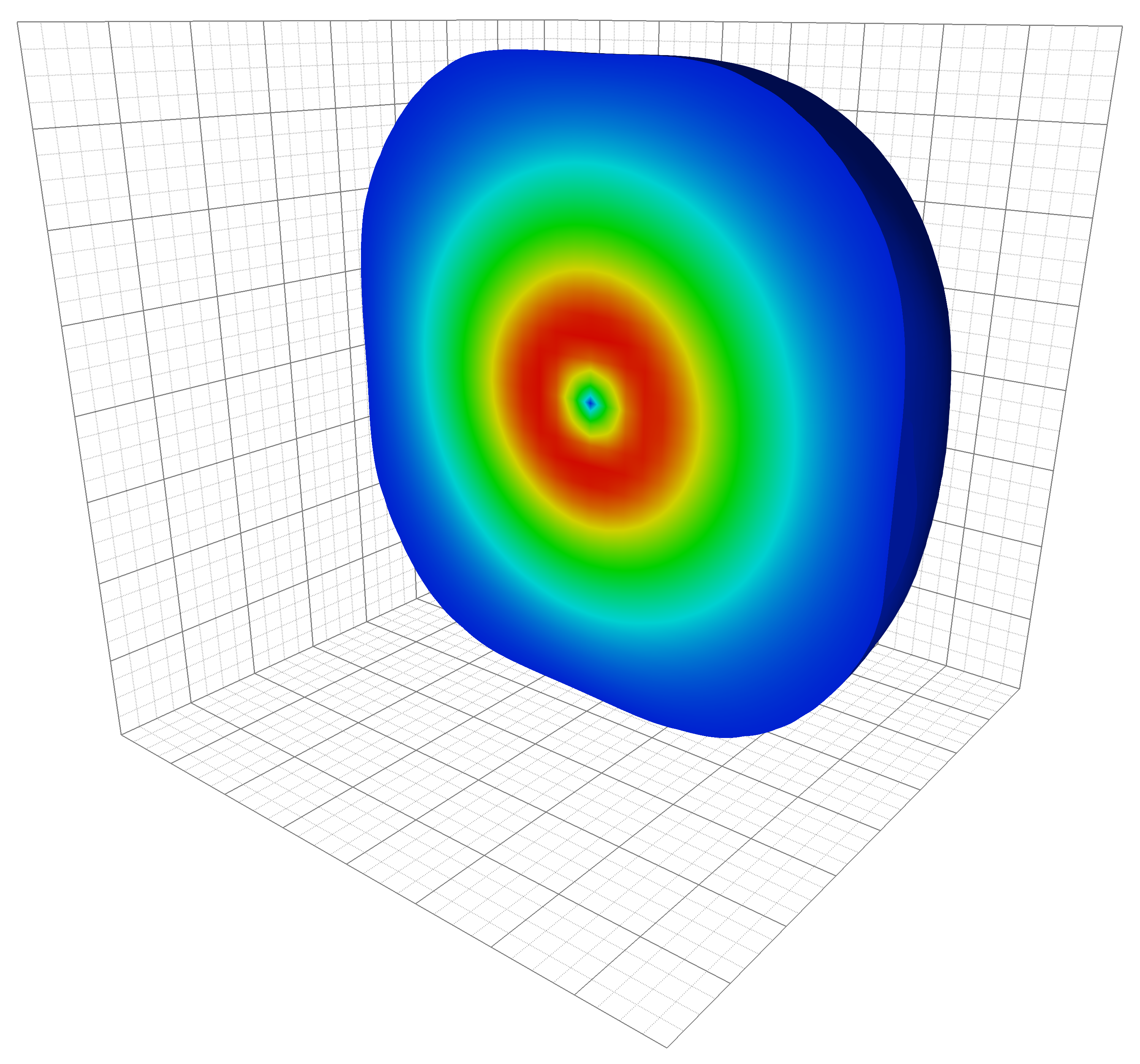}%
	\hspace{5em}
	\includegraphics[width=0.42\linewidth]{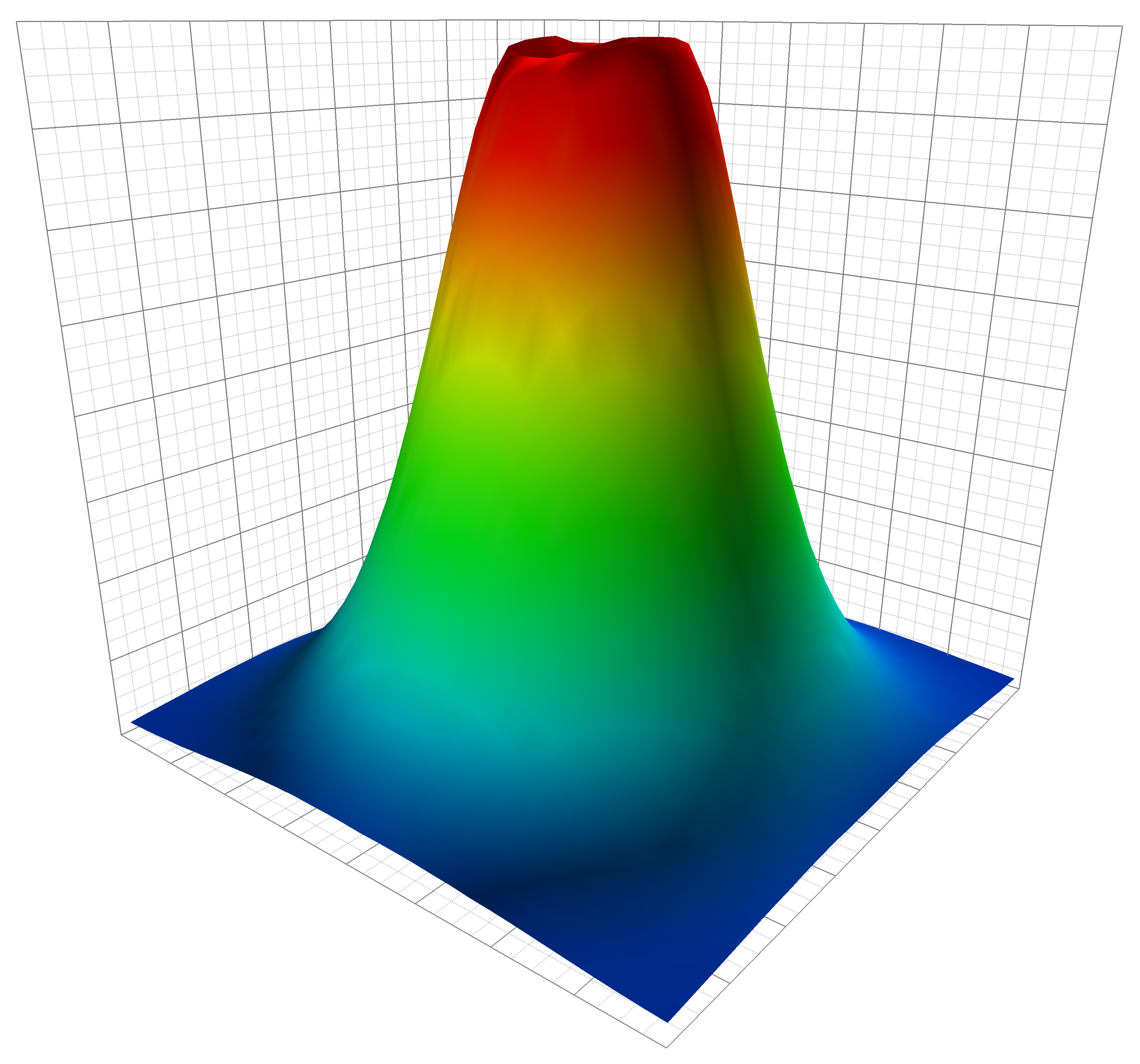}%
	\vspace{0.5em}
	\includegraphics[width=0.42\linewidth]{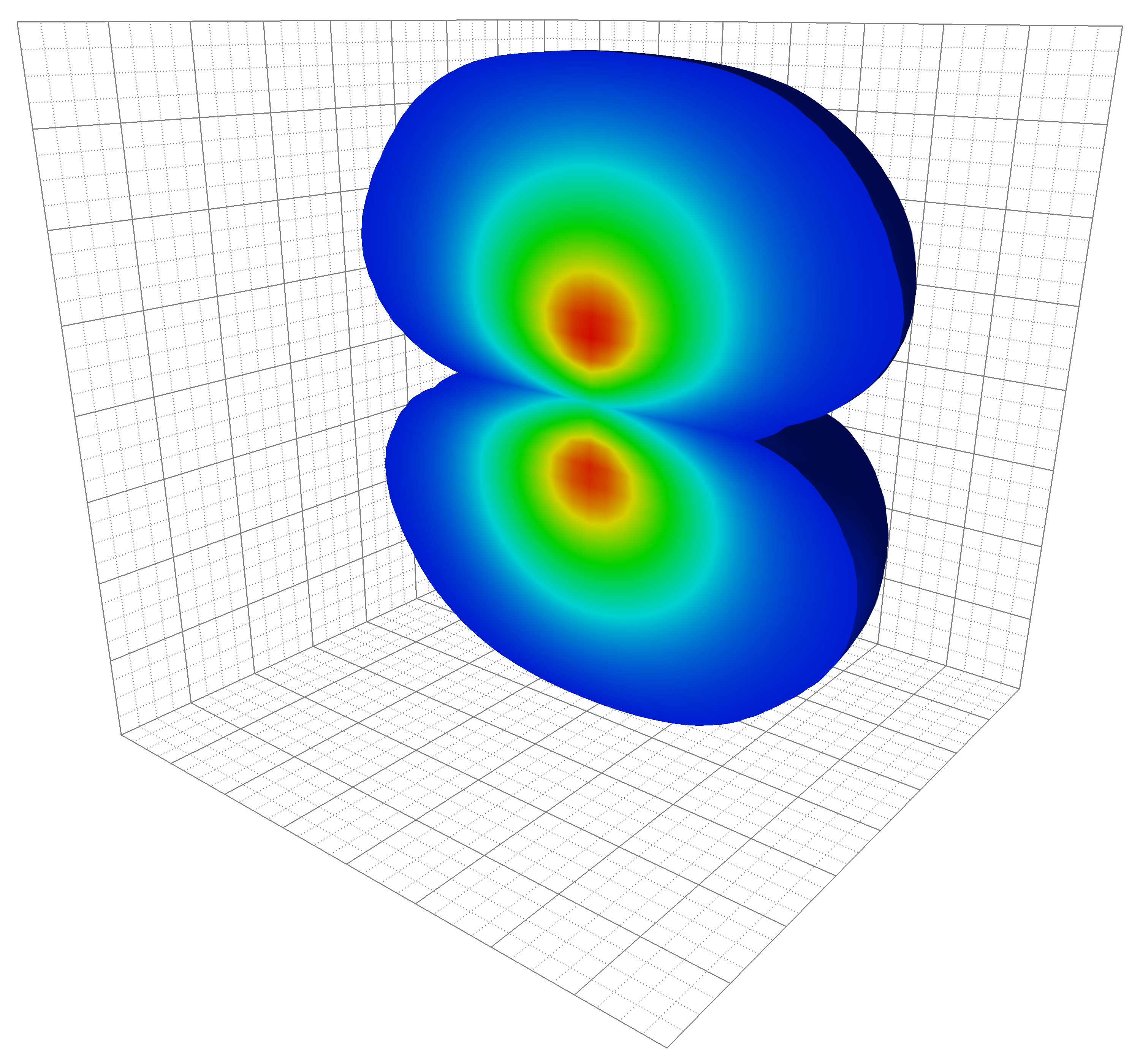}%
	\hspace{5em}
	\includegraphics[width=0.42\linewidth]{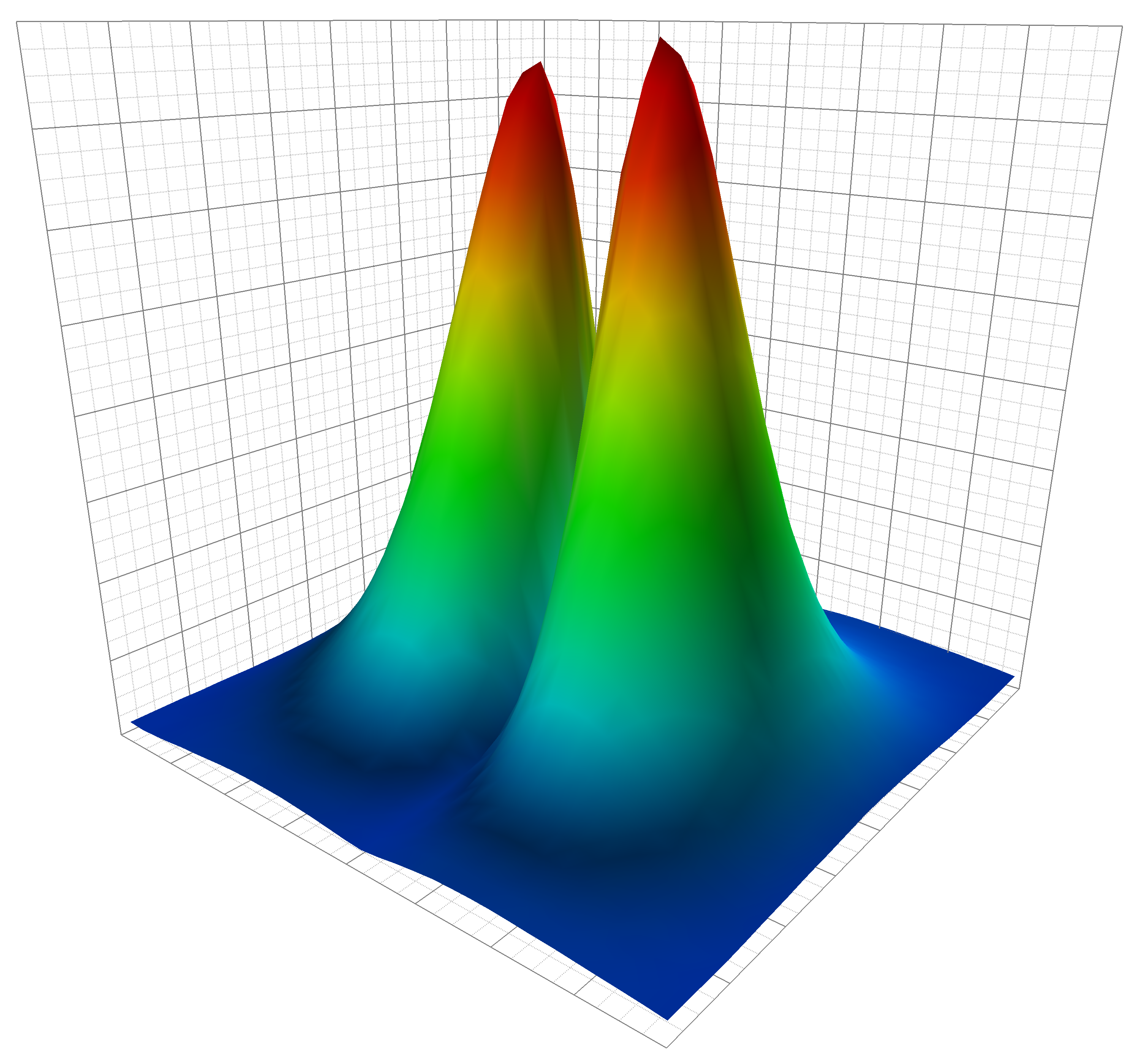}%
	\vspace{0.5em}
	\includegraphics[width=0.42\linewidth]{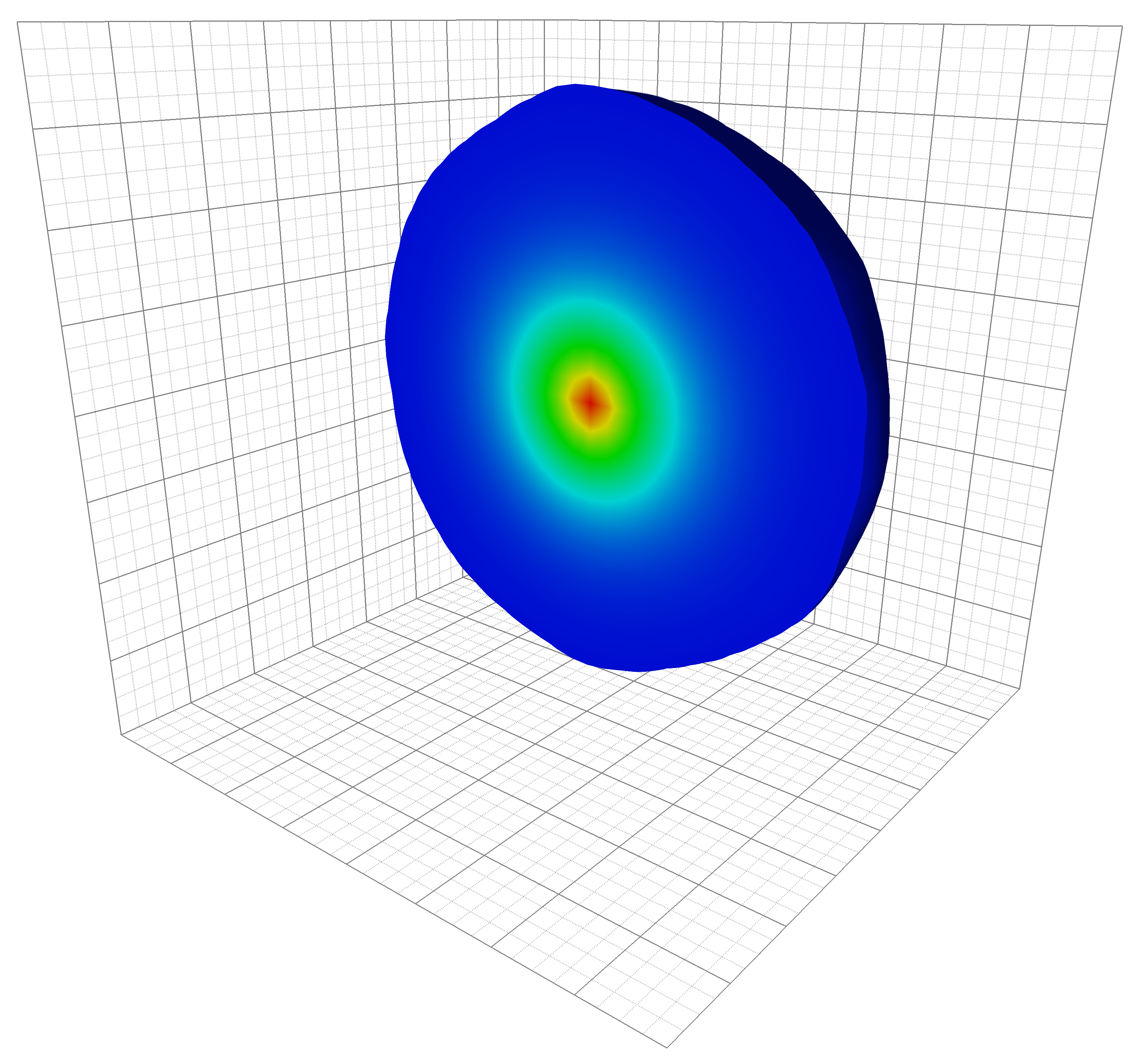}%
	\hspace{5em}
	\includegraphics[width=0.42\linewidth]{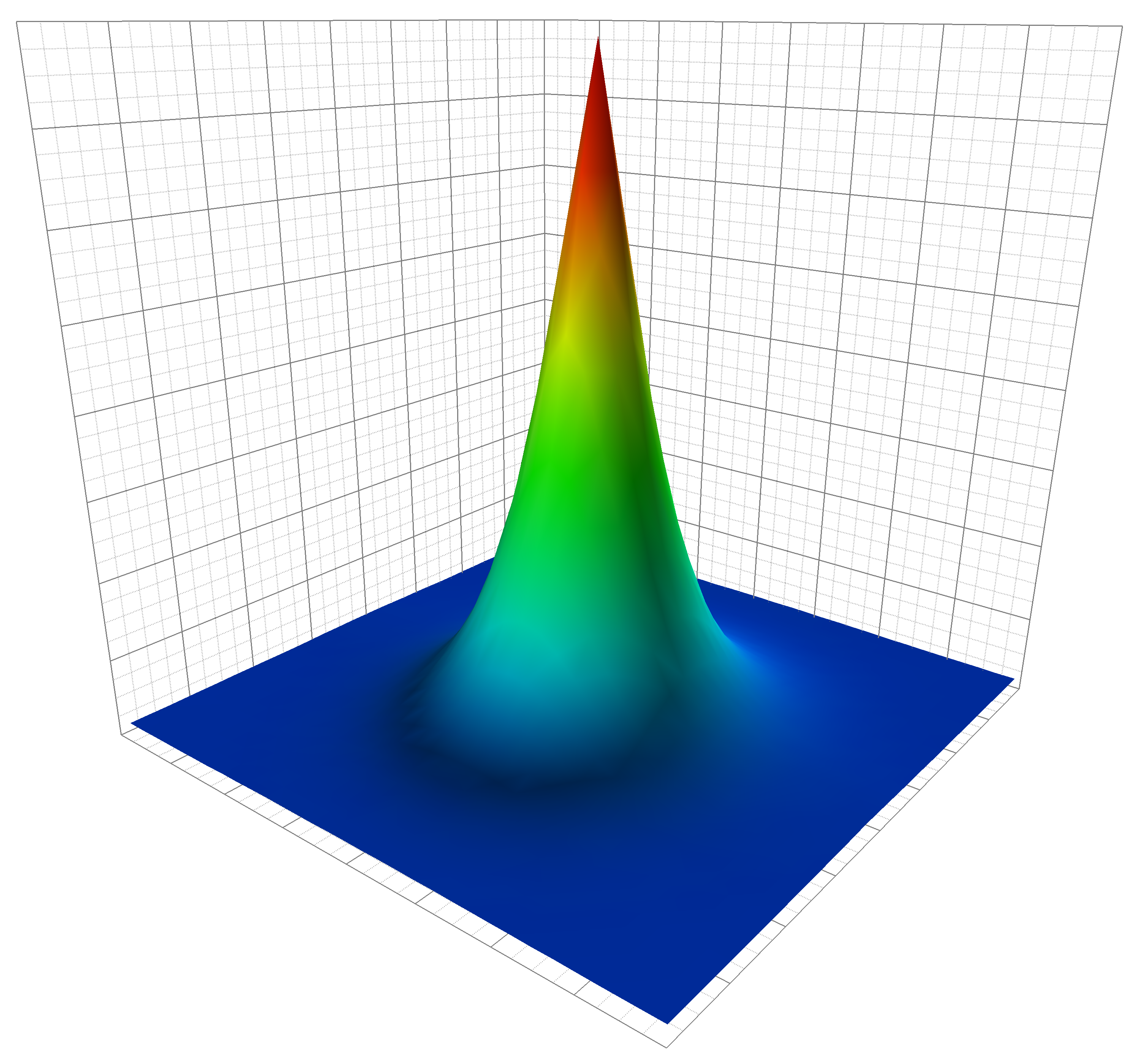}%
	\vspace{-0.5em}
	\caption{\label{fig:firstnegativeparityvis} Volume and surface renderings of the \((1,4)\) (\textbf{top}), \((2,4)\) (\textbf{middle}) and \((4,4)\) (\textbf{bottom}) components of the first negative-parity excited state's wavefunction. The \((2,4)\) and \((4,4)\) components are sliced through \(x = 0\), and the \((1,4)\) component is sliced through \(z = 0\). The expected wavefunction shapes are apparent, including the spherically symmetric \((4,4)\) component, the hourglass shape for the \((2,4)\) component and the ring in the \(xy\)-plane for the \((1,4)\) component.}
\end{figure*}

Our volume and surface renders for the first negative-parity excited state are displayed in Fig.~\ref{fig:firstnegativeparityvis}. Note that the rendering threshold chosen for the volume renders here is considerably lower than for the positive-parity visualisations (Figs.~\ref{fig:groundstatevis}--\ref{fig:secondpositiveparityvis}). The reason for this will become clear when inspecting the second negative-parity excitation. Otherwise, the visualisations are produced in an identical manner as for positive parity.

For now, we observe that there are zero nodes across all components for the first negative-parity excitation. As the lowest-lying negative-parity state, this is the expected result. The angular dependence for each component again visually agrees with the appropriate spherical harmonic. This includes the hourglass structure in the \((2,4)\) component, which has a parity mismatch between the source and sink spinor indices but no spin mismatch (\(Y_0^1\) spherical harmonic), and the ring structure in the \(xy\)-plane for the \((1,4)\) component, which has both a parity and spin mismatch (\(Y_1^1\) spherical harmonic).

\begin{figure*}
	\includegraphics[width=0.42\linewidth]{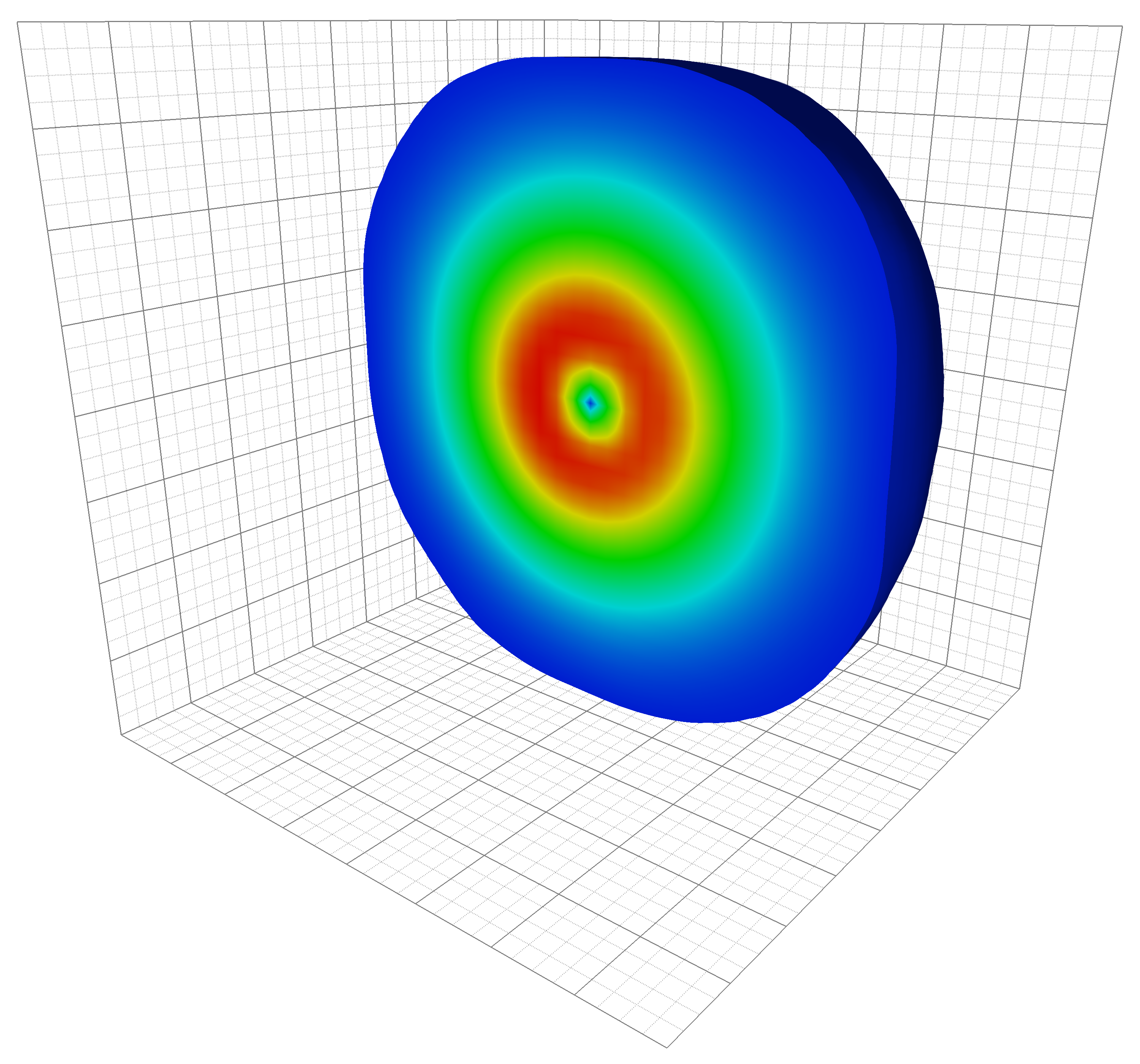}%
	\hspace{5em}
	\includegraphics[width=0.42\linewidth]{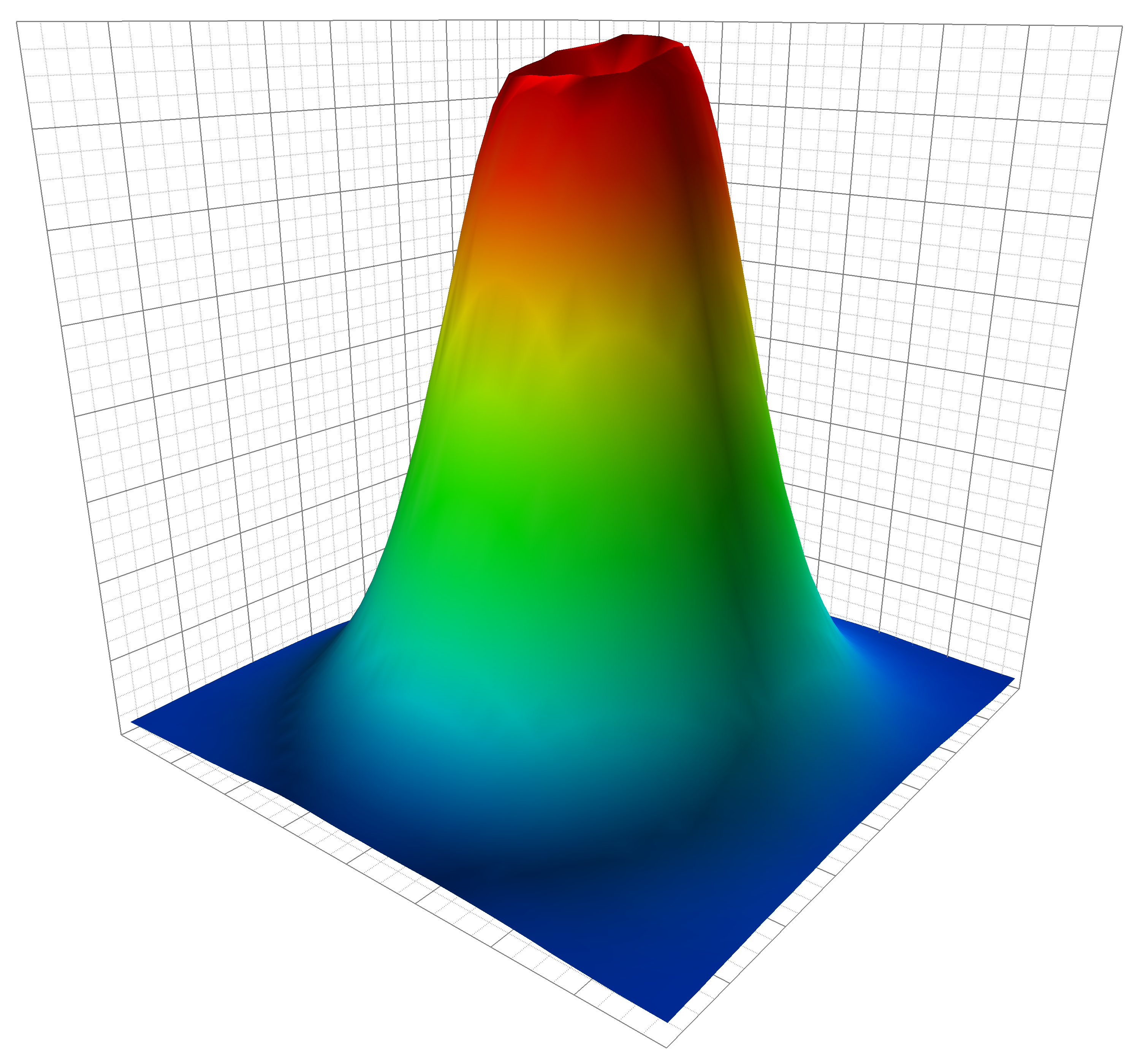}%
	\vspace{0.5em}
	\includegraphics[width=0.42\linewidth]{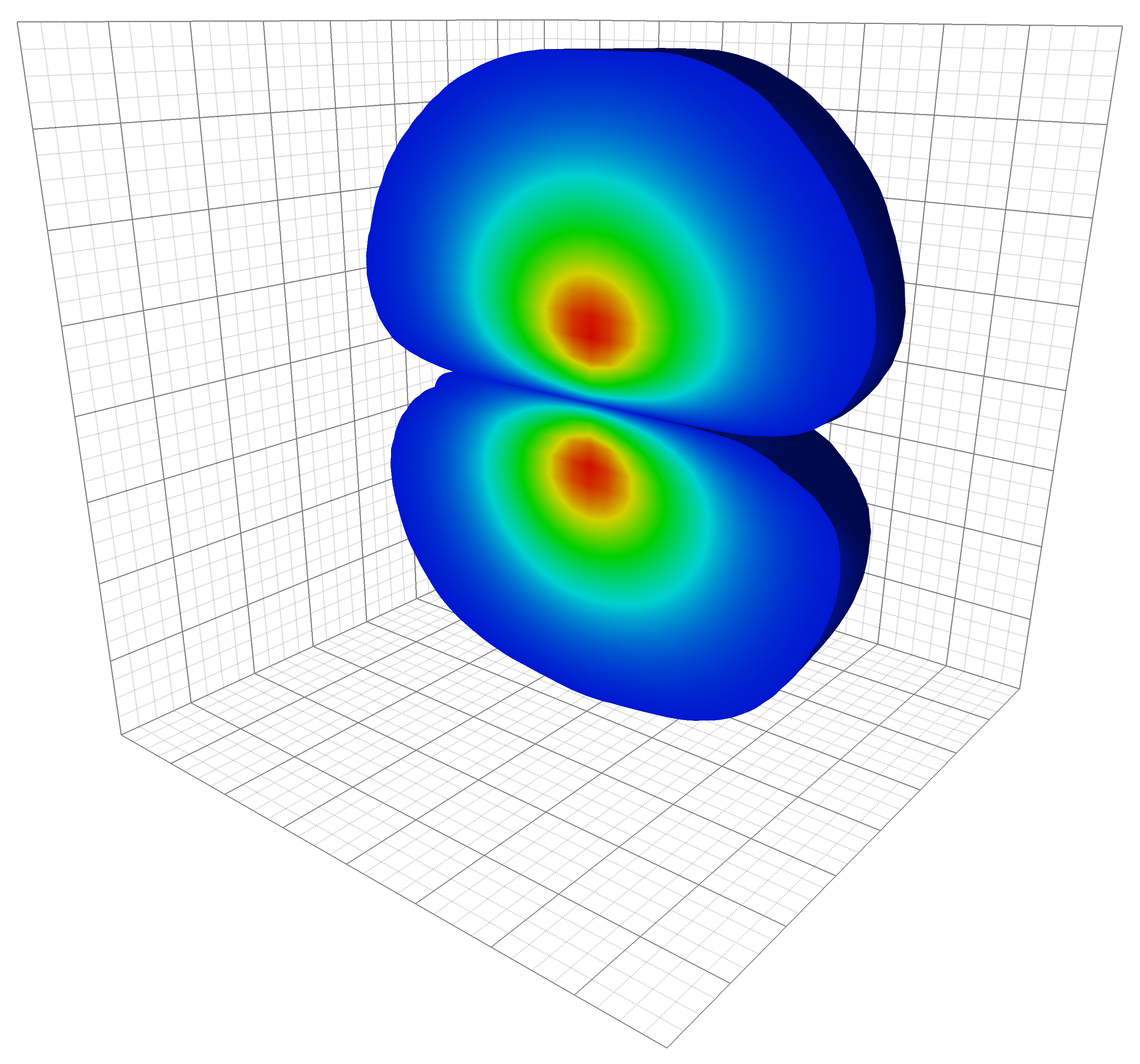}%
	\hspace{5em}
	\includegraphics[width=0.42\linewidth]{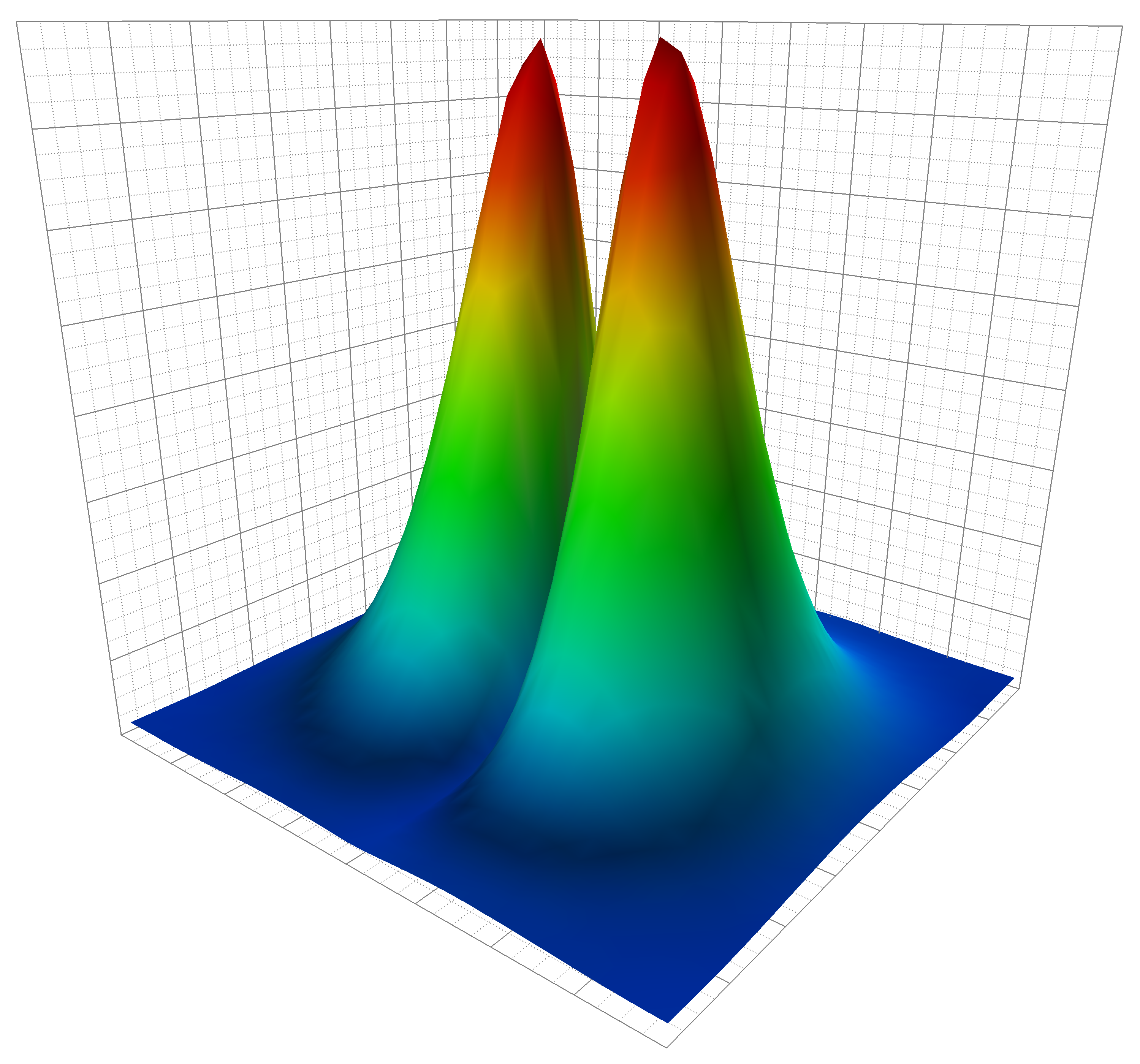}%
	\vspace{0.5em}
	\includegraphics[width=0.42\linewidth]{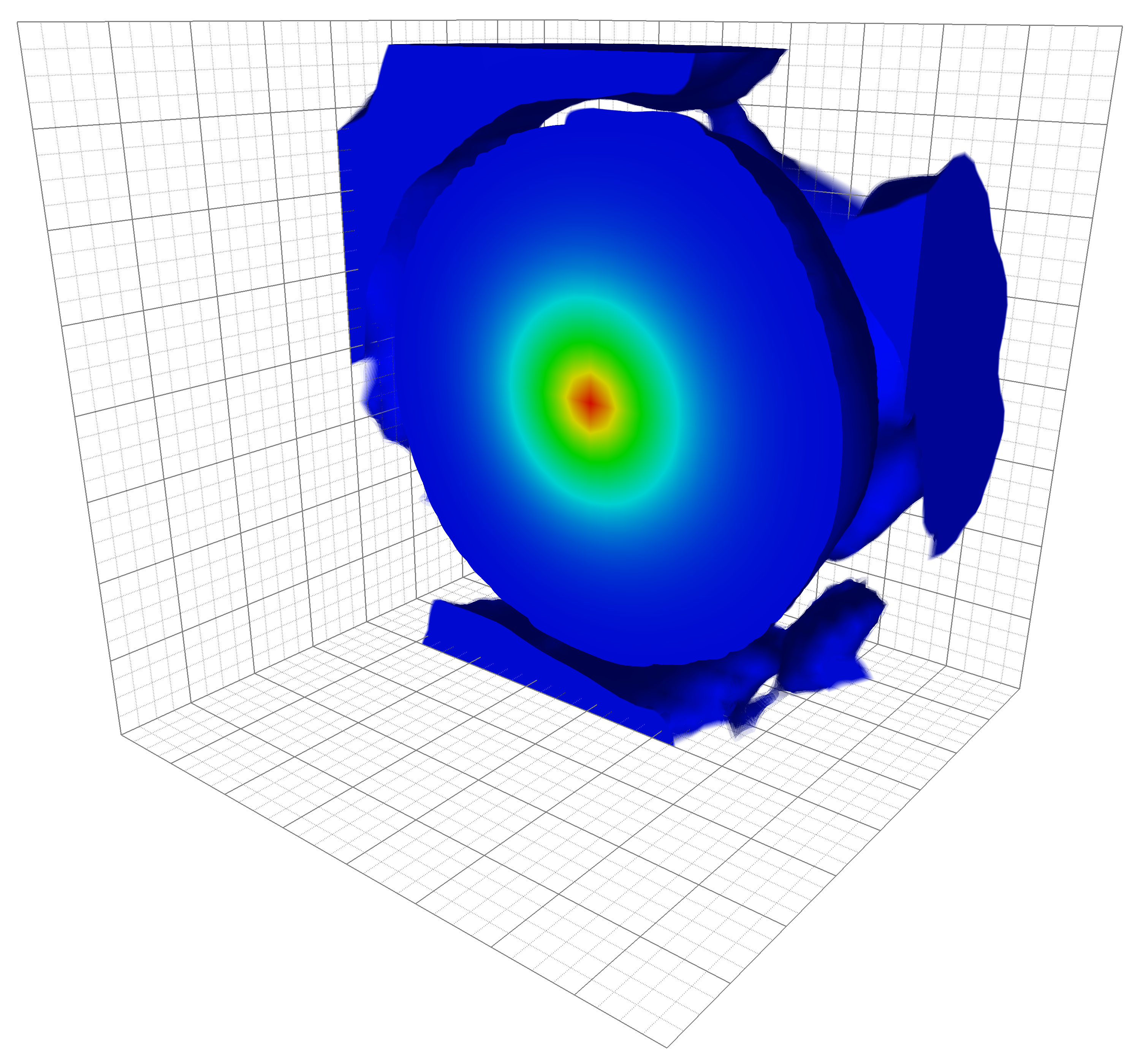}%
	\hspace{5em}
	\includegraphics[width=0.42\linewidth]{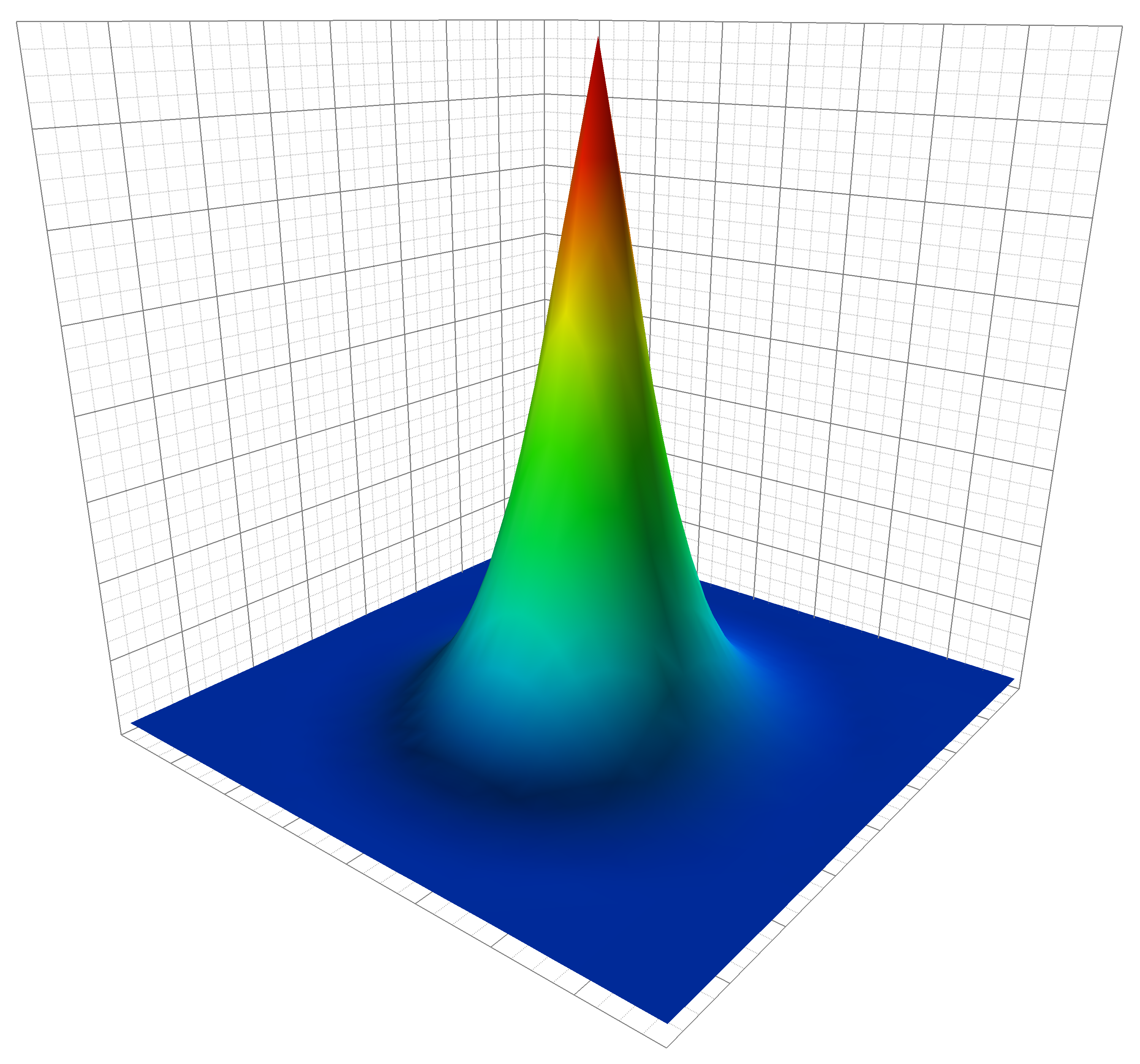}%
	\vspace{-0.5em}
	\caption{\label{fig:secondnegativeparityvis} Volume and surface renderings of the \((1,4)\) (\textbf{top}), \((2,4)\) (\textbf{middle}) and \((4,4)\) (\textbf{bottom}) components for the second negative-parity excitation. The suggestion of a node is found at large radial distances in the lower \((4,4)\) component, apparent through the volume render. Conversely, there is no evidence for a node in either upper component. This indicates an equivalent node mismatch, as found for positive-parity wavefunctions, also exists in the negative-parity spectrum.}
\end{figure*}

We next turn to the second negative-parity excited state, for which the equivalent visualisations are shown in Fig.~\ref{fig:secondnegativeparityvis}. Here, fascinating structure is present in the lower \((4,4)\) component, which displays a spherically symmetric \(s\)-wave structure. The suggestion of a node is found at large radial distances in this component, which is apparent through its volume render. No sign of the node is detectable in the surface render. This is expected, since we are looking at large enough radial distances such that the wavefunction amplitude sits too low for any node to be visible in the surface render at the displayed scale.

Even though the node cannot be unambiguously resolved here, we will see in Sec.~\ref{subsec:negativeparityradial} that it is a genuine effect. This is shown using our detailed averaging procedure for calculating the radial wavefunctions. We have therefore uncovered that, in complement to the positive-parity wavefunctions, a node mismatch also exists in the negative-parity sector, where the lower (\(s\)-wave) components of certain states possess an additional node over the upper (\(p\)-wave) components.

\subsection{Radial wavefunctions} \label{subsec:negativeparityradial}
\begin{figure*}
	\includegraphics[width=0.49\linewidth]{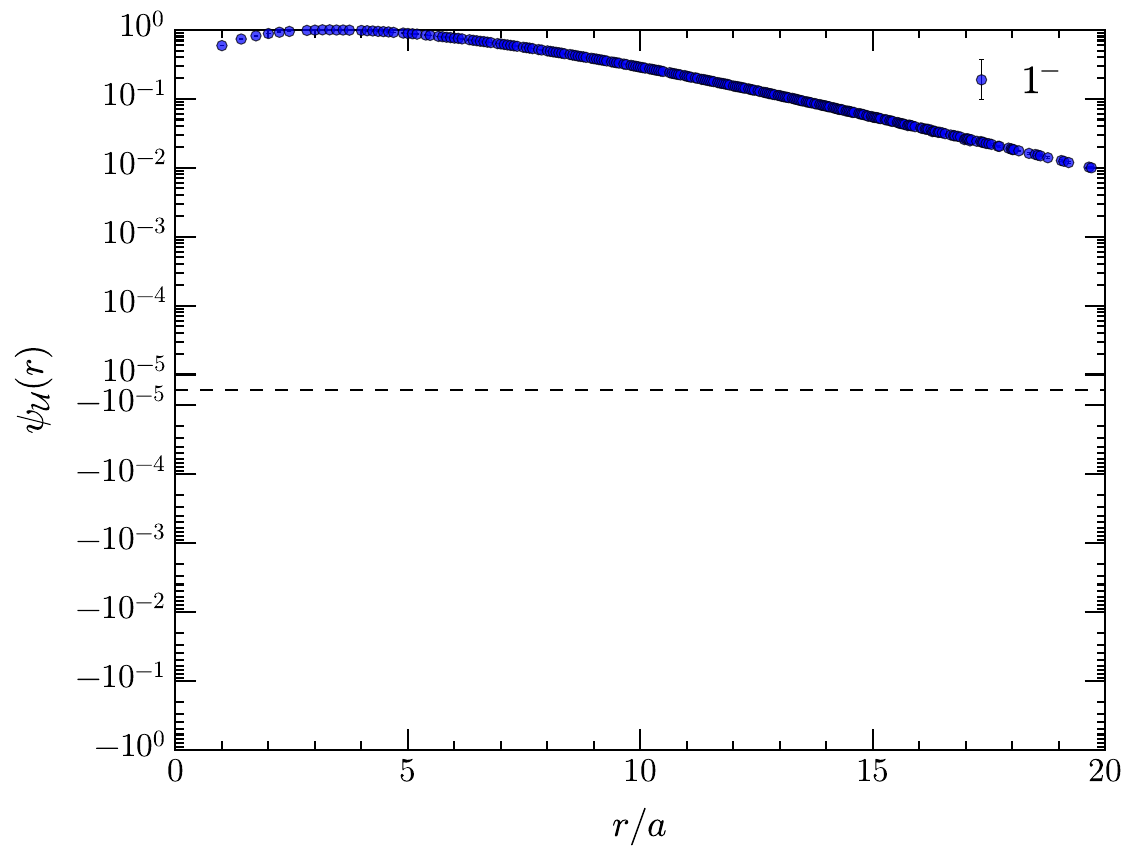}%
	\hfill
	\includegraphics[width=0.49\linewidth]{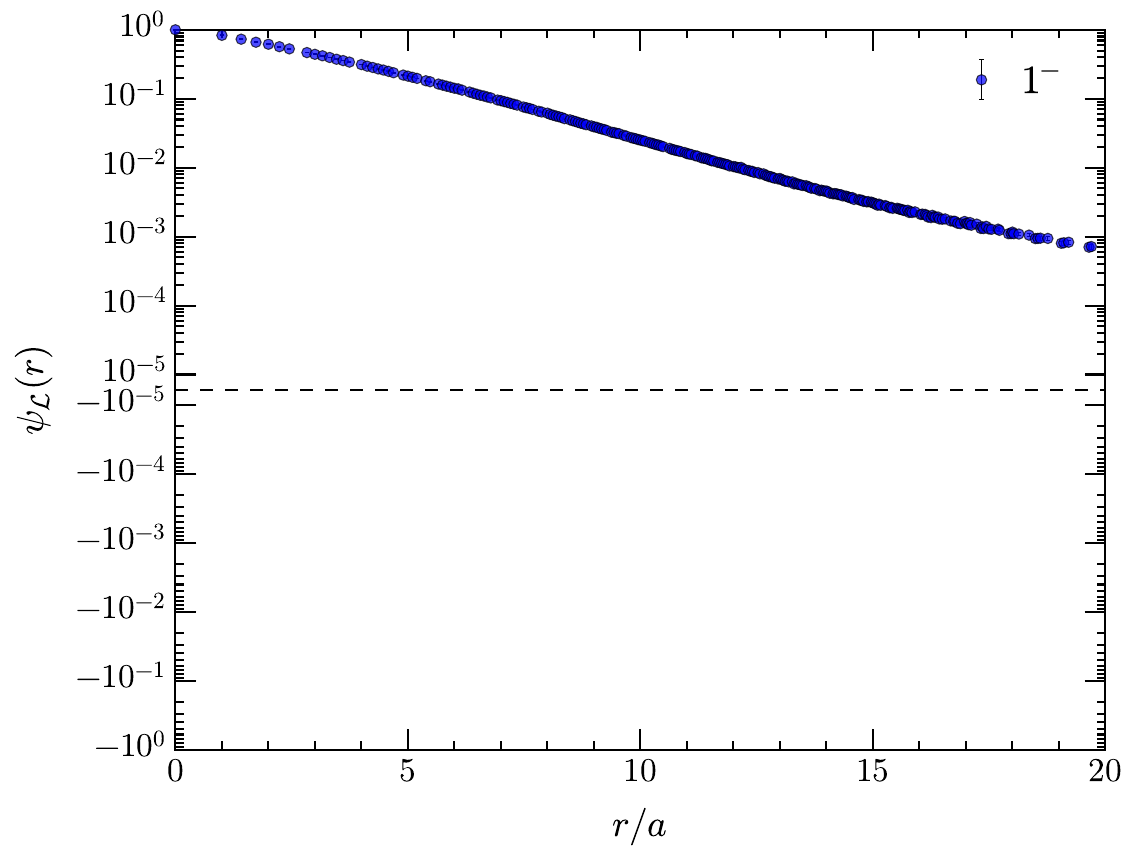}%
	\vspace{0.2em}
	\includegraphics[width=0.49\linewidth]{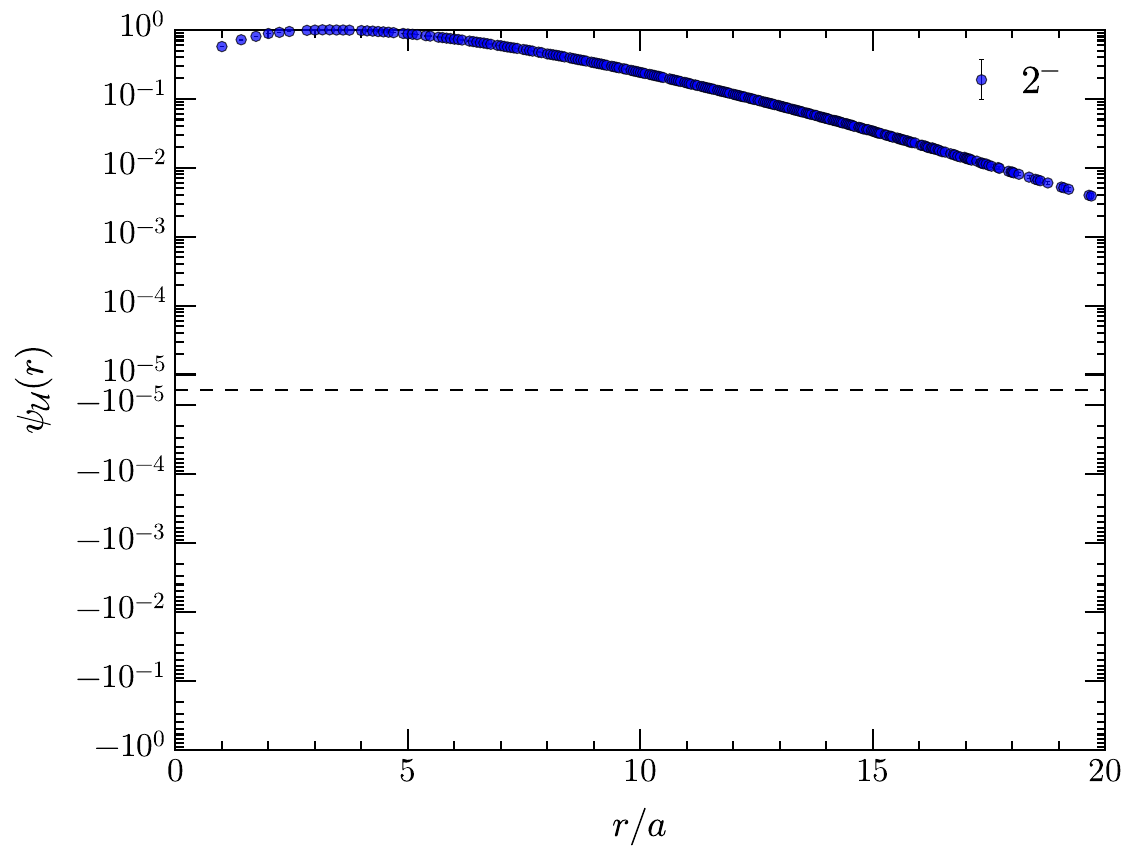}%
	\hfill
	\includegraphics[width=0.49\linewidth]{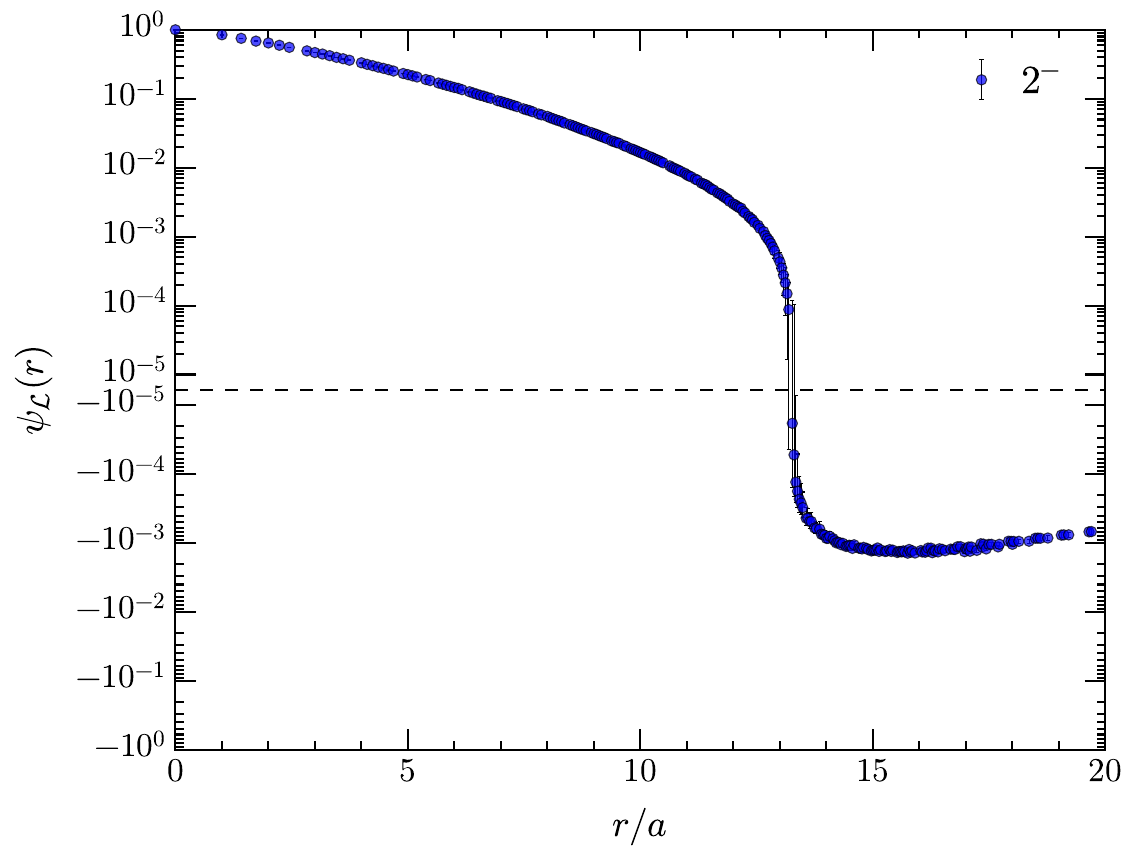}%
	\vspace{-0.5em}
	\caption{\label{fig:negativeparityradial1-2} The upper (\textbf{left}) and lower (\textbf{right}) radial wavefunctions, \(\psi_\mathcal{U}(r)\) and \(\psi_\mathcal{L}(r)\), for the two lowest-lying negative-parity excitations at Euclidean time \(\tau_\mathrm{min} = 5\) slices after the source. We find that the first negative-parity state (\textbf{top}) has zero nodes across both upper and lower wavefunction components, while the second state (\textbf{bottom}) has one node in its lower (\(s\)-wave) components but not in its upper (\(p\)-wave) components. This is similar to the node mismatch found in the positive-parity spectrum.}
\end{figure*}

\begin{figure*}
	\includegraphics[width=0.49\linewidth]{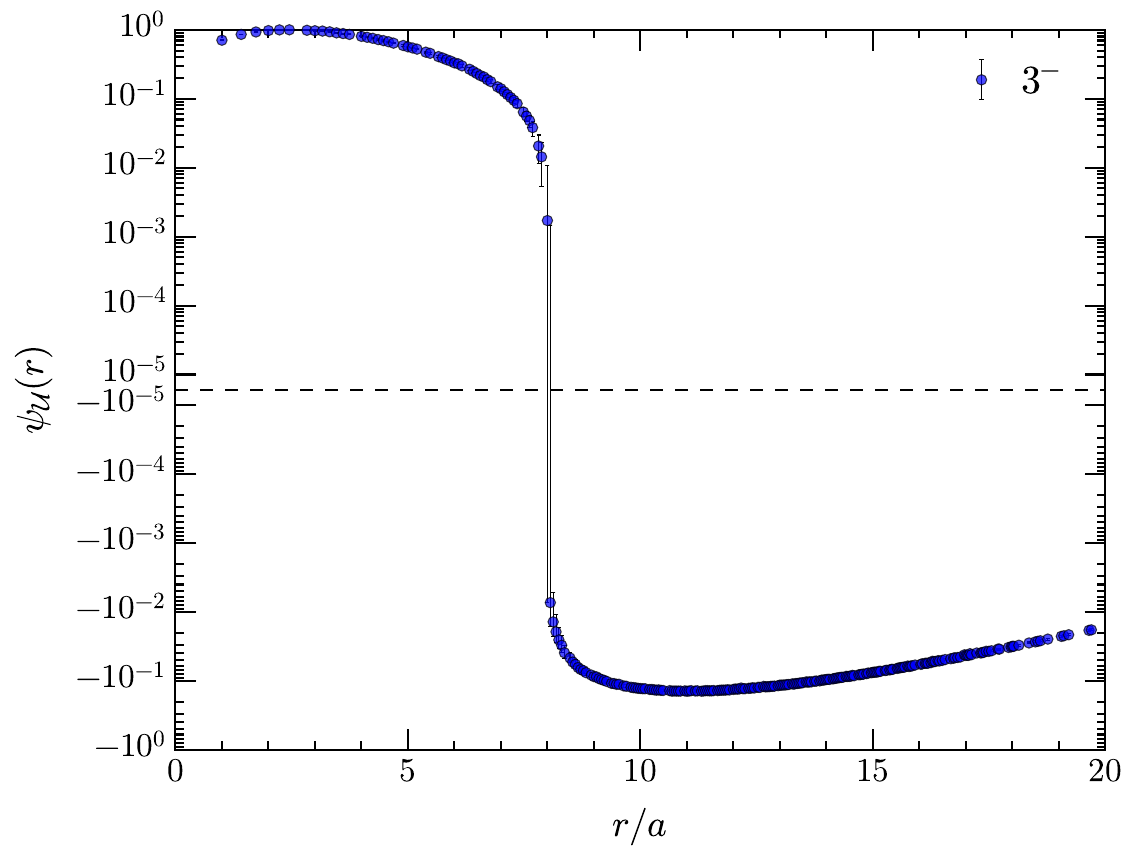}%
	\hfill
	\includegraphics[width=0.49\linewidth]{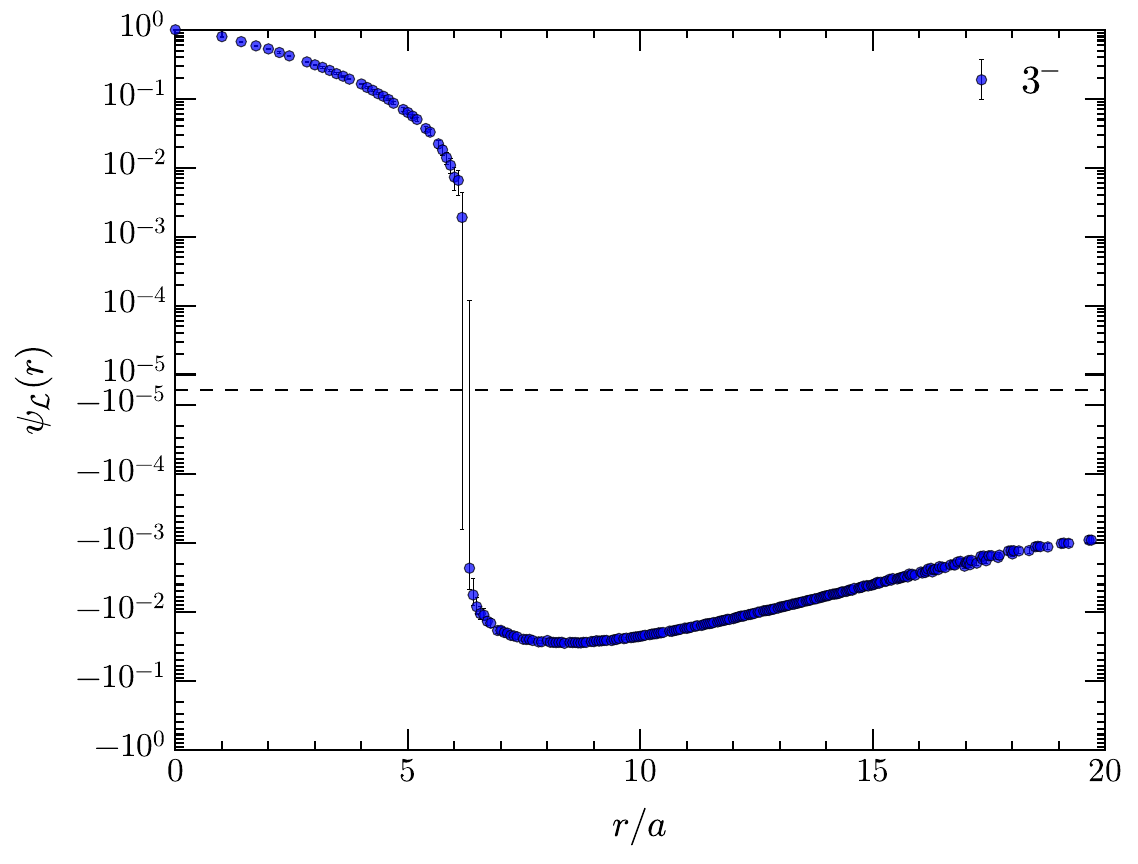}%
	\vspace{0.2em}
	\includegraphics[width=0.49\linewidth]{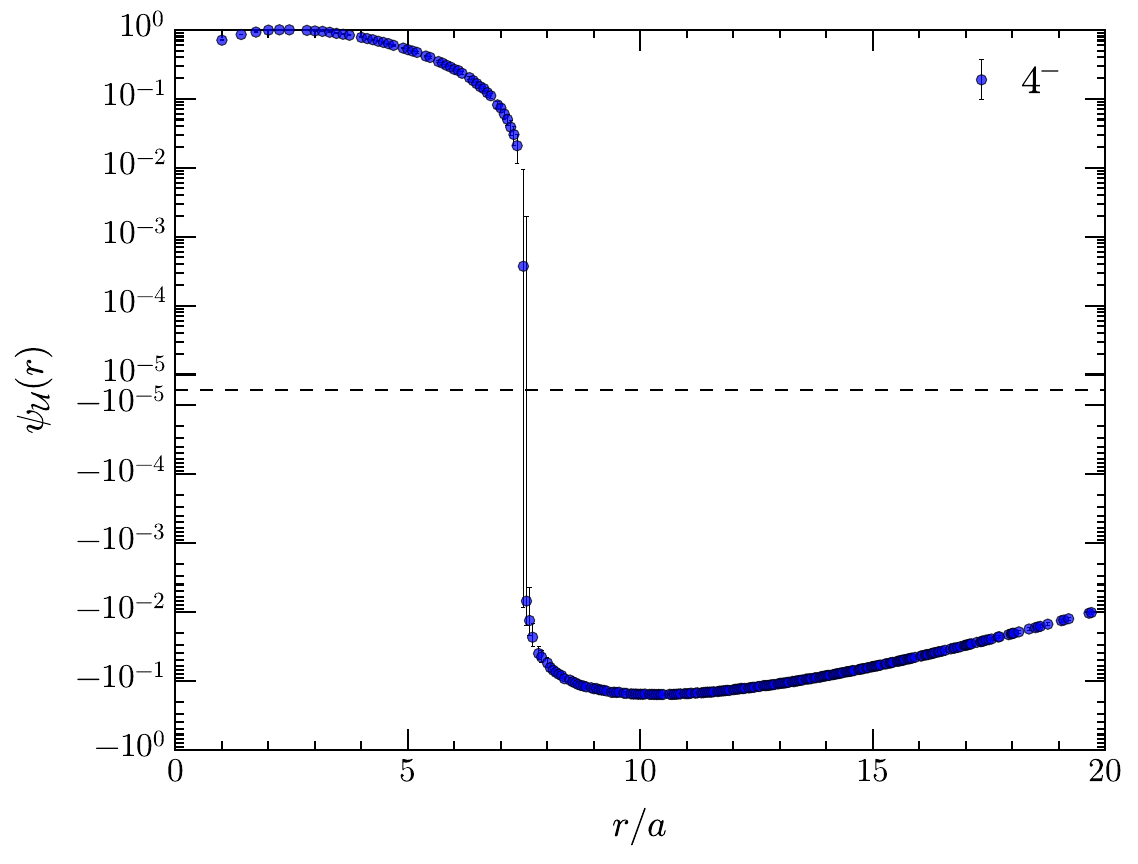}%
	\hfill
	\includegraphics[width=0.49\linewidth]{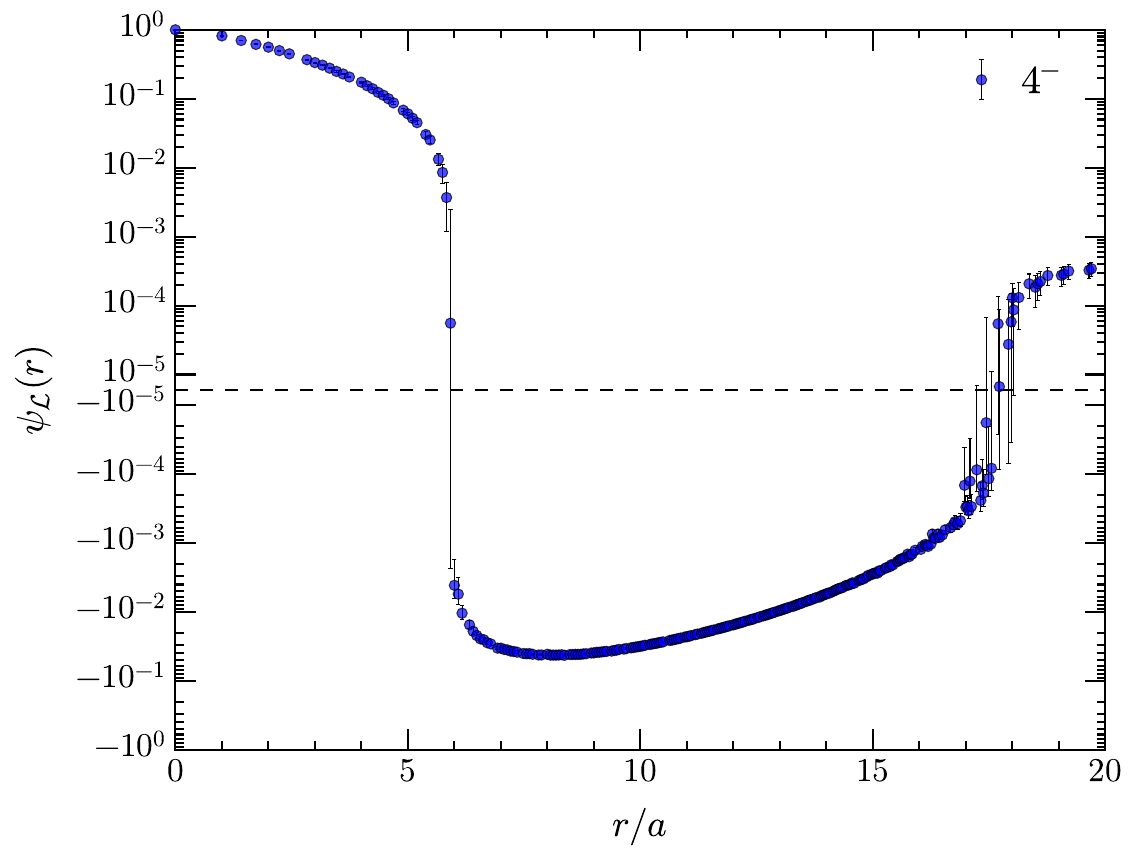}%
	\vspace{-0.5em}
	\caption{\label{fig:negativeparityradial3-4} As in Fig.~\ref{fig:negativeparityradial1-2}, but for the third (\textbf{top}) and fourth (\textbf{bottom}) negative-parity states at Euclidean time \(\tau_\mathrm{min} = 4\). The wavefunction for the third state contains one node in both its upper and lower components, while the fourth state exhibits another node mismatch, with two nodes in its lower (\(s\)-wave) wavefunction components but only a single node in its upper (\(p\)-wave) components.}
\end{figure*}

The visualisations in Figs.~\ref{fig:firstnegativeparityvis} and \ref{fig:secondnegativeparityvis} have provided an initial indication that the primary structural difference between the two lowest-lying negative-parity states is that the higher state contains a node mismatch. We now investigate this in greater detail via their radial wavefunction dependence. This is computed as outlined for positive parity in Sec.~\ref{subsec:positiveparityradial}. Prior to performing this calculation, we have again verified that within statistical precision, the angular dependence in the negative-parity sector is fully described by the appropriate spherical harmonic. The associated scatter plots are markedly similar to those for the positive-parity wavefunctions in Fig.~\ref{fig:groundstateangular} for the same angular momentum quantum numbers. We remind the reader that for negative parity, the \(s\)-wave structure is found in the lower wavefunction components, and the \(p\)-wave structure in the upper components (i.e.\ opposite to positive parity). This is apparent in the radial wavefunctions.

The radial wavefunctions of the two lowest-lying odd-parity excited states are displayed in Fig.~\ref{fig:negativeparityradial1-2}. The first excitation, like the ground state, shows zero nodes in all relativistic components. This reflects the visualisations in Fig.~\ref{fig:firstnegativeparityvis}. Moving to the second excited state, the radial wavefunctions definitively reveal a node in the lower (\(s\)-wave) components. This immediately establishes that the node suggested by Fig.~\ref{fig:secondnegativeparityvis} is a genuine effect. The node is located at large radial distances and consequently sits low in the wavefunction amplitude. Given the region around a node is noisy, it is unsurprising that this node is challenging to resolve in visualising a single wavefunction component, without any averaging.

It is noteworthy that the second excitation, with a marginally higher mass, is the one that contains a node. As previously elucidated in the context of positive parity, this node will push the probability density out to larger radial distances where the confining potential generates a higher mass for the state. However, in this case, since the node develops at large radial distances and in the small (lower) wavefunction components, it is only a soft effect. This intuitively underpins why the mass splitting between the two lowest-lying negative-parity states is as small as it is.

Furthermore, a corresponding node is absent from the upper (\(p\)-wave) components of this state's wavefunction. We have therefore discovered another mismatch in node structure between \(s\)-wave and \(p\)-wave components, this time in the negative-parity spectrum. This is not unlike the node mismatch in the positive-parity sector, where in both cases the extra node forms in the \(s\)-wave components of the wavefunction. For positive parity, these are the large (upper) components, while for negative parity they are the small (lower) components. This generates a large size discrepancy between the induced mass splittings. The extra node in this case is situated at comparatively large radial distances, whereas in positive-parity states it occurs close to the origin. Exploring the mechanism underlying these node mismatches will be a primary focus of Sec.~\ref{sec:interpolatingfields}.

We again speculatively examine the radial wavefunctions of the next two states, i.e.\ the third and fourth negative-parity excitations. These are shown in Fig.~\ref{fig:negativeparityradial3-4}. Following on from the node mismatch in the second state, the third negative-parity state now has one node in both lower and upper wavefunction components. These nodes are highly comparable to those of the first positive-parity excited state (Fig.~\ref{fig:positiveparityradial1-3}), suggesting an equivalence between these two states within their respective spectra. Finally, in the fourth negative-parity state we find another node mismatch, with two clearly resolved nodes in its lower (\(s\)-wave) components compared to one in its upper (\(p\)-wave) components. The second \(s\)-wave node is again located at a very large radial distance, just as with the \(s\)-wave node in the second negative-parity state. This only strengthens the hypothesis that there are two distinct types of nodes at play that combine to create this state.

\section{Interpolating-field analysis} \label{sec:interpolatingfields}
We have seen throughout Secs.~\ref{sec:positiveparity} and \ref{sec:negativeparity} that in both positive- and negative-parity spectra, there are two distinct categories of states that manifest:\ those whose wavefunctions have a consistent number of nodes across all spinor components, and those with a mismatch by one extra node in their \(s\)-wave components. In this section, we will illuminate this distinction by scrutinising the individual ingredients that make up each wavefunction.

This will be achieved in two parts. First, we will look at the eigenvectors \(\{u_n\}\) for each energy eigenstate. Subsequently, we will examine wavefunctions calculated for an individual interpolating field \(\chi_i\),
\begin{equation} \label{eq:interpolatorwavefunction}
	\psi_i(\mathbf{r};\tau) = \sum_\mathbf{x} \Braket{\Omega | \mathcal{T}\{\chi_1(x;\mathbf{r}) \, \bar{\chi}_i(0)\} | \Omega} \,.
\end{equation}
By considering how these ``interpolator wavefunctions'' are combined through the coefficients \(\{u_n^i\}\), we will build a comprehensive picture of the nucleon spectrum. Finally, we will compare our findings against models for nucleon structure, including the MIT bag model.

\subsection{Eigenvector components} \label{subsec:eigenvectorcomponents}
\begin{figure}
	\includegraphics[width=\linewidth]{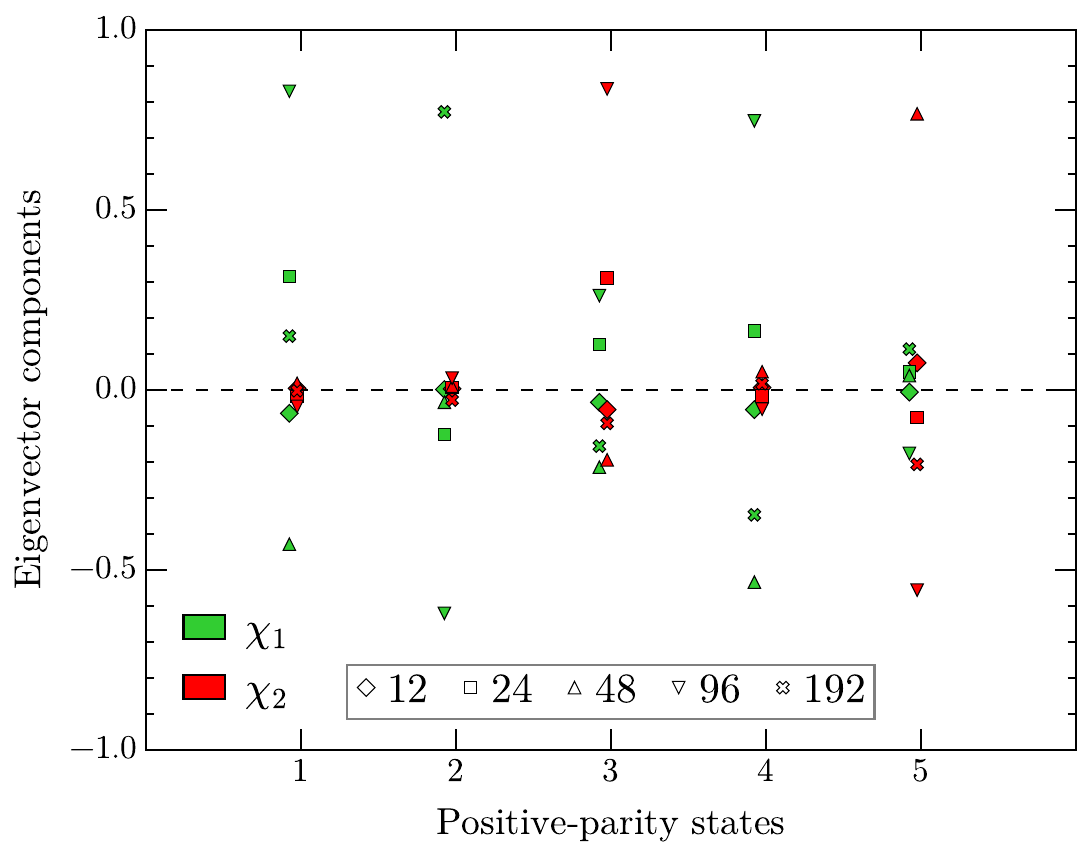}%
	\vspace{0.5em}
	\includegraphics[width=\linewidth]{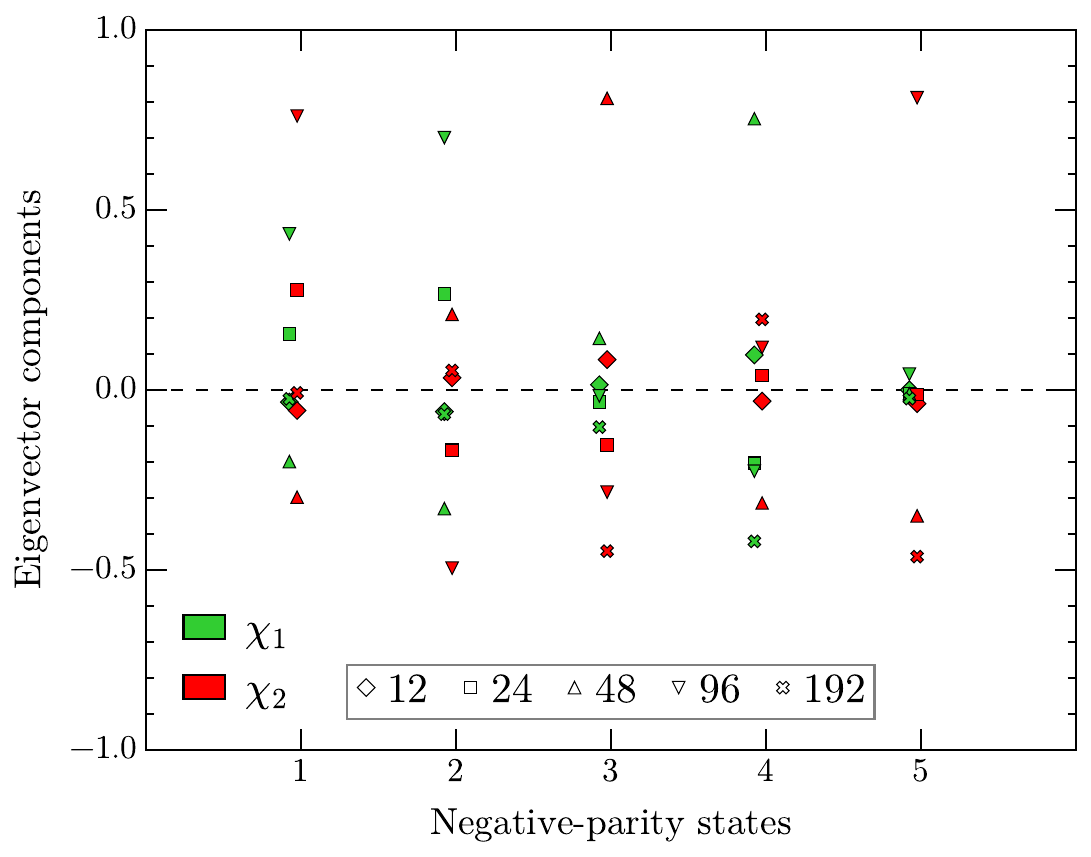}%
	\vspace{-0.5em}
	\caption{\label{fig:eigenvectorcomponents} The eigenvector components \(\{u_n^i\}\) for positive-parity (\textbf{top}) and negative-parity (\textbf{bottom}) eigenstates. Each component multiplies one of the interpolating fields \(\chi_1,\chi_2\) at the smearing levels specified in the legend (12, 24, 48, 96 or 192 sweeps). The \(\chi_1\) components are coloured in green (light grey) and \(\chi_2\) in red (dark grey). The relative signs between components generates nodes in the wavefunctions through superposition of the individual interpolating fields.}%
	\vspace{-0.5em}
\end{figure}

As mentioned, we commence by inspecting the eigenvector components \(\{u_n^i\}\) for each energy eigenstate analysed in Secs.~\ref{sec:positiveparity} and \ref{sec:negativeparity}. This will be crucial for understanding the observed node pattern of the radial excitations. These coefficients are plotted for the five lowest positive-parity and negative-parity states in Fig.~\ref{fig:eigenvectorcomponents}, and labelled by the smeared interpolating field to which they correspond.

Focusing initially on the positive-parity eigenstates, we see that the two lowest-lying states are overwhelmingly dominated by the \(\chi_1\) interpolating fields. This is consistent with prior work~\cite{Mahbub:2013ala}. Indeed, the eigenvector components multiplying \(\chi_2\) are near zero for both. Now, we know that the second state (i.e.\ first excitation) possesses a node in all components of its radial wavefunction. For this state, almost all coefficients are clustered around zero save two, which are approximately equal but opposites (\(\chi_1\) at 96 and 192 smearing sweeps). It is this relative subtraction between two interpolating fields of near-equal magnitude that generates the anticipated node in the wavefunction.

\begin{figure}
	\includegraphics[width=\linewidth]{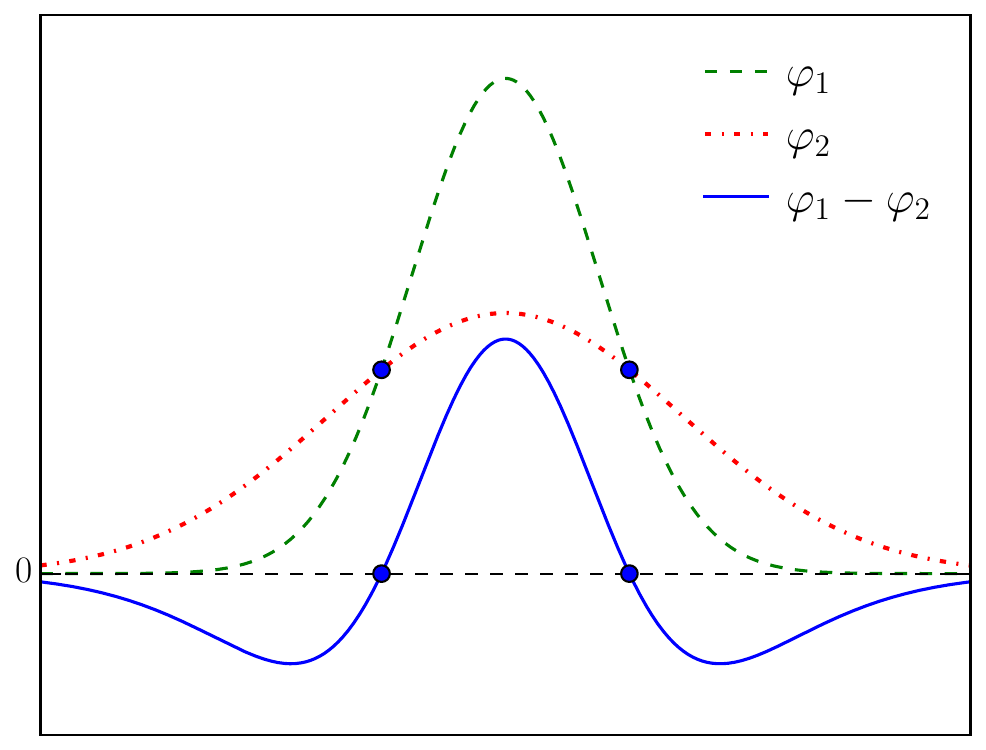}%
	\vspace{-0.5em}
	\caption{\label{fig:gaussians} An illustration of how two normalised Gaussians, \(\varphi_1\) and \(\varphi_2\), with different widths superpose to create a node. The broader Gaussian, \(\varphi_2\) (red/dash-dot), sits below the narrower Gaussian, \(\varphi_1\) (green/dashed), near their peaks. Past an intersection point, the broader Gaussian will ultimately sit higher. Consequently, in taking their difference \(\varphi_1 - \varphi_2\) (blue/solid), a node is formed at the intersection point at which the superposition changes sign.}%
	\vspace{-1em}
\end{figure}

This can be intuitively understood by imagining two (normalised) Gaussians of different widths, as illustrated in Fig.~\ref{fig:gaussians}. Around their peaks, the broader Gaussian sits below its narrower counterpart. However, beyond a certain point, the two Gaussians intersect and the situation reverses. Accordingly, in taking their difference, a node forms at the intersection point. Similar logic can be applied to the interpolating fields that superpose to construct the full wavefunction. It is clear that since the same superposition is applied to each, a node generated in this manner will appear in all four spinor components of the wavefunction.

It is also interesting to note that the ground state has a small negative component. Although this will combine with the small positive components, it will not generate a node due to the largest positive component ultimately being dominant. Still, this implies there is a degree of cancellation occurring even in the ground state to precisely produce the shape of its wavefunction.

Moving to the third positive-parity state (second excitation), this is the first state that is dominated by the \(\chi_2\) interpolating field. This was preemptively revealed back in the colour coding of the mass spectrum in Fig.~\ref{fig:massspectrum}. It appears to be a \(\chi_2\) analogue of the ground state, with no node generated through superposition of interpolating fields. However, we now know that this state features a node exclusively in its upper wavefunction components. It appears that this node mismatch can therefore be attributed to the dominance of \(\chi_2\) in the state.

This is supported by the fourth and fifth states in the spectrum. For one, the fourth state is again \(\chi_1\) dominated and has almost zero overlap with \(\chi_2\). Compared to the second state, more eigenvector components are substantially nonzero, and these arrange themselves to produce the two nodes observed in this state's wavefunction by superposition. Finally, the fifth state is back to being dominated by \(\chi_2\). This coincides with the node mismatch in the state. In this case, however, a node will also form in all wavefunction components through superposition of the \(\chi_2\) interpolators. This accounts for having two nodes in the upper (\(s\)-wave) components and one in the lower (\(p\)-wave) components.

Turning now to the negative-parity spectrum, one primary difference that arises compared to positive parity is a greater mixture of \(\chi_1\) and \(\chi_2\) components. This is especially apparent for the two lowest-lying states. We observe an interesting relationship between these two states:\ the \(\chi_1\) components of the second state are roughly equal to the \(\chi_2\) components of the first state, and the same approximate equality holds for the magnitudes of the \(\chi_2\) components of the second state and the \(\chi_1\) components of the first state but with a relative sign flip. By writing the eigenvectors in \(\chi_1/\chi_2\) blocks, this correspondence can be expressed symbolically as \((\chi_1,\chi_2) \leftrightarrow (-\chi_2,\chi_1)\), which are manifestly orthogonal to each other. This relationship also appears to hold for the next pair of states in the negative-parity spectrum (i.e.\ the third and fourth states displayed in Fig.~\ref{fig:eigenvectorcomponents}).

Despite the stronger admixture of \(\chi_1\) and \(\chi_2\) interpolating fields, the negative-parity states can still be categorised as dominated by one of \(\chi_1\) or \(\chi_2\) based purely on their largest components. For instance, we would classify the first state as \(\chi_2\) dominated and the second state as \(\chi_1\) dominated. This is how they were labelled in Fig.~\ref{fig:massspectrum}. Although more ambiguous than for positive parity, this distinction is still fundamentally connected to the node structure of these states' wavefunctions.

To be exact, the \(\chi_2\)-dominated states for negative parity are those that have the same number of nodes in all wavefunction components (first and third states). It is now the \(\chi_1\)-dominated (second and fourth) states with the node mismatch. This is the opposite correspondence than for positive parity, signifying a role reversal of the \(\chi_1\) and \(\chi_2\) interpolating fields in generating the positive- and negative-parity spectra.

\subsection{Interpolating-field wavefunctions} \label{subsec:interpolatorwavefunctions}
We have seen how the node structure of a given state has a direct correspondence to the dominant interpolating field in that state. In the positive-parity sector, a node mismatch appears in \(\chi_2\)-dominated states, whereas in the negative-parity sector, it occurs in \(\chi_1\)-dominated states. To follow on from this, we move away from the single-state projected eigenvectors from the GEVP, and now analyse wavefunctions calculated using a simple smeared interpolating field, as defined in Eq.~(\ref{eq:interpolatorwavefunction}). It is of great interest to examine each interpolating field individually, with the hope of exposing the properties that induce this observation.

These interpolating-field wavefunctions are computed at the normalisation point for our correlation matrix, two time slices after the source (Eq.~(\ref{eq:normalisation})). This admits a direct comparison between the relative magnitudes of each interpolating field. Radial wavefunctions are calculated as detailed in Sec.~\ref{subsec:positiveparityradial}, with the maximum value across all interpolators normalised to one (keeping their relative magnitudes in tact). In this discussion, ``positive parity'' refers to selection of the upper components at the source (\(\alpha = 1, 2\)), and ``negative parity'' to the lower components at the source (\(\alpha = 3, 4\)). 

\begin{figure}
	\includegraphics[width=\linewidth]{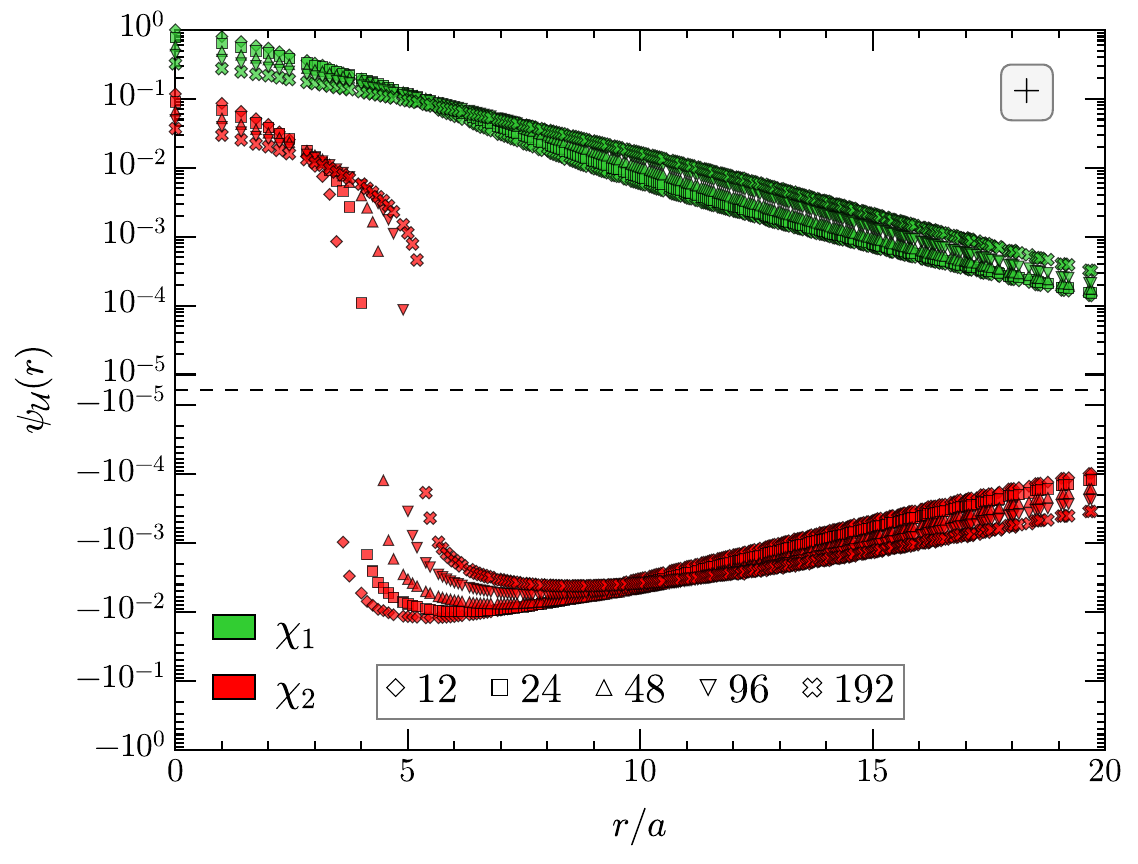}
	\includegraphics[width=\linewidth]{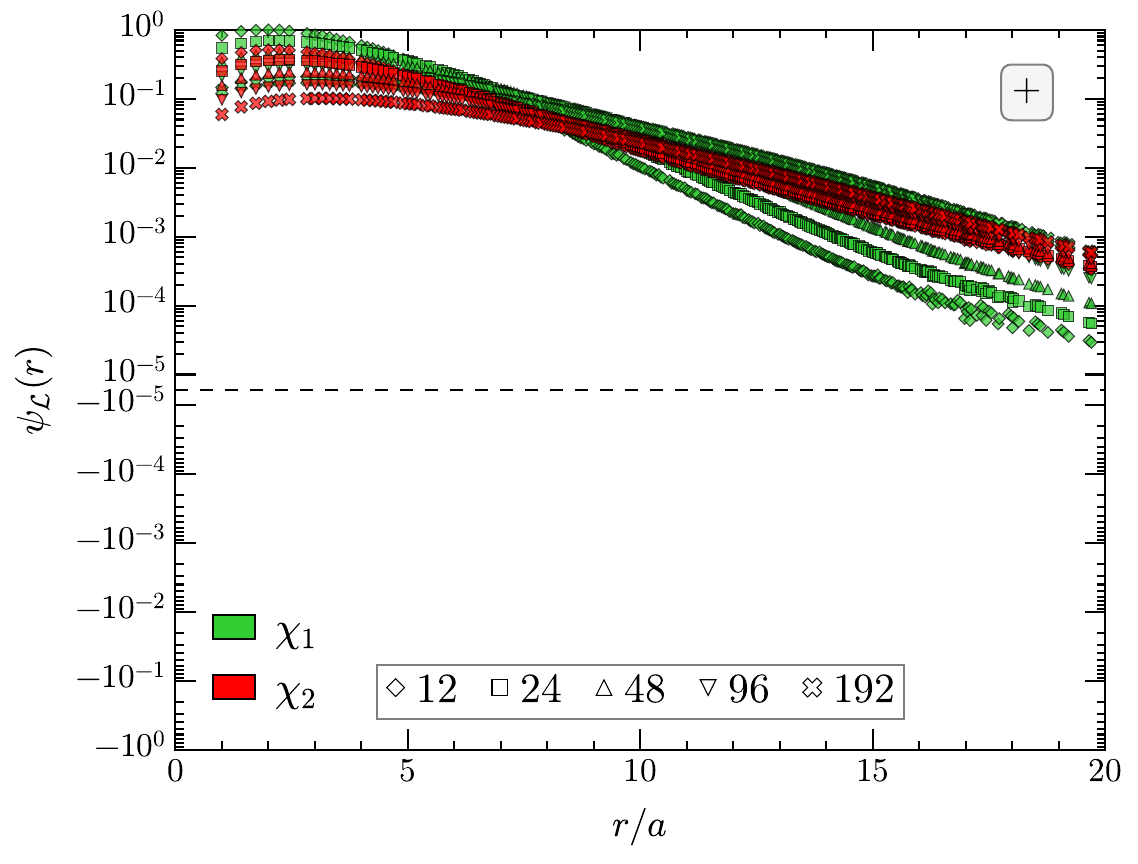}%
	\vspace{-0.5em}
	\caption{\label{fig:positiveparitywavefunctioncomponents} The upper (\textbf{top}) and lower (\textbf{bottom}) Dirac components of the positive-parity radial wavefunctions for each interpolating field \(\chi_1,\chi_2\) at the smearing levels specified in the legend (12, 24, 48, 96 or 192 sweeps). The \(\chi_1\) wavefunctions are coloured in green (light grey) and \(\chi_2\) in red (dark grey). For positive parity, the upper Dirac components display \(s\)-wave behaviour, while the lower components display \(p\)-wave behaviour. In the upper components, the \(\chi_2\) interpolators are seen to contain a built-in node, distinguishing them from \(\chi_1\). In the lower components, neither \(\chi_1\) nor \(\chi_2\) have a built-in node. This signifies that \(\chi_2\)-dominated positive-parity states will possess an extra node in their upper components over the lower components.}%
	\vspace{-0.5em}
\end{figure}

Starting with positive parity, we present the upper and lower radial wavefunctions calculated on each interpolating field in Fig.~\ref{fig:positiveparitywavefunctioncomponents}. Looking at the upper Dirac components, we immediately notice a salient feature wherein the \(\chi_2\) interpolating fields possess a node. We emphasise that this is distinct from the nodes discussed in Sec.~\ref{subsec:eigenvectorcomponents} that form through superposition of multiple interpolating fields. Instead, these nodes are \emph{built in} to the \(\chi_2\) interpolators.

Unlike the ``superposition nodes'' that manifest in all four spinor components, these ``built-in'' nodes are found to be absent from the corresponding lower Dirac components. This precisely accounts for the node mismatch observed in the positive-parity spectrum. For states that are dominated by \(\chi_1\), the node built in to \(\chi_2\) will not appear in its wavefunction, being hidden by the stronger \(\chi_1\) contribution. On the other hand, if a state is dominated by \(\chi_2\) the built-in node will be brought to the forefront. That is, \(\chi_2\)-dominated positive-parity states will contain an extra node in their \(s\)-wave (upper) components over their \(p\)-wave (lower) components.

\begin{figure}
	\includegraphics[width=\linewidth]{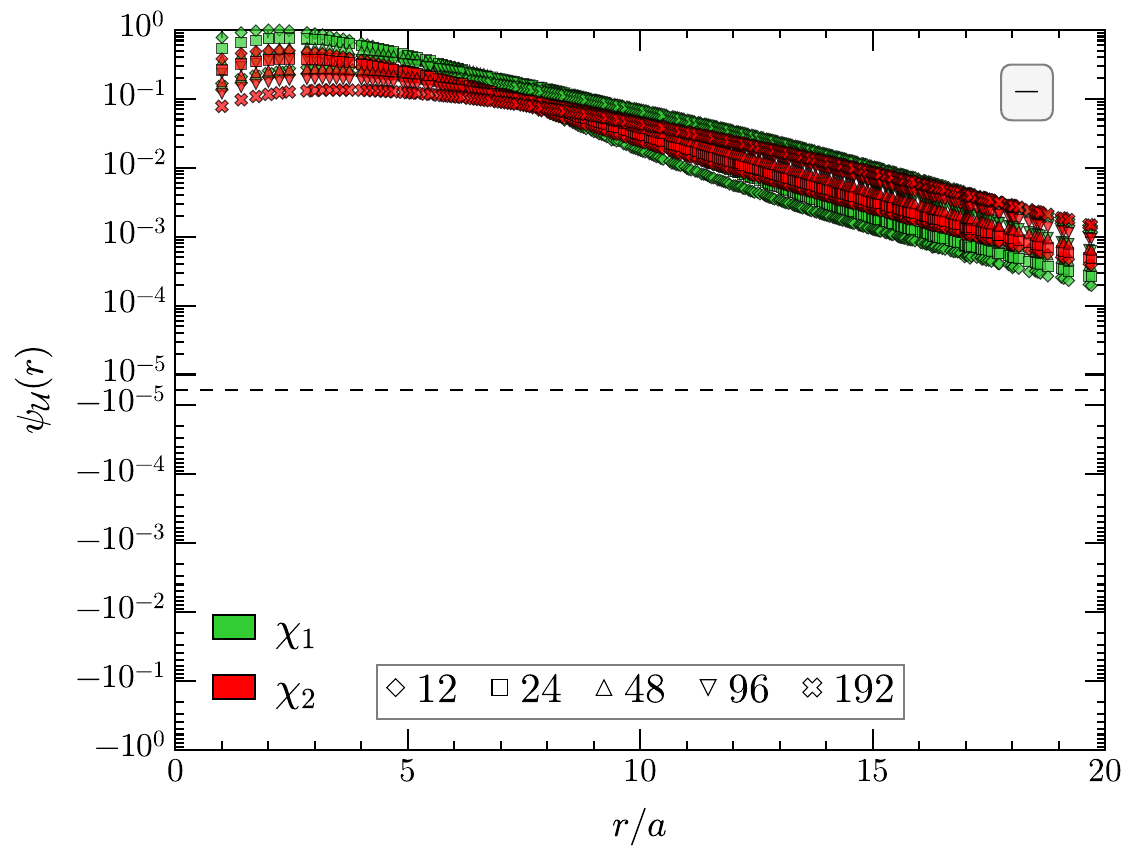}
	\includegraphics[width=\linewidth]{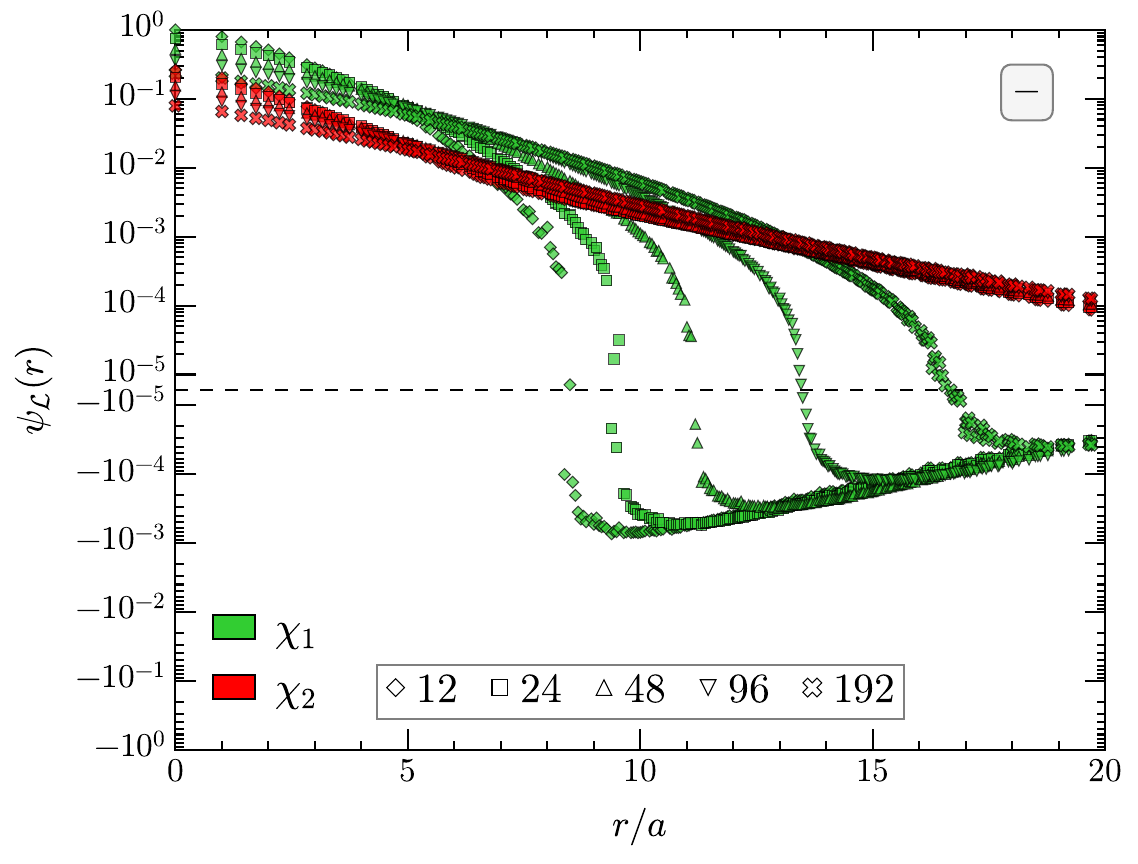}%
	\vspace{-0.5em}
	\caption{\label{fig:negativeparitywavefunctioncomponents} As with Fig.~\ref{fig:positiveparitywavefunctioncomponents}, but for negative parity. Here, it is the upper Dirac components that display \(p\)-wave behaviour, while the lower components display \(s\)-wave behaviour. In the upper components, there is no node built in to either \(\chi_1\) or \(\chi_2\) interpolating fields, in agreement with the positive-parity \(p\)-wave components (Fig.~\ref{fig:positiveparitywavefunctioncomponents}). In the lower components, \(\chi_1\) does contain a built-in node. This indicates that for negative parity, the \(\chi_1\)-dominated states will possess an extra node in their lower (\(s\)-wave) components.}%
	\vspace{-0.5em}
\end{figure}

We now shift focus to negative parity. The radial dependence of the upper and lower wavefunction components are displayed in Fig.~\ref{fig:negativeparitywavefunctioncomponents}. We again recall that opposite to positive parity, negative-parity wavefunctions display a \(p\)-wave structure in their upper components, and an \(s\)-wave structure in their lower components. We see that for negative parity, there are no built-in nodes found in the upper \(p\)-wave components of both \(\chi_1\) and \(\chi_2\). This coincides with the observation for positive parity, where the \(p\)-wave (lower) components also have no built-in nodes.

Finally, we draw our attention to the lower wavefunction components of negative-parity states. Here, the \(\chi_1\) interpolating fields are found to contain a built-in node. Similar to the discussion for positive parity, this will control the node mismatch for negative parity, where \(\chi_1\)-dominated negative-parity states will acquire an extra node in their \(s\)-wave (lower) components over their \(p\)-wave (upper) components. Relative to the positive-parity \(\chi_2\) built-in nodes (Fig.~\ref{fig:positiveparitywavefunctioncomponents}), these \(\chi_1\) nodes occur at comparatively large radial distances. This matches the observed eigenstate wavefunctions in Sec.~\ref{subsec:negativeparityradial}, where the ``extra'' nodes in the second and fourth negative-parity states are located at large radial distances.

The presence of these novel built-in nodes is fascinating, and explains the intrinsic connection between the eigenvectors \(\{u_n\}\) and node structure. This further solidifies that \(\chi_1\) and \(\chi_2\) interchange roles for positive and negative parity. There are both positive- and negative-parity states that possess an extra node in their \(s\)-wave components, but it is \(\chi_2\) that carries said node for positive parity, while it is \(\chi_1\) for negative parity.

The built-in node existing in a different interpolating field for the opposite parities has a natural explanation. To understand this, we recall the explicit forms of these interpolators in Eqs.~(\ref{eq:chi1}) and (\ref{eq:chi2}). For \(\chi_1\), when selecting the wavefunction parity through the source Dirac index, we are really identifying a spinor component of the \(u_c(x)\) quark field. Now, \(\chi_1\) carries the built-in node for negative parity. Since it is the lower source Dirac components that select negative parity, this signifies that the \(\chi_1\) built-in nodes necessarily originate from the lower Dirac components of this quark field.

Conversely, for \(\chi_2\), the source index is identifying a spinor component not of \(u_c(x)\), but of \(\gamma_5 u_c(x)\). The effect of \(\gamma_5\) is to swap the upper and lower Dirac components. In this way, \(u_c(x)\) and \(\gamma_5 u_c(x)\) have opposite parity. As a result, if we understand that \(u_c(x)\) has a node in its lower components, then \(\gamma_5 u_c(x)\) will have that same node in its upper components. This means it is the selection of the upper source Dirac components, i.e.\ the positive-parity wavefunctions, that will reveal the node for \(\chi_2\).

In summary, it is the presence or absence of \(\gamma_5\) multiplying the indexed quark field \(u_c(x)\) in the interpolating field that determines whether \(\chi_1\) or \(\chi_2\) reveals the built-in node. The former appears for negative parity, and the latter for positive parity. This allows the node pattern of all positive- and negative-parity states to be predicted according to the below prescription:
\begin{enumerate}
	\item The lowest-lying \(\chi_1\)- and \(\chi_2\)-dominated states both have zero superposition nodes.
	\item Separately add one node for each \(\chi_1\)-dominated and \(\chi_2\)-dominated radial excitation.
	\item Add an extra node to the \(s\)-wave components of \(\chi_2\)-dominated positive-parity states and \(\chi_1\)-dominated negative-parity states (the upper and lower components, respectively).
\end{enumerate}

\subsection{MIT bag model}
Although which interpolating field carries the built-in node can be easily understood, we have yet to account for its presence in the \(s\)-wave components of the wavefunctions, but absence in the \(p\)-wave components. Therefore, to conclude this section, we wish to examine whether this pattern of built-in nodes is consistent with models of nucleon structure. Perhaps the most well known such model is the MIT bag model~\cite{Chodos:1974je, Chodos:1974pn}, in which the quarks are placed in a spherical cavity of radius \(R\).

The radial wavefunctions in this model are given by the spherical Bessel functions,
\begin{align} \label{eq:MITradial}
	\psi^\pm_\mathcal{U}(r) &= j_\ell(k_n r) \,, & \psi^\pm_\mathcal{L}(r) &= \pm \, j_{\ell \pm 1}(k_n r) \,,
\end{align}
where we recall that \(\ell = j \mp 1/2\), and \(k_n\) is the wavenumber for the \(n\)th radial excitation. Without loss of generality, we assume massless quarks, \(m = 0\), for simplicity. A boundary condition is imposed such that no current can leave or enter the spherical bag,
\begin{equation}
	n_\mu j^\mu = 0,
\end{equation}
which is implemented through the following linear condition on the boundary,
\begin{equation}
	i \, n_\mu \gamma^\mu \, \psi = \psi \,.
\end{equation}
This translates to the following relation between the upper and lower components,
\begin{equation} \label{eq:MITboundarycondition}
	\psi^\pm_\mathcal{U}(R) = \psi^\pm_\mathcal{L}(R) \implies j_\ell(k_n R) = \pm \, j_{\ell\pm 1}(k_n R) \,.
\end{equation}
That is, the upper and lower radial wavefunctions are equal at the boundary. Equivalently, the two spherical Bessel functions, \(j_\ell\) and \(j_{\ell\pm 1}\), are either equal or negatives of each other, depending on the parity.

For the two Bessel functions in Eq.~(\ref{eq:MITboundarycondition}) to be equal at the boundary, as is the case for positive parity, the number of nodes must differ by a multiple of two. In actuality, since the two components share the same wavenumber, the number of nodes are necessarily equal. This implies that within the MIT bag model, the positive-parity wavefunctions will always have the same number of nodes in their upper and lower components.

\begin{figure}
	\includegraphics[width=\linewidth]{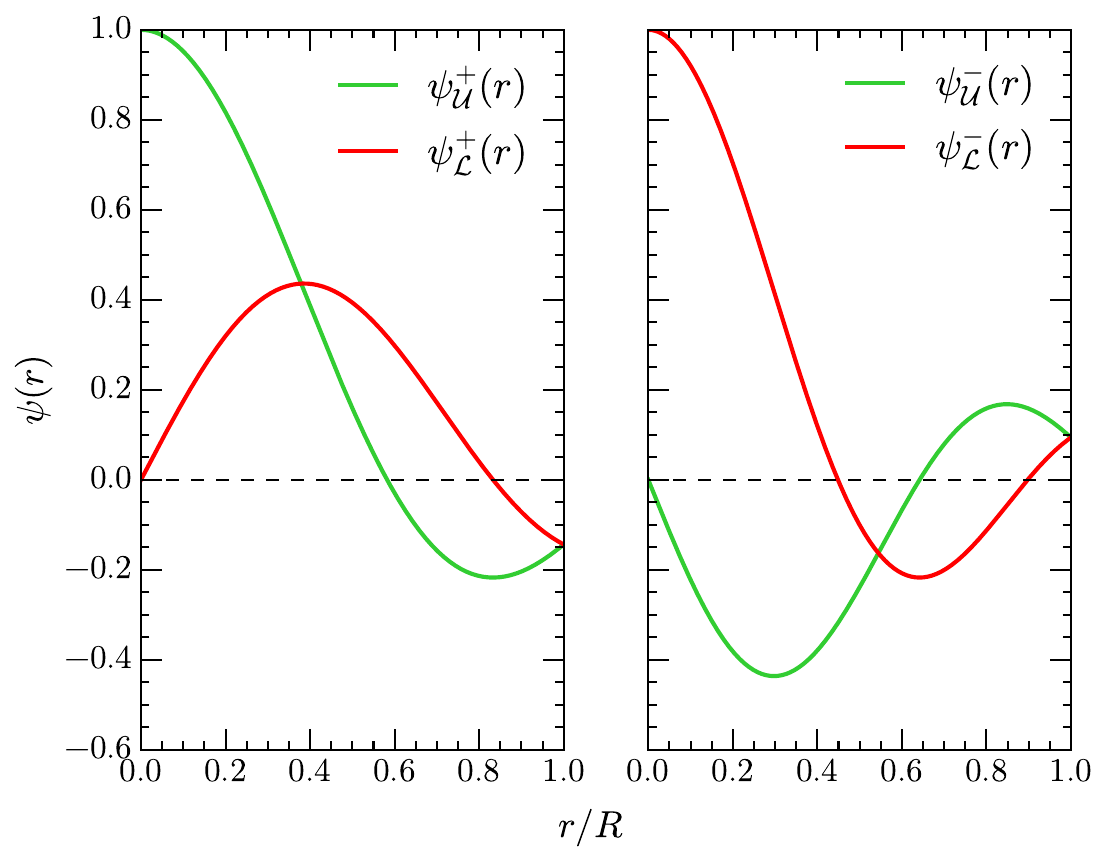}%
	\vspace{-0.5em}
	\caption{\label{fig:mitbagmodelsolutions} The upper and lower radial wavefunctions from the MIT bag model for the first excitation, \(n = 1\), for positive-parity (\textbf{left}) and negative-parity (\textbf{right}) solutions. For positive parity, the two components have the same number of nodes (\textit{not} counting that the \(p\)-wave component starts zero at \(r = 0\)). Meanwhile, the negative-parity solutions have an extra node in the \(s\)-wave (lower) components. The boundary conditions of Eq.~(\ref{eq:MITboundarycondition}) are apparent.}
\end{figure}

On the other hand, if the two Bessel functions are negatives of each other at the boundary, as for negative parity, this requires a mismatch by one node between the radial wavefunctions. For the negative-parity solutions (i.e.\ \(\ell = 1\)), it is \(j_1\) in the upper components and \(j_0\) in the lower. By examining the shapes of these functions, it is clear that \(j_0\), in the lower components, will carry the additional node. Example positive-parity and negative-parity wavefunctions for \(n = 1\) are provided in Fig.~\ref{fig:mitbagmodelsolutions} for reference, in which this pattern is evident.

We have therefore uncovered that the MIT bag model predicts an extra node in the \(s\)-wave (lower) components of the negative-parity wavefunctions. This is consistent with our previous conclusion that the \(u_c(x)\) source quark field will boast precisely such a node, which propagates through to the \(\chi_1\) and \(\chi_2\) interpolating fields as discussed in Sec.~\ref{subsec:interpolatorwavefunctions}.

Besides the MIT bag model, another approach is to model confined spectra through the introduction of a scalar potential to the Dirac equation,
\begin{align} \label{eq:scalarpotential}
	\big(i\gamma^\mu \partial_\mu - S(r)\big) \, \psi &= 0 \,,
\end{align}
which for appropriate choices of \(S(r)\) similarly produces bound states~\cite{Critchfield:1975yf, Critchfield:1976fi, Rein:1977uv, Ram:1979gs, Ram:1980dz, Ram:1980fr, Iyer:1981zx, Ravndal:1982sr, Su:1985mp}. In general, confinement occurs only when the scalar potential is stronger (to the same power in \(r\)) than a vector potential, \(V(r)\), coupled through the timelike component of a four-vector~\cite{Su:1985mp}. A particularly interesting option is a linearly rising potential, \(S(r) = \sigma r\) with \(\sigma > 0\), which has been extensively studied in the literature~\cite{Critchfield:1975yf, Critchfield:1976fi, Rein:1977uv, Ram:1980fr}.

In this case, we assume boundary conditions of \(\psi(r) \to 0\) as \(r \to \infty\), ensuring the solutions are square integrable. Although not perfect, this asymptotic behaviour is more applicable to lattice data than the boundary conditions in the MIT bag model, and it is thus worthwhile to briefly comment on these solutions.

Insight into solutions of Eq.~(\ref{eq:scalarpotential}) for various \(S(r)\) can be gleaned from the coupled differential equations for the upper and lower components,
\begin{align}
	\frac{\partial \psi^\pm_\mathcal{U}(r)}{\partial r} &= -\left(\kappa+1\right) \frac{\psi^\pm_\mathcal{U}(r)}{r} - \left(E + S(r)\right) \psi^\pm_\mathcal{L}(r) \,, \label{eq:upperradialeq} \\
	\frac{\partial \psi^\pm_\mathcal{L}(r)}{\partial r} &= \left(\kappa-1\right) \frac{\psi^\pm_\mathcal{L}(r)}{r} + \left(E - S(r)\right) \psi^\pm_\mathcal{U}(r) \,, \label{eq:lowerradialeq}
\end{align}
where \(\kappa = \mp\left(j + 1/2\right)\), and \(E\) is the energy eigenvalue.

For \(j = 1/2\) solutions, one has \(\kappa = -1\) for positive parity and the first term in Eq.~(\ref{eq:upperradialeq}) vanishes. This leads to a simplified differential equation for the upper components,
\begin{equation} \label{eq:positiveparityupperradialeq}
	\frac{\partial \psi^+_\mathcal{U}(r)}{\partial r} = - \left(E + S(r)\right) \psi^+_\mathcal{L}(r) \,.
\end{equation}
Now, provided \(E + S(r) > 0\) for all \(r\), which is certainly true if \(S(r) > 0\), Eq.~(\ref{eq:positiveparityupperradialeq}) implies that the number of turning points in the upper wavefunction, \(\psi^+_\mathcal{U}\), is equal to the number of zeros of the lower wavefunction, \(\psi^+_\mathcal{L}\). Furthermore, under the requirement that \(\psi(r) \to 0\) as \(r \to \infty\), the upper wavefunction necessarily possesses as many zeros as turning points. This leads one to conclude that for positive parity, the upper and lower components contain the same number of nodes.

For negative parity, one instead has \(\kappa = +1\) and it is the first term in Eq.~(\ref{eq:lowerradialeq}) that vanishes, leading to a simplified equation for the lower components,
\begin{equation} \label{eq:negativeparitylowerradialeq}
	\frac{\partial \psi^-_\mathcal{L}(r)}{\partial r} = \left(E - S(r)\right) \psi^-_\mathcal{U}(r) \,.
\end{equation}
Logic identical to that above applies here, except the right-hand side of Eq.~(\ref{eq:negativeparitylowerradialeq}) will now have an additional zero at the radial distance where \(E = S(r)\). The lower wavefunction, \(\psi^-_\mathcal{L}\), will correspondingly obtain an extra turning point and node. Therefore, the lower components of the negative-parity solutions will contain an extra node over the upper components. This is in line with the node structure from the MIT bag model.

This indicates that there is a general class of solutions to the Dirac equation, relevant to modelling confinement, that predict an extra node in the lower components of the negative-parity solutions. This conforms with our finding that the built-in nodes seen in Sec.~\ref{subsec:interpolatorwavefunctions} only occur in the \(s\)-wave components.

\begin{figure}
	\includegraphics[width=\linewidth]{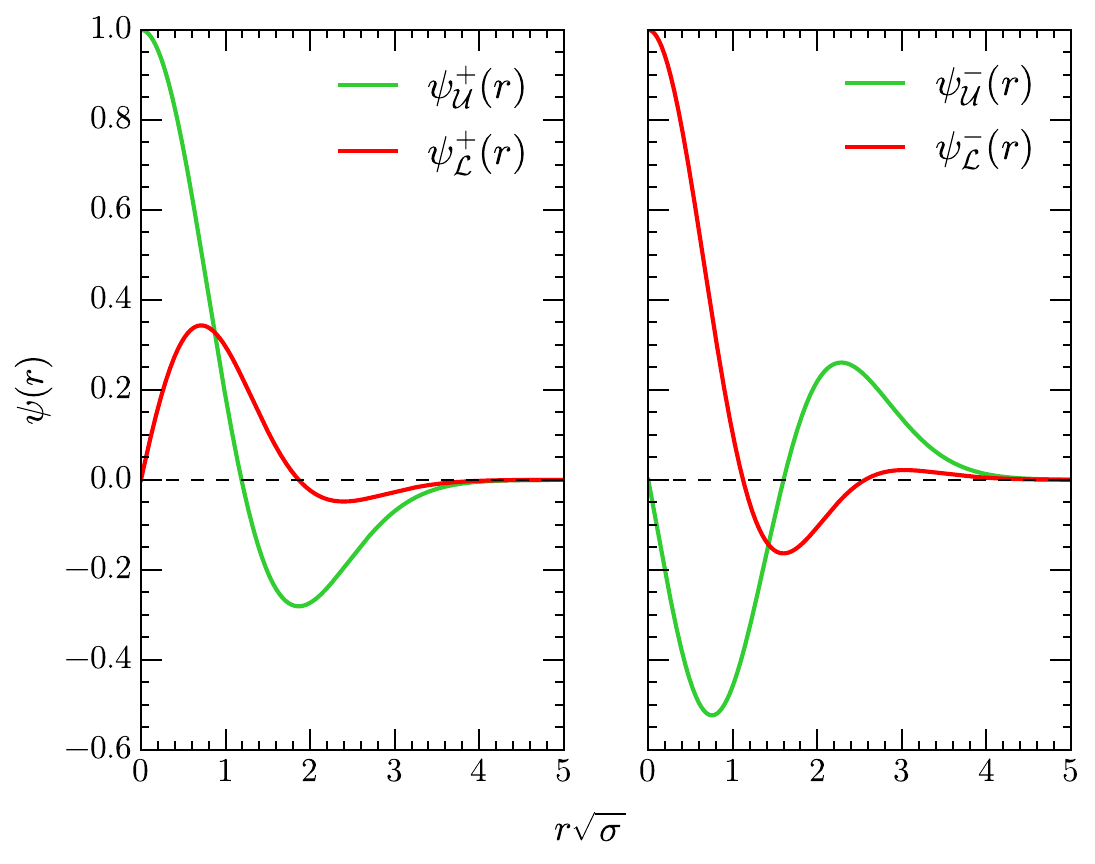}%
	\vspace{-0.5em}
	\caption{\label{fig:scalarlinearpotentialsolutions} Same as in Fig.~\ref{fig:mitbagmodelsolutions}, but for the Dirac equation with a linear scalar potential, \(S(r) = \sigma r\). The boundary conditions are such that \(\psi(r) \to 0\) as \(r \to \infty\). The extra \(s\)-wave node in the negative-parity wavefunctions is again apparent.}
\end{figure}

For comparison with Fig.~\ref{fig:mitbagmodelsolutions}, which showed example solutions for the MIT bag model, we plot in Fig.~\ref{fig:scalarlinearpotentialsolutions} the equivalent radial wavefunctions for a linear scalar potential. Since there is no known general analytic solution, these are produced by solving the Dirac equation numerically with \(m = 0\) and the boundary condition of \(\psi(r) \to 0\) as \(r\to\infty\). The predicted node pattern is evident, including the extra \(s\)-wave node for the negative-parity wavefunctions.

\section{Conclusion} \label{sec:conclusion}
In this paper, relativistic wavefunctions of the nucleon and its excited states were studied in the framework of lattice QCD. We employed the heaviest \((2+1)\)-flavour dynamical PACS-CS ensemble, which is in a regime where the quark model describes the nucleon and its excitations well and multi-particle scattering states lie above the region of interest. A \(10 \times 10\) correlation matrix was utilised by applying five levels of gauge-invariant Gaussian smearing to both the scalar-diquark and pseudoscalar-diquark proton interpolating fields, \(\chi_1\) and \(\chi_2\), of which the latter vanishes in the nonrelativistic limit. The various wavefunction components were obtained by exploiting parity and spin mismatches between the source and sink Dirac indices of the correlation matrix. This leads to a comprehensive understanding of the node structure of quark-model-like states and the inverse roles of \(\chi_1\) and \(\chi_2\).

Initially, volume and surface renderings for the three lowest-lying positive-parity states were presented. These visualisations were shown for both the upper and lower wavefunction components, which respectively exhibit \(s\)-wave and \(p\)-wave structures. The first two states display a standard node structure, with zero nodes in the ground state and one node in the first excited state across all relativistic wavefunction components. The third state, however, was found to feature a novel node mismatch between its upper and lower components, comprising one node in its \(s\)-wave (upper) components compared to zero nodes in its \(p\)-wave (lower) components.

This was followed by a quantitative analysis of node structure through calculation of the radial wavefunction dependence for positive-parity states. To achieve this, we first verified that the angular dependence of each wavefunction component is consistent with the corresponding spherical harmonic predicted by the Dirac equation. The radial wavefunctions corroborated the node structure discovered in the visualisations, including the node mismatch in the second excitation. By examining the next two positive-parity states, another pair with and without the node mismatch was uncovered. This establishes the importance of both types of states in building the spectrum.

In addition, the techniques developed for positive parity were applied to study the wavefunctions of negative-parity states for the first time. This involved a qualitative examination through visualisation, expanded upon with a comprehensive quantitative analysis. Here, the \(s\)-wave and \(p\)-wave components have switched, with the \(s\)-wave found in the lower wavefunction components and \(p\)-wave in the upper. A similar categorisation into states with and without a node mismatch in their wavefunction components was revealed for the negative-parity spectrum. For those states with the node mismatch, the extra node is again present in the \(s\)-wave components. In this case, the extra node is located at a very large radius, whereas for positive parity it is close to the origin.

This mismatch in node structure between the large and small relativistic wavefunction components brings novel insight into the origin of the mass splitting between pairs of states in the nucleon spectrum. By pushing the probability density farther out in the confining potential, the extra \(s\)-wave node generates a higher mass for those states with the node mismatch. Whereas traditionally, the small mass splitting between the first two odd-parity resonances was attributed to spin-dependent interactions between constituent quarks, we now understand there is a relativistic aspect to this.

To understand these findings, we thereafter scrutinised the individual interpolating fields that are superposed to create the energy eigenstates. First, the eigenvector components multiplying each interpolating field were examined. Based on these coefficients, it was elucidated how nodes can form in the wavefunctions through superposition of the interpolating fields. With the same superposition applied to each wavefunction component, nodes generated in this manner manifest in all four spinor components of the wavefunctions. This accounts for those states with a matching node structure in their upper and lower components.

It was further demonstrated how both positive-parity and negative-parity states tend to be dominated by one of the \(\chi_1\) or \(\chi_2\) interpolating fields. This categorisation determines the node structure of a given state. For the positive-parity spectrum, the states with matching nodes are dominated by \(\chi_1\), while those with the node mismatch are dominated by \(\chi_2\). Calculating wavefunctions on simple smeared interpolating fields exposed that for positive parity, the \(\chi_2\) interpolator possesses a novel node that is fundamentally built in to its \(s\)-wave (upper) components, but that is absent from its \(p\)-wave (lower) components. This built-in node precisely accounts for those positive-parity states with a node mismatch.

Although unambiguous for positive parity, this distinction was less clear for negative-parity eigenstates, which possess a stronger mixture of \(\chi_1\) and \(\chi_2\) contributions. Still, we discovered that it is simply the interpolating field associated to the strongest component that determines the node structure of the state. For negative parity, it is the \(\chi_1\)-dominated states with the node mismatch, while the \(\chi_2\)-dominated states now have matching nodes. This emphasises that \(\chi_1\) and \(\chi_2\) take on inverse roles for positive and negative parities. Similarly, for the negative-parity wavefunctions it is \(\chi_1\) that displays a built-in node in its \(s\)-wave (lower) components, but not its \(p\)-wave (upper) components. The complete node structure is then understood from a combination of nodes formed through superposition and built-in nodes.

Finally, we explored how certain models for confined spectra using the Dirac equation, including the MIT bag model, give rise to a node mismatch in their negative-parity solutions. From this, we inferred that the uncontracted quark field in the interpolating fields will reveal a built-in node in its negative-parity (lower) components. This explains the \(\chi_1\) built-in node for negative parity. Then, for the \(\chi_2\) interpolating field, the position of \(\gamma_5\) in front of the quark field implies that it is the positive-parity (upper) components that reveal this built-in node. That is, the interpolating field that carries the built-in node for a given parity is determined naturally from the position of \(\gamma_5\) in the interpolator.

There are several interesting avenues for future work. First and foremost, we are currently extending these calculations to nonzero momenta. Due to parity mixing, this requires using the parity-expanded variational analysis technique~\cite{Stokes:2013fgw}. Another direct extension would be to separate the \(u\)-quarks along a common axis, thereby building a complete picture of the nucleon wavefunctions. This has previously been carried out for the upper wavefunction components of positive-parity states~\cite{Roberts:2013oea}, but remains unexplored for the lower components.

In addition, it is of interest to study wavefunctions on centre-vortex-removed fields. Centre vortices~\cite{tHooft:1977nqb, tHooft:1979rtg, Nielsen:1979xu, Greensite:2003bk, Greensite:2016pfc} are a prime candidate for underpinning many nonperturbative phenomena of QCD, including confinement and chiral symmetry breaking. Removal of centre vortices has previously been demonstrated to result in restoration of chiral symmetry~\cite{OMalley:2011aa, Trewartha:2015nna, Trewartha:2017ive}. A natural next step is to extend this to wavefunctions, where it will be interesting to observe how the quark probability distributions are affected by the removal of centre vortices.

The hadronic correlation functions in this work were constructed using the COLA software library, developed at Adelaide University~\cite{Kamleh:2022nqr}.

\begin{acknowledgments}
We thank the PACS-CS Collaboration for making their \(2+1\) flavour configurations available and the ongoing support of the International Lattice Data Grid (ILDG). This work was supported by the Pawsey Supercomputing Centre through the Pawsey Centre for Extreme Scale Readiness (PaCER) program. This work was supported with supercomputing resources provided by the Phoenix HPC service at Adelaide University. This research was undertaken with the assistance of resources and services from the National Computational Infrastructure (NCI), which is supported by the Australian Government. This research was supported by the Australian Research Council through Grant No.\ DP210103706.
\end{acknowledgments}

\bibliography{main}

\end{document}